\documentclass[journal=dce]{CUP-JNL-DTM}%

\usepackage{graphicx}
\usepackage{multicol,multirow}
\usepackage{amsmath,amssymb,amsfonts}
\usepackage{mathrsfs}
\usepackage{amsthm}
\usepackage{rotating}
\usepackage{appendix}
\usepackage{ifpdf}
\usepackage[T1]{fontenc}
\usepackage{newtxtext}
\usepackage{newtxmath}
\usepackage{textcomp}
\usepackage{xcolor}
\usepackage{lipsum}
\usepackage[colorlinks,allcolors=blue]{hyperref}
\usepackage{cancel}
\usepackage{bm}
\usepackage{listings}
\usepackage{subcaption}
\usepackage{cancel}
\usepackage{float}
\usepackage{physics}
\usepackage{comment}
\usepackage{xparse}
\usepackage[normalem]{ulem}
\usepackage{soul}
\usepackage{tikz}
\usepackage{cleveref}

\definecolor{Gray}{gray}{0.9}

\DeclareMathOperator*{\argmin}{arg\,min}

\newcommand{\bea}{\begin{eqnarray}} 
\newcommand{\eea}{\end{eqnarray}}

\def\Rd{\bm{R}}

\newcommand{\T}{^{\mathsf{T}}}
\newcommand{\mT}{^{-\mathsf{T}}}

\def\CC{{C\nolinebreak[4]\hspace{-.05em}\raisebox{.2ex}{\small\bf ++} }}

\definecolor{codegreen}{rgb}{0,0.6,0}
\definecolor{codegray}{rgb}{0.5,0.5,0.5}
\definecolor{codepurple}{rgb}{0.58,0,0.82}
\definecolor{backcolour}{rgb}{0.96,0.96,0.96}

\newlength\wantedwidth
\newcommand{\fakewidth}[2]{%
  \settowidth{\wantedwidth}{\ensuremath{#2}}%
  \makebox[\wantedwidth]{\ensuremath{#1}}%
}

\lstdefinestyle{mystyle}{
    backgroundcolor=\color{backcolour},   
    commentstyle=\color{codegreen},
    keywordstyle=\color{magenta},
    numberstyle=\tiny\color{codegray},
    stringstyle=\color{codepurple},
    basicstyle=\ttfamily\footnotesize,
    breakatwhitespace=false,         
    breaklines=true,                 
    captionpos=b,                    
    keepspaces=true,                 
    numbers=left,                    
    numbersep=5pt,                  
    showspaces=false,                
    showstringspaces=false,
    showtabs=false,                  
    tabsize=2
}

\DeclareRobustCommand{\review}[2]{%
  {%
    \ifcase#1\relax
    \or\color{blue}
    \or\color{green!50!black}
    \or\color{red!60!black}
    \else\color{black}%
    \fi
    #2%
  }%
}

\newif\ifreviewcolors
\reviewcolorstrue 

\newcommand{\ReviewColor}[1]{%
  \ifcase#1\relax
  black%
  \or blue
  \or green!50!black
  \or red!60!black
  \else black%
  \fi
}

\soulregister\cite7
\soulregister\citep7
\soulregister\citet7
\soulregister\ref7
\soulregister\eqref7
\soulregister\autoref7
\soulregister\cref7
\soulregister\Cref7
\soulregister\url7

\newcommand{\reviewstrike}[2]{%
  \begingroup
  \ifmmode
    \ifcase#1\relax
      #2
    \or
      {\renewcommand{\CancelColor}{\color{blue}}\cancel{#2}}
    \or
      {\renewcommand{\CancelColor}{\color{green!50!black}}\cancel{#2}}
    \or
      {\renewcommand{\CancelColor}{\color{red}}\cancel{#2}}
    \else
      #2%
    \fi
  \else
    \ifcase#1\relax
      #2
    \or
      {\setstcolor{blue}\st{#2}}
    \or
      {\setstcolor{green!50!black}\st{#2}}
    \or
      {\setstcolor{red}\st{#2}}
    \else
      #2%
    \fi
  \fi
  \endgroup
}

\theoremstyle{definition}

\numberwithin{equation}{section}

\jname{Data-Centric Engineering}
\articletype{RESEARCH ARTICLE}
\jyear{2026}

\begin{document}

\begin{Frontmatter}

\title[Article Title]{An End-to-End PyTorch Interface for Differentiable PDE Solvers: A RANS Model-Correction Study}

\author[1,2]{Luca Saverio}
\author[1]{Michele Alessandro Bucci}
\author[2]{Gianmarco Farro}
\author[2]{Cédric Content}
\author[3]{Denis Sipp}

\authormark{Luca Saverio \textit{et al}.}

\address[1]{\orgdiv{Digital Sciences \& Technologies Department}, \orgname{Safran Tech}, \orgaddress{\city{Magny-Les-Hameaux}, \postcode{78114}, \state{France}}}

\address[2]{\orgdiv{DAAA}, \orgname{ONERA, Institut Polytechnique de Paris}, \orgaddress{\city{Meudon}, \postcode{92190}, \state{France}}}

\address[3]{\orgdiv{MONHADE, équipe INRIA-ONERA, DSG}, \orgname{ONERA, Institut Polytechnique de Paris}, \orgaddress{\city{Palaiseau}, \postcode{91120}, \state{France}}}

\address{\textbf{Corresponding author:} Luca Saverio; \email{luca.saverio@safrangroup.com}}

\authormark{Luca Saverio et al.}


\abstract{This work presents an end-to-end strategy for solving inverse problems constrained by Partial Differential Equations within a fully differentiable Machine Learning framework.
The proposed formulation provides a unified and user-friendly methodology applicable to a wide range of problems, from data assimilation to closure modeling.
Our approach combines a baseline differentiable PDE solver, which predicts the state $\bm{w}$ from the nonlinear system $\Rd(\bm{w}) = 0$, with a generic additive, parametrized and differentiable correction $\bm{f}_{{\bm{\vartheta}}}(\bm{w})$, with trainable parameters ${\bm{\vartheta}}$.
We show how to optimize ${\bm{\vartheta}}$ within a fully differentiable Python workflow by reformulating the PDE as an implicit layer, enabling its integration into arbitrary objective functions, while leveraging PyTorch's automatic differentiation graph.
The method is demonstrated on the Reynolds-Averaged Navier-Stokes equations for compressible flows, where the closure term, or a portion of it, is modeled using trainable parameters or a Neural Network.
The first application considers the 2D NASA Wall-Mounted Hump test case, where a production-term parameter is optimized against time-averaged LES data. 
A second application is carried out on the VKI LS-59 turbine blade, where the Spalart-Allmaras eddy viscosity field is reconstructed through the optimization of a trainable spatial field.
A dataset is generated starting from the VKI LS-59 turbine blade geometry using the differentiable BROADCAST solver with the Spalart-Allmaras turbulence model.
The results highlight the flexibility of the framework, showing its applicability beyond turbulence modeling to a broader class of physics-informed PDE-constrained problems with data-driven components.}

\end{Frontmatter}

\section*{Impact Statement}
This work introduces a fully differentiable framework that integrates data-driven closure models directly into the solution of nonlinear PDE systems. By treating the equations as implicit layers and enabling end-to-end optimization, the approach unifies simulation and learning within a single workflow. This reduces computational duplication and ensures that physical constraints remain embedded throughout training. Applied to the Reynolds-Averaged Navier-Stokes equations for compressible flows, the method demonstrates that differentiable solvers and Neural Networks can be combined in a stable and scalable manner. Beyond this specific case, the framework is broadly applicable to PDE-driven problems with data-informed components, supporting the development of next-generation hybrid modeling strategies at the interface between scientific computing and Machine Learning.


\newpage

\section{Introduction}
\label{sec:Intro}

Partial Differential Equations (PDEs) are at the heart of mathematical modeling in physics and engineering. They describe how physical quantities such as temperature, pressure, and velocity in fluids evolve in space and time, encapsulating conservation laws such as those of mass, momentum, and energy. The complexity of these equations often renders analytical solutions intractable, especially in realistic geometries or in the presence of non-linearities. As a result, numerical methods have become indispensable tools for solving PDEs, giving rise to a wide range of simulation frameworks across scientific disciplines.

From an engineering perspective, the development of Computational Fluid Dynamics (CFD) codes is crucial for predicting the forces needed to size and manufacture components, especially in high-performance applications such as turbines in jet engines. This process relies on advanced computational and experimental techniques to optimize performance.

The design of aerodynamic profiles is an iterative process, starting with data acquisition, followed by simulations and prototype testing to refine the design. Modern methods integrate optimization algorithms, more recently Machine Learning (ML)~\citep{Wu2018, End-to-end}, to balance aerodynamics, thermodynamics, and structural mechanics, improving the overall design.

In recent years, ML has gained increasing attention as a powerful tool in applied mathematics, particularly for problems traditionally addressed through numerical analysis. Recent works have investigated the integration of ML directly into the numerical discretization of Finite Volume (FV) solvers. Data-driven learned discretizations for hyperbolic conservation laws have been proposed to replace classical gradient or flux reconstructions with Neural Networks (NNs) trained on high-resolution data, achieving improved accuracy on coarse meshes and in the presence of shocks~\citep{deromémont2024datadrivenlearneddiscretizationapproach,deromémont2025datadrivenadaptivegradientrecovery}. These approaches effectively modify the internal structure of the CFD solver itself and are typically implemented (including the PDE solver) in dedicated ML-based frameworks. More generally, structure-preserving neural architectures such as volume-preserving transformers have been introduced to ensure stability and long-term consistency in learned dynamical systems~\citep{brantner2024volumepreservingtransformerslearningtime}.

ML's flexibility in approximating complex functions has led to the development of surrogate models \citep{Kutz2017, casenave2025physicslearningaidatamodelplaid, Catalani2025} that can either completely replace or be integrated into and used to accelerate classical solvers, especially when repeated evaluations are required, such as in (shape) optimization or uncertainty quantification \citep{chu2024physicsconstraineddeeplearning}. Beyond surrogate modeling, ML techniques have also been used to correct or augment existing numerical models, by learning discrepancies between model predictions and high-fidelity data. In the context of PDEs, NNs have been employed both to represent solutions directly, as in Physics-Informed Neural Networks (PINNs) \citep{RAISSI2019686}, and to infer hidden physical quantities or closure terms, offering new paradigms for solving inverse problems or discovering governing laws from data \citep{End-to-end}. These approaches open new perspectives for hybrid modeling strategies that combine domain knowledge with data-driven learning.

CFD mainly focuses on solving the Navier-Stokes (NS) equations to study the behavior and interactions of fluids. Specifically, for turbulent flows, the Reynolds-Averaged Navier-Stokes (RANS) equations introduce turbulence models to close the system, each model representing a compromise between fidelity and computational feasibility. The choice and accuracy of these turbulence models critically influence the reliability of CFD simulations, especially in the design and optimization of high-performance components.

The accurate simulation of turbulent flows remains one of the major challenges in CFD. Traditional RANS models provide a cost-efficient approach by introducing closure relations for the Reynolds stresses, yet their performance strongly depends on empirical assumptions and model calibration.
Classical turbulence closures, such as the $k$–$\varepsilon$, $k$–$\omega$ and Spalart-Allmaras (SA) models \citep{nasaTurbulenceModeling}, while robust and computationally efficient, are known to fail in the prediction of separated or highly anisotropic flows.

The closure modeling problem is a generic feature of nonlinear multiscale PDE systems, arising whenever a reduced or averaged description is sought. A closely related viewpoint emerges in data assimilation, where the full state of a nonlinear multiscale system is reconstructed from partial observations by solving an inverse problem, typically through the introduction of correction or control terms. In variational approaches, these corrections are adjusted to enforce consistency between the model and the data \citep{talagrand1997assimilation}. From this perspective, data assimilation can be interpreted as an implicit closure strategy, in which the inferred corrections encode the effects of unresolved scales or model errors; once learned or parameterized as functions of the resolved state, they can be recast as explicit closure models, thereby bridging inverse problems and reduced-order modeling. \cite{duraisamy2019turbulence} provided a comprehensive overview of data assimilation (field-inversion) and ML-augmented RANS modeling, highlighting the opportunities and difficulties in integrating data-driven corrections within physics-based solvers.
Furthermore, a more recent view made by \cite{Girimaji2023} emphasized that, despite the promising results, ML-driven turbulence modeling must still tackle issues of generalization, interpretability, and physical consistency.

In the literature, various ML-based approaches have been proposed to enhance RANS closures \citep{ZHAO2020109413, Zhu2021}. For instance, \cite{Sanhueza2022} employed NNs to model the effects of variable fluid properties on turbulence transport, while \cite{Chen2024} trained data-driven closures for the $k$–$\omega$ SST model, learning discrepancies between RANS and high-fidelity data to improve mean-field predictions. \cite{Aly2024} introduced a Deep Learning (DL) model to predict the eddy viscosity field, training directly on the eddy viscosity, yet still achieving improved accuracy in bluff-body simulations, opening the way for turbulent viscosity prediction using NNs. Surrogate modeling has also been applied to turbulence closure, for instance \cite{Schlichter2024} benchmarked surrogate models for RANS closures based on the $k$–$\omega$ SST model.

Despite the progress, most of these methods treat the ML model as an external correction or surrogate, rather than embedding it within an end-to-end differentiable simulation pipeline, as done in \cite{End-to-end}. In addition, most ML-RANS studies rely on incompressible solvers. This preference arises because, in the incompressible regime, the Reynolds stress tensor admits an exact decomposition following \citet{Pope_1975}, as commonly exploited in recent works \citep{ZHANG2023108632, End-to-end}. Furthermore, the incompressible NS equations admit a continuous adjoint formulation, which substantially simplifies the derivation and computation of adjoint-based gradients.

A complementary line of research has explored the use of Graph Neural Networks (GNNs) as data-driven correction models defined directly on unstructured computational meshes. Recent works by Quattromini et al.\ have proposed physics-constrained GNN architectures for mean-flow reconstruction and data assimilation in CFD~\citep{quattromini2025activelearningdataassimilationclosures,quattromini2025meanflowdataassimilation}. In these methods, the GNN is trained to predict the forcing or closure term that brings a baseline RANS or mean-flow solution into agreement with high-fidelity or experimental data. The graph-based formulation naturally accommodates arbitrary mesh connectivity and allows the use of solver-derived sensitivities during training.

Industrial CFD codes are typically developed over time spans of 10 to 20 years within High-Performance Computing (HPC) environments, often consisting of Python interfaces wrapped around legacy Fortran or \CC kernels to ensure high performance on large clusters. It is therefore desirable to design a pipeline that enables seamless integration of such CFD solvers with modern ML frameworks like PyTorch, as similarly done by \cite{Atkinson2025}.

In the context considered in this work, the PDE solver—specifically a CFD solver—must be treated as a standard PyTorch module, supporting both forward and backward differentiation. At the same time, we aim to preserve the mature, highly optimized numerical algorithms already available in industrial CFD codes—for example, Newton-based nonlinear relaxation schemes and preconditioned GMRes methods \citep{GMRes} for large-scale linear systems. This motivates handling the CFD solver as an implicit layer within PyTorch, with full support for automatic differentiation in both directions.

Furthermore, we aim to introduce ML operators, such as NNs, in an end-to-end structure inside the CFD-PyTorch module. Therefore, once the parameters, or the NN, have been trained, we would still like to rely on the dedicated algorithms embedded in the CFD code to solve the resulting hybrid PDE–ML system efficiently. This requires incorporating the contributions of the trainable model component directly into the Jacobian used for implicit time integration or steady-state solution. The present paper introduces a pipeline that fulfills these objectives.

Here, we demonstrate the approach by coupling PyTorch with the compressible CFD solver BROADCAST~\citep{Poulain2023}, which typifies industrial CFD codes built around a Python interface wrapping Fortran kernels. Algorithmic Differentiation (AD) is performed using the software Tapenade~\citep{Tapenade}, enabling efficient evaluation of vector–Jacobian products at a reasonable computational cost~\citep{maugars:hal-03759125}. This capability is essential for the practical implementation of adjoint-based optimization.

NNs are adopted as ML models due to the universal approximation theorem \citep{nishijima2021universalapproximationtheoremneural}, which guarantees their ability to approximate a broad class of nonlinear functions. In addition, NNs can be efficiently trained using gradient-based optimization methods. Their capability to handle heterogeneous datasets, together with their flexibility and generality, makes them a robust and widely applicable strategy for processing and predicting data arising from numerical simulations. This approach bridges the gap between physics-based solvers and data-driven learning, extending ML-RANS methodologies towards fully integrated optimization pipelines for aerodynamic design. 

This pipeline is applied to the improvement of turbulence models for the RANS equations. In this setting, we introduce an end-to-end differentiable framework that couples the BROADCAST compressible RANS solver with PyTorch, enabling the optimization of NN weights directly within the solver. This approach allows turbulence quantities, such as the eddy viscosity, to be optimized through algorithmic differentiation, thereby providing a fully consistent and automated route for data-driven closure modeling.

The pipeline is first tested on the 2D NASA Wall-Mounted Hump separated flow \citep{jespersen2016, nasaNASAWallMounted} to optimize a control parameter in the production term of the transport equation of the turbulent variable of the SA model \citep{FANIZZA2025115984, CATO2023106054}. Subsequently, to further explore the capabilities of the framework, it is applied to the VKI LS-59 turbine blade case~\citep{Hercus2011} with transonic flow to perform an optimization directly on the eddy viscosity, and finally on a dataset generated starting from the benchmark case. This comprehensive approach aims to enhance the predictive accuracy of turbulence models, thereby contributing to more efficient and reliable aerodynamic simulations.

The structure of this paper is as follows. First, the general formulation of the tackled PDEs is introduced, together with a discussion of their abstract representation and closure requirements (\S~\ref{sec:MLPDEs}). Next, the proposed optimization framework is described, outlining the possible objective functions and distinguishing between cases where the full state variable is available and cases where it is unknown and must be computed (\S~\ref{sec:Opt}). The specific problem under investigation is then presented as a benchmark to demonstrate the capabilities of the developed framework, followed by a detailed presentation of the CFD equations in use and closure models (\S~\ref{sec:RANS}). Subsequently, the technical implementation is discussed, with particular emphasis on the PyTorch-based routines and the design of custom differentiable modules (\S~\ref{sec:PyTorchB}). The application cases are presented by first describing the considered geometries and then the adopted optimization strategies (\S~\ref{sec:Applications}). The generation of the training dataset is described, with particular emphasis on the variations in geometry and flow conditions (\S~\ref{sec:DATAGen}). Each case is followed by a detailed presentation and analysis of the results obtained with the different approaches. Finally, conclusions are drawn and perspectives for future developments are outlined.

\section{Enhancing Partial Differential Equations resolution using Machine Learning}
\label{sec:MLPDEs}

We start by considering a baseline model $\mathcal{R}(\bm{\omega}) = 0$, where $\bm{\omega}$ denotes the state variables of interest. Discretization on a mesh using Finite Difference (FD), Finite Element (FE), or FV methods leads to a system $\Rd(\bm{w}) = 0$, where $\bm{w}$ represents the discretized state.
For example, with a FV method with $ m $ variables on a structured mesh of size $ n_i \times n_j $, the total number of degrees of freedom of the state $N=m \times n_in_j$. 
To enhance the model accuracy, a discrete additive correction term $\bm{f}(\bm{w}) = \bm{f}_{\bm{\vartheta}}(\bm{w})$, parameterized by coefficients ${\bm{\vartheta}}$, is introduced. These parameters are determined by minimizing an objective functional $\mathcal{J}(\bm{w})$, defined, for instance, from sparse experimental observations or high-fidelity reference data.

Then, the hybrid discrete model can be written as
\begin{equation}
\label{eq:generic}
    \bm{R}(\bm{w}) + \bm{f}_{\bm{\vartheta}}(\bm{w}) = 0 \, ,
\end{equation}
where $\bm{w} \in \mathbb{R}^{N}$,  $\bm{R}(\bm{w}): \mathbb{R}^{N} \rightarrow \mathbb{R}^{N}$ and $\bm{f}_{{\bm{\vartheta}}}(\bm{w}): \mathbb{R}^{N} \rightarrow \mathbb{R}^{N}$. In the following, the implicit time discretization scheme of the hybrid model is assumed to be:

\begin{equation}
\label{eq:timeint}
    \frac{\bm{w}_{n+1} - \bm{w}_{n}}{\Delta t}
    + \bm{R}(\bm{w}_{n+1})
    + \bm{f}_{{\bm{\vartheta}}}(\bm{w}_{n+1}) = 0 \, .
\end{equation}

On the other hand, the correction $\bm{f}_{\bm{\vartheta}}(\bm{w})$ may represent either a local or a non-local closure. In the local case, it can be written as $ \bm{f}_{\bm{\vartheta}}(\bm{w}) = \bm{f}(\bm{w},\bm{\alpha}_{\bm{\vartheta}} \circ \bm{\phi}(\bm{w}))$, where $\bm{\phi}(\bm{w})$ denotes a set of local features and $\bm{\alpha}_{\bm{\vartheta}}$ is a parametric mapping that returns a local correction field $\bm{\alpha}$. In the non-local case, the correction takes the form $\bm{f}_{\bm{\vartheta}}(\bm{w}) = \bm{f}\big(\bm{w}, \bm{\alpha}_{\bm{\vartheta}}(\bm{w})\big)$, where $\bm{\alpha}_{\bm{\vartheta}}$ depends on the global or neighborhood state and can therefore capture non-local interactions between cells. In practice, $\bm{\alpha}_{\bm{\vartheta}}$ may be instantiated, for instance, as a MultiLayer Perceptron (MLP) in the local setting or as a GNN in the non-local one, although the present formulation remains independent of the specific choice of parametrization.

In the following, we address two algorithmic problems that combine a PDE solver and trainable parameters.
First (\S~\ref{sec:solve}), for given parameters  ${\bm{\vartheta}}$, we seek the state $\bm{w}$ that solves \eqref{eq:generic}, or:
\begin{equation}
\bm{w} = \argmin_{\bm{w}} \; || \Rd(\bm{w}) + \bm{f}_{\bm{\vartheta}}(\bm{w}) ||^2.
\end{equation}
This requires adapting the solution strategies of the PDE solver underlying $\Rd$ so that the correction term $\bm{f}_{\bm{\vartheta}}(\bm{w}) $ is properly taken into account.

Second (\S~\ref{sec:Opt}), for a given cost functional $\mathcal{J}(\bm{w},{\bm{\vartheta}})$, we seek parameters ${\bm{\vartheta}} $ such that
\begin{equation}
\begin{aligned}
{\bm{\vartheta}} &= \argmin_{\bm{\vartheta}} \; \mathcal{J}(\bm{w},{\bm{\vartheta}}) \\
&\text{s.t. } \Rd(\bm{w}) + \bm{f}_{\bm{\vartheta}}(\bm{w}) = 0.
\end{aligned}
\label{eq:min}
\end{equation}
The objective is to optimize the weights ${\bm{\vartheta}} $ (e.g., within PyTorch), and the key question becomes: what functionalities must the PDE code $ \Rd(\bm{w}) $ provide to enable transparent and smooth forward and backward propagations through the PDE solver?

\subsection{Solving for the state for fixed parameters} \label{sec:solve}

For context, it is important to stress that in DL, models are typically constructed as compositions of elementary explicit functions, called layers, where each layer produces its output from its input through a closed-form evaluation. In contrast, the solution of PDEs often relies on the numerical solution of implicit nonlinear systems, where the state is defined as the solution of a residual equation rather than by an explicit formula. This fundamental difference motivates the introduction of implicit operators within differentiable learning frameworks, as done in \cite{Bai2019}. Having defined the concepts of explicit and implicit layers, given a fixed set of parameters, the hybrid problem \eqref{eq:generic} can be solved either by a residual minimization technique (an explicit operation) or a Newton-like relaxation method (an implicit system).

\subsubsection{Residual minimization with an explicit layer} \label{sec:residual}

The first strategy to solve problem \eqref{eq:generic} avoids forming the Jacobian explicitly by recasting the state solve as a residual minimization problem. In this formulation, the state is obtained by minimizing the norm of the residual, typically through an iterative method that requires only the evaluation of the residual itself and back-propagation through the PDE solver $(\partial_{\bm{w}} \Rd)\T$ and through the closure term $(\partial_{\bm{w}} \bm{f}_{\bm{\vartheta}})\T$. Then, the functional to be minimized is fixed as:
\begin{equation}
    \mathcal{J}(\bm{w})=\|\Rd(\bm{w})+\bm{f}_{\bm{\vartheta}}(\bm{w})\|^2_Q = \left[\Rd(\bm{w}) + \bm{f}_{\bm{\vartheta}}(\bm{w})\right]\T \bm{Q}\left[\Rd(\bm{w}) + \bm{f}_{\bm{\vartheta}}(\bm{w})\right],
\end{equation}
and the differential of $\mathcal{J}(\bm{w})$ can be computed as:
\begin{align}
    \delta \mathcal{J} &= 2\left[\Rd(\bm{w}) + \bm{f}_{\bm{\vartheta}}(\bm{w})\right]\T \bm{Q}\left[\partial_{\bm{w}} \Rd(\bm{w})+\partial_{\bm{w}} \bm{f}_{\bm{\vartheta}}(\bm{w})\right] \delta \bm{w}, \notag \\
    &= \left(2\left[(\partial_{\bm{w}} \Rd(\bm{w})+\partial_{\bm{w}} \bm{f}_{\bm{\vartheta}}(\bm{w}))\T \right]\bm{Q}\left[\Rd(\bm{w})+\bm{f}_{\bm{\vartheta}}(\bm{w})\right]\right)\T \bm{Q}^{-1}\bm{Q} \delta \bm{w},  \notag \\
    &=\left<2\bm{Q}^{-1}\left[(\partial_{\bm{w}} \Rd(\bm{w}) +\partial_{\bm{w}} \bm{f}_{\bm{\vartheta}}(\bm{w}))\T \right]\bm{Q}\left[\Rd(\bm{w})+\bm{f}_{\bm{\vartheta}}(\bm{w})\right], \delta \bm{w} \right>_Q \label{eq:dJ}
\end{align}
where $ \| \cdot \|_Q^2=\left<\cdot,\cdot \right>_Q$ and 
$\left<\bm{w}_1,\bm{w}_2 \right>_Q=\bm{w}_1\T \bm{Q}\bm{w}_2$ being the discrete inner product, and where $\bm{Q}\in\mathbb{R}^{N\times N}$ is a symmetric positive-definite matrix defining this discrete inner-product in the state space. In practice, $\bm{Q}$ involves the continuous $ L^2 $ inner-product and accounts for the mesh discretization and for the relative scaling of the different state variables. For instance, in a FV discretization, $\bm{Q}$ can be chosen as a diagonal matrix containing the cell volumes repeated for each state variable. This definition ensures that the residual norm corresponds to a physically meaningful weighted $L^2$ norm over the computational domain.

The forward and backward 
operators can be sketched as:
\begin{alignat}{3}
    \bm{w} \; &\longrightarrow& \fakewidth{\fcolorbox{black}{white}{$ \|\Rd(\bm{w})+\bm{f}_{\bm{\vartheta}}(\bm{w})\|^2_Q $}}{\fcolorbox{black}{white}{$2\bm{Q}^{-1}\leftarrow \left(\partial_{\bm{w}} \bm{f}_{{\bm{\vartheta}}} +\partial_{\bm{w}} \Rd\right)\T \leftarrow \bm{Q}\leftarrow\left(\Rd(\bm{w})+\bm{f}_{\bm{\vartheta}}(\bm{w})\right)$}} &\longrightarrow &\mathcal{J} \\
        \delta \bm{w} \; &\longleftarrow& \; \fcolorbox{black}{white}{$2\bm{Q}^{-1}\leftarrow \left(\partial_{\bm{w}} \bm{f}_{{\bm{\vartheta}}} +\partial_{\bm{w}} \Rd\right)\T \leftarrow \bm{Q}\leftarrow\left(\Rd(\bm{w})+\bm{f}_{\bm{\vartheta}}(\bm{w})\right)$} \; &\longleftarrow \delta &\mathcal{J},
\end{alignat}
where all boxes, handled by a combination of the PDE solver and the automatic differentiation framework (e.g., PyTorch), are explicit since only explicit evaluation operations are involved.

Such an approach, which is similar to PINNs \citep{RAISSI2019686}, reduces the memory footprint (there is no matrix to invert) and simplifies the implementation, making it attractive when the Jacobian is difficult or expensive to compute. However, it may converge more slowly than Newton-type schemes, and its efficiency depends on the choice of minimization algorithm and the conditioning of the residual landscape.

\subsubsection{Relaxation method with an implicit layer} \label{sec:newton}

As an alternative, problem \eqref{eq:generic} may be solved iteratively by the following relaxation method:
\begin{equation}\label{eq:Newtonstep}
    \bm{w}_{n+1}=\bm{w}_{n}+\delta \bm{w}, \;\;\;\; \text{with}\;\;\;\; \left(\frac{\mathbf{I}}{\Delta t}+ \partial_{\bm{w}}\Rd + \partial_{\bm{w}}\bm{f}_{\bm{\vartheta}} \right)\delta \bm{w} = -\left(\Rd(\bm{w}_{n}) + \bm{f}_{\bm{\vartheta}}(\bm{w}_{n})\right).
\end{equation}
The pseudo time-step $\Delta t$ is small at the beginning of the iterations to accommodate for the initial condition, then it increases to reach $ \Delta t \rightarrow \infty$, where we recover the quadratic convergence properties of the Newton method \citep{Newton}. The pseudo time-step $ \Delta t $ can be derived from an adaptive stability criterion that evolves with the residual norm, see for example \citet{crivellini2011implicit} in the case of advection dominated systems with the local $CFL = c\Delta t/\Delta x$ condition ($\Delta x$ being the local mesh size and $ c $ the speed of the fastest waves). Note that using a Newton strategy requires the full Jacobian $\partial_{\bm{w}} \Rd + \partial_{\bm{w}} \bm{f}_{\bm{\vartheta}} $ to be invertible at the targeted fixed point, which excludes being in the vicinity of saddle-node, pitchfork or transcritical bifurcation thresholds.

This method requires assembling and inverting the sparse Jacobian by exploiting the stencil-based independence of the numerical scheme. The sparsity pattern of the Jacobian is inferred from the discretization stencil, allowing the use of directional test vectors to probe multiple independent degrees of freedom simultaneously. As a consequence, several Jacobian entries are computed at once using the same test vector, and the corresponding matrix-vector products are performed in parallel, resulting in an efficient sparse assembly procedure.

The Jacobian can be decomposed into two contributions: the part associated with the PDE solver, $\partial_{\bm{w}} \Rd$, and the component arising from the linearization of the ML model, $\partial_{\bm{w}} \bm{f}_{{\bm{\vartheta}}}$. When $\bm{f}_{{\bm{\vartheta}}}$ is represented by, or contains, a NN, it is essential that the spatial support (or receptive field) of its linear operations does not exceed the stencil width of the underlying numerical scheme. Otherwise, additional couplings would be introduced beyond the assumed sparsity pattern, leading to an incorrect Jacobian structure. This aspect is further discussed in Appendix~\ref{sec:CNN}. To relax the constraints on the NN architecture, matrix-free methods could be considered. However, their convergence is typically more challenging without ad hoc preconditioning. This aspect could be explored in future works.

Both methods, explicit and implicit, are compatible with the differentiable framework used in this work, and the choice between them depends on the trade-off between computational cost, implementation complexity, and convergence properties.

\subsection{Optimization framework}
\label{sec:Opt}

In the following, we aim at optimizing the parameters ${\bm{\vartheta}}$ of a correction term $ \bm{f}_{\bm{\vartheta}}(\bm{w}) $.
Similarly to how we previously introduced $\bm{Q}$, we introduce a symmetric positive-definite matrix $\bm{M}$ defining the natural inner-product in the parameter space:
\begin{align}
    \left< {\bm{\vartheta}}_1, {\bm{\vartheta}}_2 \right>_M&={\bm{\vartheta}}_1\T \bm{M} {\bm{\vartheta}}_2.
\end{align}

Two classical situations arise:

\begin{enumerate}
\item In the case of data assimilation, we assume 
\begin{equation} \label{eq:da}
    \bm{f}_{{\bm{\vartheta}}=\bm{\alpha}}(\bm{w})=\bm{f}(\bm{w},\bm{\alpha}),
\end{equation}
meaning that the correction $\bm{f}_{\bm{\vartheta}}$ is a function of a volumetric source term $ \bm{\alpha} $ defined on the computational mesh and handled by the discretization of a continuous operator. In such a case: $ {\bm{\vartheta}}=\bm{\alpha}$ and the scalar product $\bm{M}$ may be chosen to discretize the continuous $ L^2 $ inner product, consistently with the inner product used for the state variables.
\item In the case of closure, we assume that
\begin{equation} \label{eq:closure}
    \bm{f}_{\bm{\vartheta}}(\bm{w})=\bm{f}(\bm{w},\bm{\alpha}_{\bm{\vartheta}}(\bm{w})),
\end{equation}
meaning that the correction $\bm{f}_{\bm{\vartheta}}(\bm{w})$ is a function of the above-introduced parameter $ \bm{\alpha} $, which now depends on the state $ \bm{w}$ via a parametric model with parameters ${\bm{\vartheta}}$, $\bm{\alpha}_{\bm{\vartheta}}(\bm{w}) $. In this case, the natural choice for the scalar product is $\bm{M}=\bm{I}$, since these parameters are not associated with a spatial discretization. 
Note that:
\begin{align}
        \partial_{\bm{w}} \bm{f}_{\bm{\vartheta}}(\bm{w})&=\partial_{\bm{w}}\bm{f}(\bm{w},\bm{\alpha}_{\bm{\vartheta}}(\bm{w}))+\partial_{\bm{\alpha}}\bm{f}(\bm{w},\bm{\alpha}_{\bm{\vartheta}}(\bm{w})) \; \partial_{\bm{w}} \bm{\alpha}_{\bm{\vartheta}}(\bm{w}), \\
    \partial_{\bm{\vartheta}} \bm{f}_{\bm{\vartheta}}(\bm{w})&=\partial_{\bm{\alpha}}\bm{f}(\bm{w},\bm{\alpha}_{\bm{\vartheta}}(\bm{w})) \; \partial_{\bm{\vartheta}} \bm{\alpha}(\bm{w}),
\end{align}
in which all derivatives are handled by the PDE solver, except for $\partial_{\bm{\vartheta}} \bm{\alpha}(\bm{w})$, which is treated by the automatic differentiation framework (e.g., PyTorch), as it corresponds to the differentiation of the ML model.
\end{enumerate}

\subsubsection{Full-state \texorpdfstring{$\bm{w}=\bm{w}_m$}{w=wm} and explicit layer}

In this section, the full state variable $\bm{w}$ is assumed to be known at the training stage, $\bm{w}=\bm{w}_m$. For example, this would be the case if a high-fidelity solver were used to obtain the training data.
The optimization problem \eqref{eq:min} may then be simplified and replaced by minimizing a cost-functional based on the residue:
\begin{align}
    \mathcal{J}({\bm{\vartheta}}) &:= \| \Rd(\bm{w}_m)+\bm{f}_{{\bm{\vartheta}}}(\bm{w}_m) \|^2_Q.
\end{align}

The gradient of the cost-functional can then be obtained straightforwardly:
\begin{align}
    \delta\mathcal{J}&=\left<2\bm{M}^{-1}(\partial_{\bm{\vartheta}} \bm{f}_{\bm{\vartheta}})\T \bm{Q}\left(\Rd(\bm{w}_m)+\bm{f}_{\bm{\vartheta}}(\bm{w}_m)\right), \delta {\bm{\vartheta}} \right>_{\bm{M}}.
\end{align}
When using an optimizer based on the $L^2$ norm, it is essential that the gradient supplied to it lives in a Euclidean space. This inconsistency may result in suboptimal convergence, unstable optimization steps, or biased updates of the parameters. Hence, if $ \bm{M}\neq I$, we need to introduce the variable $\tilde{{\bm{\vartheta}}} = \bm{N}{\bm{\vartheta}}$, where $\bm{M} = \bm{N}\T \bm{N}$ (Cholesky decomposition of a 
diagonal positive definite matrix), so that 
\begin{align}
    \left< {\bm{\vartheta}}_1,{\bm{\vartheta}}_2\right>_M=\left< \tilde{{\bm{\vartheta}}_1},\tilde{{\bm{\vartheta}}_2}\right>_2,
\end{align}
where the subscript $ _2 $ refers to the Euclidean norm.
The correct gradient to be supplied to the Euclidean optimizer may then be obtained through:
\begin{align}
    \delta\mathcal{J} &=\left[2\bm{M}^{-1}(\partial_{\bm{\vartheta}} \bm{f}_{\bm{\vartheta}})\T \bm{Q}\left(\Rd(\bm{w}_m)+\bm{f}_{\bm{\vartheta}}(\bm{w}_m)\right)
    \right]\T \bm{N}\T \bm{N} \delta {\bm{\vartheta}} \notag \\
    &=\left[2\bm{N}^{-1}(\partial_{\bm{\vartheta}} \bm{f}_{\bm{\vartheta}})\T \bm{Q}\left(\Rd(\bm{w}_m)+\bm{f}_{\bm{\vartheta}}(\bm{w}_m)\right)
    \right]\T \delta \tilde{{\bm{\vartheta}}} \notag \\
    &=\left<2\bm{N}^{-1}(\partial_{\bm{\vartheta}} \bm{f}_{\bm{\vartheta}})\T \bm{Q}\left(\Rd(\bm{w}_m)+\bm{f}_{\bm{\vartheta}}(\bm{w}_m)\right)
    ,  \delta \tilde{{\bm{\vartheta}}}\right>_2. \label{eq:dJN}
\end{align}

In the present case, if $ {\bm{\vartheta}}$ represents the weights of a neural operator, the forward and backward operators can be sketched as:

\begin{alignat}{3}
    {\bm{\vartheta}} &\longrightarrow& \fakewidth{\fcolorbox{black}{white}{$ \|\Rd(\bm{w}_m)+\bm{f}_{\bm{\vartheta}}(\bm{w}_m)\|^2_Q $}}{\fcolorbox{black}{white}{$2\bm{N}^{-1}\leftarrow \left(\partial_{\bm{\vartheta}} \bm{f}_{{\bm{\vartheta}}}\right)\T \leftarrow \bm{Q}\leftarrow\left(\Rd(\bm{w}_m)+\bm{f}_{\bm{\vartheta}}(\bm{w}_m)\right)$}} &\longrightarrow &\mathcal{J} \label{opt:forward_explicit} \\
        \delta {\bm{\vartheta}} &\longleftarrow& \; \fcolorbox{black}{white}{$2\bm{N}^{-1}\leftarrow \left(\partial_{\bm{\vartheta}} \bm{f}_{{\bm{\vartheta}}}\right)\T \leftarrow \bm{Q}\leftarrow\left(\Rd(\bm{w}_m)+\bm{f}_{\bm{\vartheta}}(\bm{w}_m)\right)$} \; &\longleftarrow \delta &\mathcal{J}, \label{opt:backward_explicit}
\end{alignat}
where the boxes are handled by a combination of the PDE solver and the automatic differentiation framework. All layers are explicit in the present case since simple (explicit) evaluation operations are involved in all boxes.

\subsubsection{Partial-state and implicit layer}
\label{sec:FixedPoint}

In the case where the full state variable $\bm{w}$ is not known but only partial state information collected in a cost functional $ \mathcal{J}(\bm{w})$ is available, we can directly tackle the optimization problem in \eqref{eq:min} by considering the Lagrangian:
\begin{equation}\label{eq:Lagrangian}
    \mathcal{L}({\bm{w}},\bm{\lambda}, {\bm{\vartheta}}) = \mathcal{J}({\bm{w}},{\bm{\vartheta}}) + \bm{\lambda}\T (\Rd({\bm{w}}) + \bm{f}_{{\bm{\vartheta}}}(\bm{w}))\, ,
\end{equation}

The Karush-Kuhn-Tucker (KKT) optimality conditions are obtained by setting the derivatives with respect to the direct state $\bm{w}$ and the adjoint state $\bm{\lambda} \in \mathbb{R}^N$ of the Lagrangian to zero:
\begin{align}
    \partial_{\bm{\lambda}} \mathcal{L} \delta \bm{\lambda} &= 0 ~\Rightarrow~ \Rd(\bm{w}) + \bm{f}_{\bm{\vartheta}}(\bm{w}) = 0 \,,
    \label{eq:KKT1} \\
    {\partial_{\bm{w}} \mathcal{L}} \delta{\bm{w}}&= 0 ~\Rightarrow~ \bm{\lambda}\T \left( 
        \partial_{\bm{w}} \Rd
        + \partial_{\bm{w}} \bm{f}_{{\bm{\vartheta}}}
    \right) 
    = -\partial_{\bm{w}} \mathcal{J} \,,
    \label{eq:KKT2}\\
    \partial_{\bm{\vartheta}} \mathcal{L} \delta {\bm{\vartheta}} & = \delta \mathcal{J} ~\Rightarrow~ \left(\partial_{\bm{\vartheta}} \mathcal{J} + \bm{\lambda}\T \partial_{\bm{\vartheta}} \bm{f}_{{\bm{\vartheta}}}\right) \delta {\bm{\vartheta}} = \delta \mathcal{J} \,, \label{eq:KKT3}
\end{align}
then, having computed the solution $\bm{w}$ of \eqref{eq:KKT1}, we can use it to compute the Lagrange multiplier from \eqref{eq:KKT2}. Finally, by substituting $\bm{\lambda}$ in \eqref{eq:KKT3}, it is possible to solve for the optimal value of ${\bm{\vartheta}}$ with a gradient descent strategy, where the gradients are computed as:

\begin{equation}\label{eq:dL}
    \delta\mathcal{J}=\left< \bm{M}^{-1}\left(\left(\partial_{\bm{\vartheta}} \mathcal{J}\right)\T -  \left(\partial_{\bm{\vartheta}} \bm{f}_{{\bm{\vartheta}}}\right)\T \left(\partial_{\bm{w}} \Rd+\partial_{\bm{w}} \bm{f}_{\bm{\vartheta}}\right)\mT \left(\partial_{\bm{w}} \mathcal{J}\right)\T \right),\delta {\bm{\vartheta}} \right>_M\,,
\end{equation}
where $ \bm{M} $ is again the natural inner-product depending on the nature of the optimized parameters $ {\bm{\vartheta}} $.

In the present case, the forward and backward operators of the considered implicit layer can be sketched as:
\begin{alignat}{5}
    {\bm{\vartheta}} &\longrightarrow \fakewidth{\fcolorbox{black}{lightgray}{$ \Rd(\bm{w})+\bm{f}_{\bm{\vartheta}}(\bm{w})=0$, see \S~\ref{sec:solve}}}{\fcolorbox{black}{lightgray}{$\bm{M}^{-1}\leftarrow\left(\partial_{\bm{\vartheta}} \bm{f}_{{\bm{\vartheta}}}\right)\T \leftarrow \bm{\lambda} \leftarrow  - \left(\partial_{\bm{w}} \Rd+\partial_{\bm{w}} \bm{f}_{\bm{\vartheta}}\right)\mT $}} &\longrightarrow& \fakewidth{\bm{w}}{\; \delta\bm{w}} &\longrightarrow \fakewidth{\fcolorbox{black}{white}{$ \mathcal{J}(\bm{w}) $}}{ \fcolorbox{black}{white}{$ \left(\partial_{\bm{w}}\mathcal{J}\right)\T$}} &\longrightarrow &\mathcal{J} \label{opt:forward_implicit} \\
        \delta {\bm{\vartheta}} &\longleftarrow  \fcolorbox{black}{lightgray}{$\bm{M}^{-1}\leftarrow\left(\partial_{\bm{\vartheta}} \bm{f}_{{\bm{\vartheta}}}\right)\T \leftarrow \bm{\lambda} \leftarrow  - \left(\partial_{\bm{w}} \Rd+\partial_{\bm{w}} \bm{f}_{\bm{\vartheta}}\right)\mT $} \; &\longleftarrow& \; \delta\bm{w} &\longleftarrow \fcolorbox{black}{white}{$ \left(\partial_{\bm{w}}\mathcal{J}\right)\T$} &\longleftarrow \delta &\mathcal{J}. \label{opt:backward_implicit}
\end{alignat}
The boxes are handled by the PDE solver and the automatic differentiation framework, the ones in light-gray containing implicit operations (involving complex operations such as a Newton solver or a large-scale matrix inverse), the others being completely explicit.
We insist on the fact that the light-grey box in the backward operator is implicit: the operator $(\partial_{\bm{w}}\Rd+\partial_{\bm{w}}\bm{f}_{{\bm{\vartheta}}})\mT$ should not be interpreted as the application of the explicit inverse of the adjoint matrix (which is generally dense and cannot be stored). Instead, it corresponds to solving the linear system involving the adjoint matrix:
\begin{equation}
-(\partial_{\bm w}\Rd+\partial_{\bm w}\bm{f}_{{\bm{\vartheta}}})\T \bm{\lambda} = \delta{\bm{w}},
\end{equation}
which can be done with either direct methods \citep[for example with sparse LU methods][]{MUMPS:1, MUMPS:2} or iterative strategies \citep{FANIZZA2025115984}.
The adjoint system inherits the same sparsity pattern as the Jacobian arising from the linearization of the baseline PDE operator and its eigenvalues and condition number remain unchanged with respect to the original system. Hence, from a mathematical point of view, the complexity in solving the adjoint system should be equivalent to solving the direct one.

In the case where $ {\bm{\vartheta}} $ designates a volumetric source term in the PDE solver, one needs, as in the previous section, to move to a Euclidean space by introducing the variable $ \tilde{{\bm{\vartheta}}}$ (see \S~\ref{sec:grads}).

\subsection{The end-to-end differentiable learning pipeline}
\label{sec:PyTorch}

Thanks to its flexibility, native support for automatic differentiation, and strong integration with scientific computing libraries, PyTorch~\citep{ADPyTorch} was selected as the core framework for implementing the computational routines developed in this work. The overall pipeline is organized around a modular structure composed of three fundamental components responsible for (i) enforcing boundary conditions, (ii) evaluating the discrete residual of the governing equations, and (iii) applying external or model-derived forcing terms. This layered design provides a clean interface between high-level Python code and low-level numerical kernels, while ensuring that each stage of the computation remains physically consistent and computationally efficient.

A central requirement of this architecture is full differentiability. To achieve this, each component is implemented by overriding the \texttt{forward} and \texttt{backward} operations, through PyTorch's \texttt{torch.autograd.Function} interface\footnote{It is also possible to define the Jacobian-vector product (\texttt{jvp}) inside a \texttt{torch.autograd.Function}, although it is not needed for optimization purposes; it may, however, be useful to construct the Jacobian for a Newton method.}. This mechanism allows complex, optimized numerical routines to be embedded inside a differentiable computational graph, while remaining fully compatible with PyTorch's automatic differentiation engine.

The computational flow proceeds by sequentially applying the boundary-condition layer to the input state, evaluating the discrete residual of the governing equations, and finally incorporating the forcing contribution. By composing these custom layers into a standard PyTorch \texttt{nn.Module}, the framework automatically manages gradient propagation and the definition of the backward pass, enabling seamless integration with optimization algorithms and ML architectures.

To solve the nonlinear system associated with the discretized equations, the full computation is embedded within an outer layer implementing a Fixed-Point (FP) or Newton-type iteration strategy. This external layer overrides PyTorch’s default backward behavior by solving an associated adjoint system to compute exact gradients with respect to the trainable parameters. This ensures consistent gradient flow across iterative solves and enables end-to-end learning constrained by general nonlinear governing equations.

While the present work focuses on turbulence modeling within the RANS framework, as detailed in the following section, the abstract structure introduced in \eqref{eq:generic} is in fact representative of a much broader class of physical problems. In reactive flows, for instance, the correction term $\bm{f}(\bm{w})$ may account for source terms arising from chemical reactions. In porous media, it can represent Darcy-type drag effects, while in radiation or multiphase transport, it may encode closure relations for interfacial fluxes or effective material properties. Comparable formulations also emerge in solid mechanics, particularly in the constitutive modeling of complex materials. More generally, this work addresses the modeling of multiscale systems, where similar challenges arise across a wide range of physical contexts. In kinetic theory, such difficulties appear through moment hierarchies derived from the Boltzmann equation \citep{chapman1990mathematical}. They are also central to many-body statistical physics via the BBGKY hierarchy \citep{huang2008statistical}, and to solid mechanics through homogenization techniques and the derivation of effective constitutive laws \citep{torquato2002random}. From this perspective, the methodology proposed here can be viewed as a general strategy for incorporating modeling closures into PDE-based systems in a systematic and flexible manner. By leveraging differentiable solvers together with adjoint-based optimization, it enables a consistent and scalable coupling between physics-based models and trainable components. As such, it provides a natural framework for integrating ML into scientific computing workflows governed by PDEs, with potential applications well beyond turbulence modeling.

\section{Data assimilation and closure with compressible RANS equations}

In this section, we present the PDE problem under investigation and illustrate how the proposed pipeline can be applied to it.
First (\S \ref{sec:RANS}), we present the compressible RANS equations, before we introduce a data assimilation and closure problem with two possible tunable parameters: the eddy viscosity $ \mu_t $ (\S \ref{sec:mut}) and the production term $ \beta $ of the turbulent variable equation (\S \ref{sec:beta}).

\subsection{Compressible RANS equations with Boussinesq hypothesis} \label{sec:RANS}

The steady compressible RANS equations can be written as:
\begin{equation}
\label{eq:Rans}
    \begin{cases}
        \nabla \cdot (\rho \bm{U}) = 0 \\
        \nabla \cdot (\rho \bm{U} \bm{U}+p\bm{I}) -  \nabla \cdot \bm{\tau} - \nabla \cdot \bm{\tau}^R=0 \\
        \nabla \cdot \left[ (\rho E + p) \bm{U} \right] - \nabla \cdot \left( \bm{\tau} \cdot \bm{U} - \bm{q} \right) - \nabla \cdot \left(  \bm{\tau}^R \cdot \bm{U} - \bm{q_t} \right)=0
    \end{cases},\,
\end{equation}
where the density $\rho$, pressure $p$, viscous stress tensor $\bm{\tau}$ and heat flux $\bm{q}$ are Reynolds-averaged quantities, while the velocity vector $\bm{U} = [U, \, V]$ and the total energy $E$ are Favre-averaged quantities~\citep{bookWilcox}. Finally, $\bm{\tau}^R$ and $\bm{q}_t$ denote the Reynolds stress tensor and the turbulent heat flux, respectively.

To close the system, models relating $\bm{\tau}^R$ and $\bm{q}_t$ to the state variables are required. Most commonly used approaches rely on the Boussinesq eddy viscosity assumption \citep{boussinesq1877}, which states that the Reynolds stress tensor $\bm{\tau}^R$ is proportional to the deviatoric part of the mean strain-rate tensor $\bm{S}$, with proportionality coefficient given by the eddy viscosity $\mu_t$. This yields:
\begin{equation}\label{eq:Boussinesq}
    \bm{\tau}^R = 2 \mu_t \left(\bm{S} - \frac{1}{3} (\nabla \cdot \bm{U}) \bm{I} \right), \;\;\;\; \bm{q}_t = \frac{\mu_t}{Pr_t} \nabla T,
\end{equation}
where $Pr_t$ is the (constant) turbulent Prandtl number and $T$ the Reynolds-averaged temperature. 
As discussed in \cite{bookWilcox}, a constant turbulent Prandtl number, $Pr_t = 0.9$, is used in the present simulations and is considered an acceptable baseline choice for the adiabatic configurations investigated here. Then, by incorporating \eqref{eq:Boussinesq} in \eqref{eq:Rans}, the system reads:

\begin{equation}
\label{eq:RansBussi}
\underbrace{\left\{\begin{array}{l}
        \nabla \cdot (\rho \bm{U}) \\
        \nabla \cdot (\rho \bm{U} \bm{U}+p\bm{I}) -  \nabla \cdot \bm{\tau} \\
        \nabla \cdot \left[ (\rho E + p) \bm{U} \right] - \nabla \cdot \left( \bm{\tau} \cdot \bm{U} - \bm{q} \right) 
\end{array}\right.}_{(*)}
\underbrace{\begin{array}{l}
      \\
     -\nabla \cdot \left[2 \mu_t \left(\bm{S} - (\nabla \cdot \bm{U}) \bm{I}/3 \right) \right]  \\
     -\nabla \cdot \left[  2 \mu_t \left(\bm{S}\cdot \bm{U} - (\nabla \cdot \bm{U}) \bm{U}/3 \right)  - \mu_t \nabla T / Pr_t \right]
\end{array}}_{(**)}
\begin{array}{l}
   =0   \\
=0  \\
     =0
\end{array}
\end{equation}

In the following, after discretizing these equations, we consider two possible choices for the corrected model, characterized by different state variables $\bm{w}$, governing equations $\Rd(\bm{w})$ and correction terms $ \bm{f}_{\bm{\vartheta}}(\bm{w})$.

\subsection{Data assimilation and closure with \texorpdfstring{$\mu_t$}{mt}} \label{sec:mut}

The first approach consists of defining the baseline model with the state vector $\bm{w}=[\rho, \, \rho U, \, \rho V, \, \rho E ]$ and the discrete version of $ (*) $ in \eqref{eq:RansBussi} for $\Rd(\bm{w})$. The correction term $\bm{f}(\bm{w},\bm{\alpha})$ in \eqref{eq:da} then represents the left-over part of the equation, the discretization of term $ (**) $, with $ \bm{\alpha}=\mu_t$. The corrected model finally reads:
\begin{equation}
\label{eq:NS}
    \Rd(\bm{w})+\bm{f}(\bm{w},{\mu_t}) = 0
\end{equation}
In the data assimilation setting, $\mu_t$ is treated as a spatially varying tunable parameter, $\mu_t(\bm{x})$. In contrast, the closure problem \eqref{eq:closure} consists of identifying a functional relationship between the eddy viscosity and the state, $\mu_{t,{\bm{\vartheta}}}(\bm{w})$. In Section~\ref{sec:Dataset}, this function will be parametrized using a NN with trainable parameters.
\subsection{Data assimilation and closure with \texorpdfstring{$\beta$}{b}} \label{sec:beta}

An alternative approach consists of incorporating a turbulence model directly into $\Rd$. Here, we consider the SA model, in which the eddy viscosity $\mu_t$ depends analytically on an additional variable $\tilde{\nu}$ (i.e., $\mu_t(\tilde{\nu})$) governed by \citep{spalart1992one}:
\begin{equation} \label{eq:turb}
    \nabla \cdot (\rho \tilde{\nu} \bm{U})-\nabla\cdot (\sigma^{-1}(\mu+\rho\tilde{\nu})\nabla \tilde{\nu}) =P(\bm{w})+D(\bm{w})+C(\bm{w}).
\end{equation}
Here, $P(\bm{w})$, $D(\bm{w})$, and $C(\bm{w})$ denote the production, destruction, and cross-diffusion terms, respectively \citep{spalart}. In this framework, the state vector is augmented as $\bm{w}=[\rho, \, \rho U, \, \rho V, \, \rho E, \, \rho \tilde{\nu} ]$, and the baseline operator $\Rd$ now includes the full $ (*)+(**) $ equation \eqref{eq:RansBussi}, and the turbulence variable equation \eqref{eq:turb}, together with the analytic relation $ \mu_t(\tilde{\nu}) $. 

The correction term $\bm{f}_{\bm{\vartheta}}(\bm{w})$ is introduced as an additional source term acting on the turbulence equation of the form $\beta P(\bm{w})$ \citep{FANIZZA2025115984, CATO2023106054}, where $\beta$ is a scalar parameter and $P(\bm{w})$ is the production term appearing on the right-hand side of \eqref{eq:turb}. The corrected model reads:
\begin{equation}
\label{eq:NS_SA}
    \Rd(\bm{w})+\bm{f}(\bm{w},\beta) = 0 \text{, where } \bm{f}(\bm{w},\beta) \text{ is the discrete version of } \left( \begin{array}{c}
         0 \\ 0 \\ 0 \\ -\beta P(\bm{w})
    \end{array} \right). 
\end{equation}

In the data assimilation framework \eqref{eq:da}, $\beta$ is treated as a spatially varying parameter, $\beta(\bm{x})$. The associated closure problem \eqref{eq:closure} then consists of identifying a functional relation $\beta_{\bm{\vartheta}}(\bm{w})$.

\subsection{The end-to-end pipeline applied to BROADCAST}
\label{sec:PyTorchB}

The BROADCAST solver \citep{Poulain2023} was chosen as core CFD solver because of the possibility of automatically differentiating \citep{AutomaticD} its Fortran (F90) routines thanks to the software Tapenade \citep{Tapenade}. Indeed, every routine can be linearized to obtain the Jacobian-vector product (JVP) and the vector-Jacobian product (VJP). The (JVP) and (VJP) are used to construct the sparse representations of the Jacobian and of its transpose, making use of the independent stencil defined by the order of the solver. The solution of the linear systems can then be obtained by a sparse LU-inverse.

The BROADCAST solver is a cell-centred FV solver. Boundary conditions are handled via ghost cells. The full state vector therefore consists of components associated with the physical cells $ \bm{w}_i $ and components associated with the ghost cells $ \bm{w}_o $. The boundary conditions are enforced by filling the ghost cells ($\bm{w}_o=\bm{B}(\bm{w}_i)$) prior to application of the model $ \Rd_i(\bm{w}_i,\bm{w}_o) $ and the correction $ \bm{f}_{\bm{\vartheta}}(\bm{w}_i,\bm{w}_o)$.
The discrete equations, which are equivalent to Eq. \eqref{eq:timeint}, may be written as:
\begin{equation}
    \bm{M}\frac{\bm{w}_{n+1}-\bm{w}_{n}}{\Delta t} =\Rd(\bm{w}_{n+1})+\bm{P}\bm{f}_{\bm{\vartheta}}(\bm{w}_{n+1}),
\end{equation}
where
\begin{equation}
    \bm{w}=\left(\begin{array}{c}  \bm{w}_i  \\ \bm{w}_o  \end{array}\right), \;\;\;\; \bm{M}=\left(\begin{array}{cc}  \bm{V} & \bm{0} \\ \bm{0} & \bm{0} \end{array}\right), \;\;\;\; \Rd(\bm{w})=\left(\begin{array}{c}  \Rd_i(\bm{w}_i,\bm{w}_o) \\
    \bm{w}_o-\bm{B}(\bm{w}_i)
    \end{array}\right), \;\;\;\; \bm{P}=\left(\begin{array}{c}  \bm{I}  \\ \bm{0}  \end{array}\right),
\end{equation}
In this equation, $ \bm{V} $ is a diagonal matrix representing the volume of each cell.

The current implementation of the BROADCAST solver employs a FV formulation with the high order Flux-Extrapolated MUSCL (FE-MUSCL) reconstruction described in \citep{CINNELLA20161}. Moreover, it relies on structured meshes, either single- or multi-block. This choice is primarily driven by the formulation of the numerical schemes and their implementation in the code, where the connectivity between control volumes is implicitly defined through the $(i,j)$ indexing. Such a structure enables efficient stencil operations, predictable sparsity patterns in the Jacobian matrix, and optimized memory access patterns, which results in good computational performance and facilitates the implementation of high-order discretization schemes.

It should be noted, however, that the methods used in this work, applied here to structured meshes using the BROADCAST solver, could also be implemented for unstructured ones. To do so, one would need to provide the JVP and the sparsity pattern of the Jacobian. The strategy remains general and could be applied in the future to different unstructured solvers. The use of structured meshes therefore represents an implementation choice that favors computational efficiency, regular stencil operations, and accurate boundary-layer resolution, at the cost of reduced geometric flexibility and more complex mesh generation for highly intricate geometries.

Finally, within the BROADCAST solver, the governing equations are adimensionalized using freestream quantities as reference scales, i.e., $\rho_\infty$, $U_\infty$, and either the static temperature $T_s$ or the freestream temperature $T_\infty$.

The computational flow follows the sketch introduced in Section~\ref{sec:PyTorch}. The pipeline is structured around a main module delivering $\Rd(\bm{w}) $, \textbf{RANS}, composed of three internal layers: \textbf{BC}, \textbf{NS} and \textbf{FORCE}, which respectively handle the imposition of boundary conditions ($ \bm{w}_o=\bm{B}(\bm{w}_i)$), the evaluation of the baseline model (applying $ \Rd(\bm{w}_i,\bm{w}_o) $), and of the correction term (adding $ \bm{f}_{\bm{\vartheta}}(\bm{w}_i,\bm{w}_o) $). Each layer serves as an interface to underlying Fortran routines from the BROADCAST solver~\citep{Poulain2023}, ensuring that the operations performed at each stage remain physically faithful and computationally efficient.

The custom layers (\textbf{BC}, \textbf{NS}, and \textbf{FORCE}) can be composed into a PyTorch \texttt{nn.Module}: \textbf{RANS}, which automatically handles the chaining of gradients and the definition of the backward pass, enabling seamless integration with optimization algorithms and NN architectures.

\begin{figure}
    \centering
    \includegraphics[width=\textwidth]{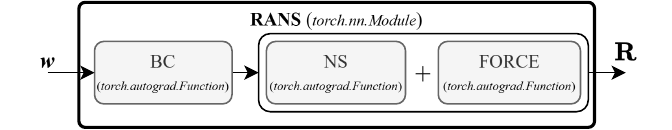}
    \caption{Diagram representing the explicit RANS layer.}
    \label{fig:Scheme1}
\end{figure}

Figure~\ref{fig:Scheme1} provides a sketch of the \textbf{RANS} module, which combines the \textbf{BC}, \textbf{NS}, and \textbf{FORCE} operators.

To solve the nonlinear system arising from the RANS equations with turbulent closure, the entire computation is wrapped in an outer layer that implements a FP iteration strategy based on Newton’s method.

\begin{figure}
    \centering
    \includegraphics[width=\textwidth]{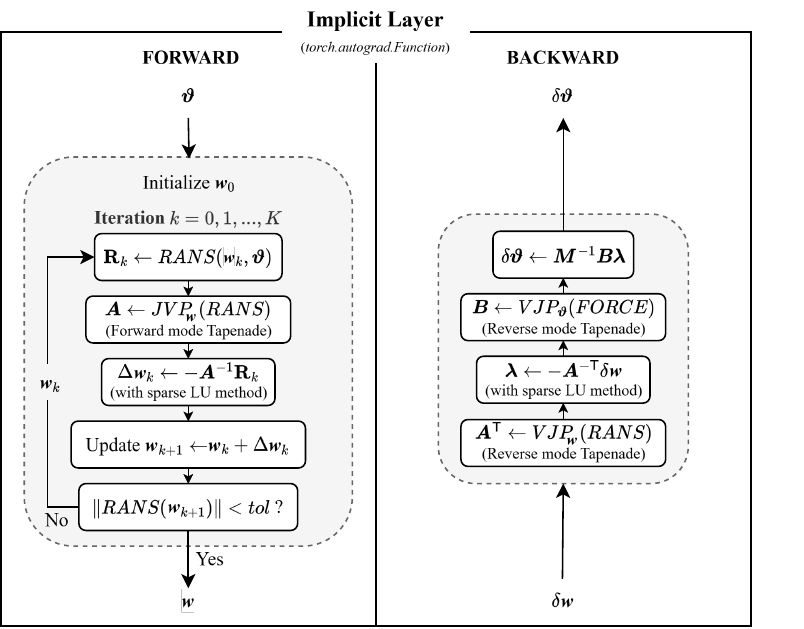}
    \caption{Diagram representing the operations inside of the implicit layer (light-gray boxes) appearing in \eqref{opt:forward_implicit} and \eqref{opt:backward_implicit}.}
    \label{fig:Scheme2}
\end{figure}

Figure~\ref{fig:Scheme2} depicts the details of the two implicit layers appearing in the forward \eqref{opt:forward_implicit} and backward \eqref{opt:backward_implicit} operators, highlighting the hierarchical structure of the implicit solver.

Appendix~\ref{sec:code} provides a detailed overview of the PyTorch implementation developed in this work, along with the strategy used to interface the Fortran routines with the Python execution flow. The connection with the Deep Equilibrium Models (DEQ) framework~\citep{Bai2019} is also discussed. The overall workflow is illustrated through several pseudocode listings.

We emphasize that the development of the implicit layer was the most critical part of this work, as defining a robust and efficient implicit solver involves several aspects that must be carefully considered. Throughout this process, multiple strategies were explored to achieve correct gradient incorporation within the routine. Ultimately, the adopted approach required a complete redefinition of the backpropagation mechanism, allowing for a more accurate and reliable computation of gradients, as detailed in Section~\ref{sec:FixedPoint} and Appendix~\ref{sec:grads}.

In the case of the turbulent closure problem with the eddy viscosity $ \mu_{t}(\bm{w})$, a Convolutional Neural Network (CNN) \citep{lecun:hal-05083427, oshea2015introductionconvolutionalneuralnetworks} was implemented to represent this relation. Its architecture combines convolutions with varying kernel sizes, with the output excluding ghost cells. This design captures multi-scale features while preserving output consistency and ensuring geometric invariance. The kernel sizes of the convolutions have been chosen so that the sparsity pattern of the correction term $\bm{f}(\bm{w},\mu_{t,{\bm{\vartheta}}}(\bm{w}))$ appearing in \eqref{eq:closure} matches the one of $ \Rd(\bm{w})$. This aspect and the proposed CNN are further discussed in Appendix~\ref{sec:CNN}.

\section{Results}

\label{sec:Applications}

The routines are first (\S~\ref{sec:WH}) validated on the NASA 2D Wall-Mounted Hump for data assimilation of the $ \beta $ parameter (using both explicit and implicit strategies). We then consider a more ambitious configuration and progressively move toward the development of an algebraic turbulence model based on a CNN:
\begin{itemize}
\item First (\S~\ref{sec:VKI}), we verify that the BROADCAST code, combined with the baseline SA model, yields consistent results for the standard VKI LS-59 turbine blade cascade.
\item Then (\S~\ref{sec:ResultsVKI}), we demonstrate that, given the SA solution for the VKI LS-59 case, $ \bm{w}^{SA}=[\rho,\rho \bm{U}, \rho E]$, we are able to satisfactorily reconstruct the SA eddy viscosity field $\mu_t^{SA}$ using the explicit residual minimization strategy by tuning $ \bm{\alpha} = \mu_t$.
\item In \S~\ref{sec:Dataset}, we first introduce a custom dataset of flow fields obtained with the SA model by varying both flow and geometric parameters. We then construct a CNN-based algebraic turbulence model $ \mu_{t,\bm{\vartheta}}(\bm{w})$ that reproduces the SA eddy viscosity $ \mu_t^{SA} $ over this dataset. To this end, the CNN weights $ \bm{\vartheta}$ are optimized using the explicit strategy by minimizing the residual of a corrected model on the custom dataset.
\end{itemize}

Building an algebraic turbulence model capable of reproducing the results of a one-equation SA model is an extremely challenging task, even when using a CNN-based formulation (here with a limited kernel width). It is worth recalling that classical algebraic turbulence models, such as the model by \citet{baldwin1978thin}, are generally well suited for attached turbulent boundary layers, but perform poorly in separated flows and in wake regions behind profiles, where the turbulent viscosity is significantly mispredicted. In such models, non-locality is only partially accounted for through the distance to the closest wall; however, this does not capture streamwise non-local effects, which are particularly important in wake regions. This limitation notably motivated the introduction of transport-equation-based models, in which streamwise non-locality is represented through the advection of turbulence variables. In the present approach, the CNN-based algebraic eddy viscosity model can be regarded as strongly local, since the distance to the closest wall is not included and the kernel width of the CNN is limited to that of the spatial scheme, i.e., a few grid cells in each spatial direction (see \S~\ref{sec:CNN} for more details). As a result, the CNN learns local features based on the resolved state $ \bm{w}=[\rho,\rho \bm{U}, \rho E]$ (and its derivatives) to predict the eddy viscosity $\mu_t$.

In the future, learning a CNN-based $\beta$ turbulence production correction term from high-fidelity Large Eddy Simulations (LES) data appears to be a promising direction, as it naturally captures streamwise non-locality while identifying local features. Increasing the kernel width provides additional flexibility but may violate the stencil locality of the numerical scheme, introducing extra couplings and modifying the sparsity structure of the associated operators.

\subsection{2D NASA Wall-Mounted Hump separated flow}
\label{sec:WH}

\begin{figure}
    \centering
    \includegraphics[width=\textwidth]{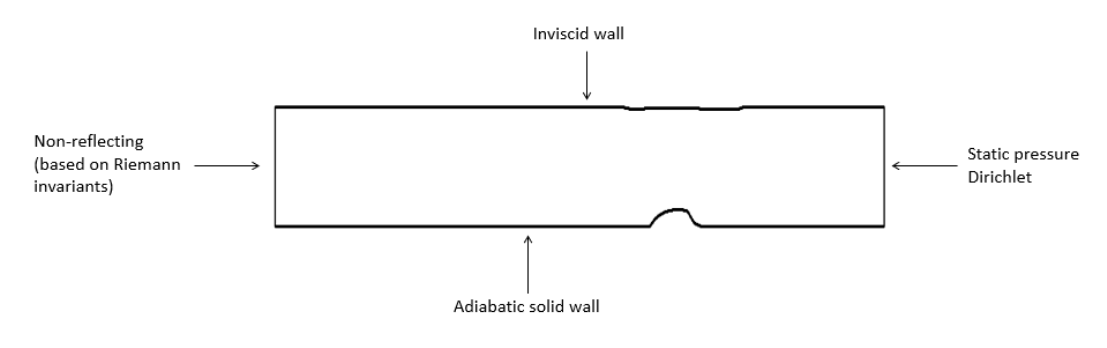}
    \caption{Outline of the geometry of the 2D NASA Wall-Mounted Hump with the corresponding boundary conditions.}
    \label{fig:geomWH}
\end{figure}

The first case selected for the application of the framework is NASA's 2D Wall-Mounted Hump (WH) (no plenum), shown in Figure~\ref{fig:geomWH}. This case is characterized by a separation of the flow from the body with subsequent reattachment and boundary-layer recovery.

It is a well-established benchmark configuration developed at NASA Langley Research Center for the validation of turbulence models in separated flows. It is widely used within the CFD community because it combines a relatively simple two-dimensional geometry with complex physical phenomena, including adverse pressure gradients, boundary-layer separation, recirculation, and reattachment. The case was designed to provide high-quality
reference data, such as LES, against which RANS and hybrid turbulence models can be systematically assessed.

The geometry, as depicted in Figure~\ref{fig:geomWH}, consists of a two-dimensional channel-like domain in which a smooth hump is mounted on the lower wall. Upstream of the hump, the lower wall is flat, allowing a turbulent boundary layer to develop under nearly equilibrium conditions. As the flow approaches and passes over the hump crest, it experiences a strong adverse pressure gradient that leads to boundary-layer separation downstream of the crest. This separation produces a recirculation bubble whose size and reattachment location are highly sensitive to turbulence modeling assumptions. Further downstream, the flow reattaches and gradually recovers toward a redeveloping turbulent boundary layer.

The canonical operating conditions correspond to a low-speed subsonic flow with a freestream Mach number of approximately $M \approx 0.1$ and a Reynolds number on the order of $Re \approx 9.36 \times 10^{5}$, typically based on a reference length associated with the hump geometry. Under these conditions, compressibility effects remain weak.
At the inlet, a turbulent boundary layer consistent with the experimental measurements must be prescribed. The lower wall, including the hump, is treated as a no-slip boundary. The outlet is generally modeled as a pressure outlet with a specified static pressure to ensure the correct overall mass flow.

The primary validation quantities for this case include the surface pressure coefficient $C_p$, the skin-friction coefficient $C_f$, and mean velocity profiles at selected streamwise stations. The skin-friction distribution is especially important because it clearly identifies the separation and reattachment points.

A distinctive feature of the distributed two-dimensional grids is the non-flat upper boundary, which exhibits a mild contour rather than being straight. Although the hump is physically located only on the lower wall, the shaped upper wall is intentionally introduced to account for wind-tunnel confinement and blockage effects present in the original experiments. In the experimental setup, the test section had finite height and side plates, which influenced the global pressure distribution within the tunnel. A purely two-dimensional simulation with a flat and distant upper boundary would not reproduce these confinement effects accurately, leading to discrepancies in the pressure gradient imposed on the boundary layer over the hump. Since the separation process is driven by the adverse pressure gradient, even small differences in the global pressure field can significantly alter the predicted separation and reattachment behavior. To avoid performing a fully three-dimensional simulation of the entire wind tunnel, the two-dimensional benchmark geometry incorporates a carefully contoured upper wall that effectively mimics the experimental blockage. In this sense, the upper-wall shape acts as a surrogate for three-dimensional tunnel effects, allowing a two-dimensional computation to reproduce more faithfully the pressure distribution and separated-flow characteristics observed in the experiments.

More information on the case and the physical parameters imposed to run the RANS simulation can be found in \citet{jespersen2016, nasaNASAWallMounted}.
Note that this case is part of the ERCOFTAC Database (Classic Collection). It is listed as Case C.83: Wall-mounted two-dimensional hump with oscillatory zero-mass-flux jet or suction through a slot.

\subsubsection{Data assimilation with \texorpdfstring{$\beta$}{b}: partial-state measurements and implicit layer}
\label{sec:Beta}

In this section, we use the corrected model presented in \eqref{eq:NS_SA} and the implicit optimization framework described in \crefrange{opt:forward_implicit}{opt:backward_implicit}. The forward implicit solution in the red-framed light-grey box is obtained by iteratively solving \eqref{eq:Newtonstep} with a direct-LU method. The backward implicit solution of the adjoint system uses this same inversion technique. 
 For the definition of the cost functional to optimize, we use as a reference solution the LES simulation run by \cite{doi:10.2514/1.J056397}. Here we consider partial-state measurements based solely on the streamwise and wall-normal velocity to define
\begin{equation}
    \mathcal{J} = \dfrac{1}{2} \norm{\bm{U}^{LES}-\dfrac{\left(\rho\bm{U}\right)^{RANS}}{\rho^{RANS}}}^2_Q \, ,
\end{equation}
and the data assimilation process aims at reconstructing the full state $\bm{w}=[\rho, \, \rho \bm{U}, \, \rho E, \, \rho \tilde{\nu} ]$, i.e., in particular the density, the temperature and the turbulence variable which are considered unknown.

In this benchmark test, the routine was initialized with $\beta$ set to zero throughout the domain. The BROADCAST solver is launched on a mesh of $n_i \times n_j = 408 \times 108$ cells in the $x$ and $y$ directions respectively ($n = n_i n_j$). 

\begin{figure}
    \centering
    \includegraphics[width=0.5\textwidth]{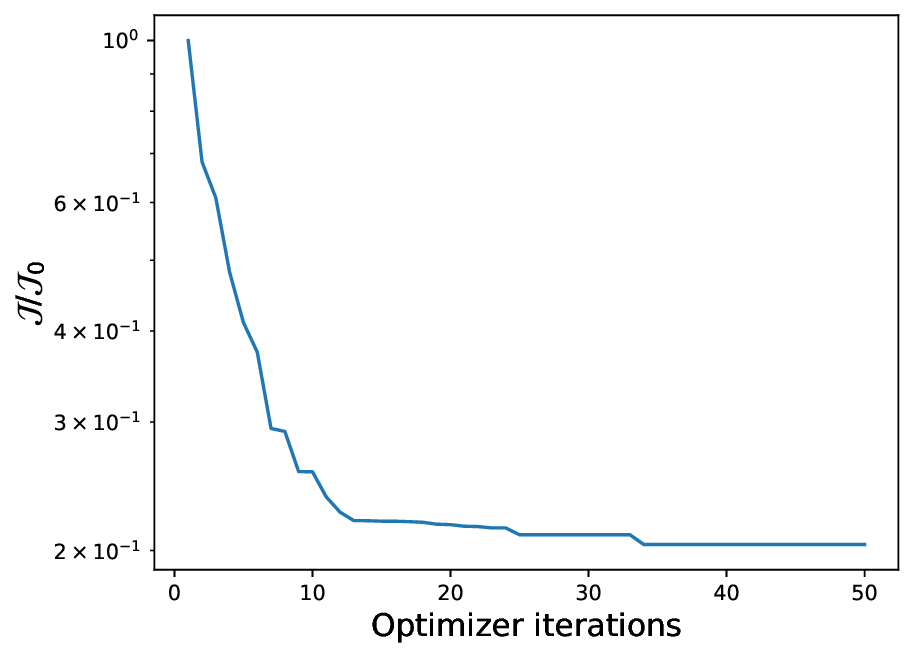}
    \caption{Data assimilation with $ \beta $ on Wall-Mounted Hump case handled with implicit layer: behavior of the normalized cost-function $\mathcal{J}/\mathcal{J}_0$ for 50 optimizer iterations using the L-BFGS PyTorch optimizer. The procedure is halted once the variation in the loss function satisfies a specified tolerance threshold. Specifically, the norm of the loss function starts at $3.245\times 10^{-4}$, and at the end of the computation the reached value is $6.611 \times 10^{-5}$.}
    \label{fig:loss_beta_implicit}
\end{figure}

The optimization algorithm uses the Limited-memory Broyden–Fletcher–Goldfarb–Shanno (L-BFGS) optimizer~\citep{lbfgs}. The stopping criterion is reached when the variation in the normalized loss function satisfies a specified tolerance threshold ($10^{-6}$). Figure~\ref{fig:loss_beta_implicit} shows the behavior of the normalized objective function.

\begin{figure}
    \centering
    \begin{subfigure}[b]{0.495\textwidth}
        \centering
        \includegraphics[width=\textwidth]{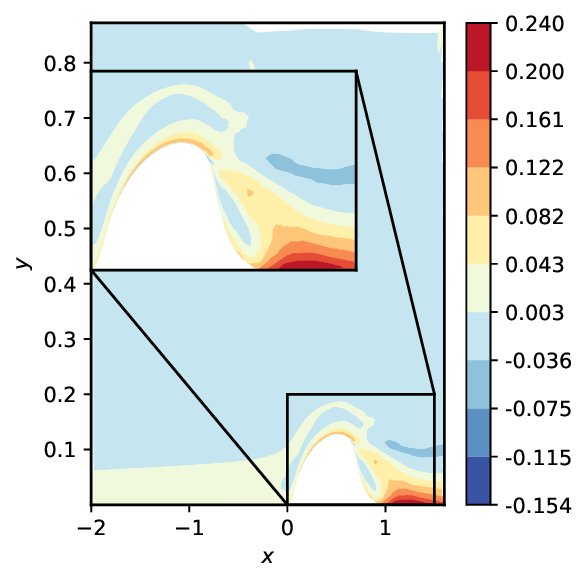}
        \caption{Error field for $U$ ($(U^{LES}(\bm{x})-U^{SA}(\bm{x}))/U_\infty$).}
        \label{fig:u_wh_sa}
    \end{subfigure}
    \begin{subfigure}[b]{0.495\textwidth}
        \centering
        \includegraphics[width=\textwidth]{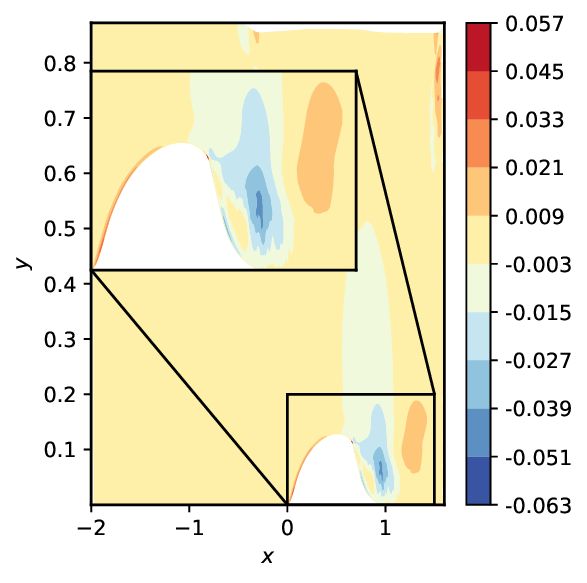}
        \caption{Error field for $V$ ($(V^{LES}(\bm{x}) - V^{SA}(\bm{x}))/U_\infty$). }
        \label{fig:v_wh_sa}
    \end{subfigure}
    \caption{Data assimilation with $ \beta $ on Wall-Mounted Hump case handled with implicit layer: representation of the error fields for the velocities computed with the SA turbulence model (the baseline) with respect to the LES solutions, normalized by the reference freestream velocity $U_\infty$. It is not possible to show the whole computational domain depicted in Figure~\ref{fig:geomWH} since the LES solution is available only in the central region of the computational mesh. As the region of primary interest is the recirculation zone downstream of the hump, the figure shows the error profiles with a zoom on this area.}
    \label{fig:vel_wh_sa}
\end{figure}

\begin{figure}
    \centering
    \begin{subfigure}[b]{0.495\textwidth}
        \centering
        \includegraphics[width=\textwidth]{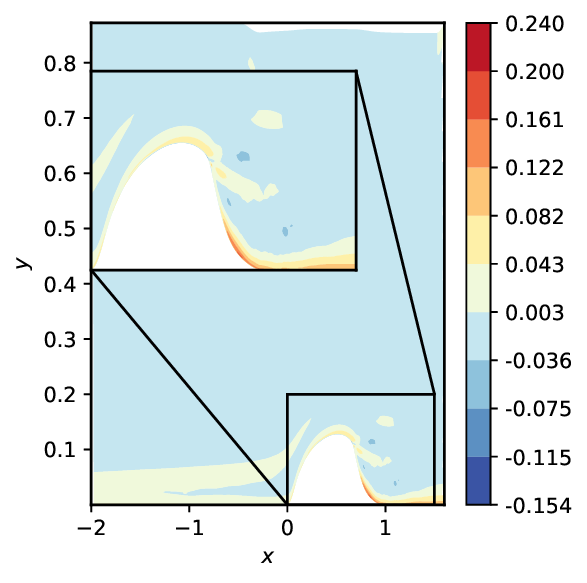}
        \caption{Error field for $U$ ($(U^{LES}(\bm{x})-U^{opt}(\bm{x}))/U_\infty$).}
        \label{fig:u_wh_impli}
    \end{subfigure}
    \begin{subfigure}[b]{0.495\textwidth}
        \centering
        \includegraphics[width=\textwidth]{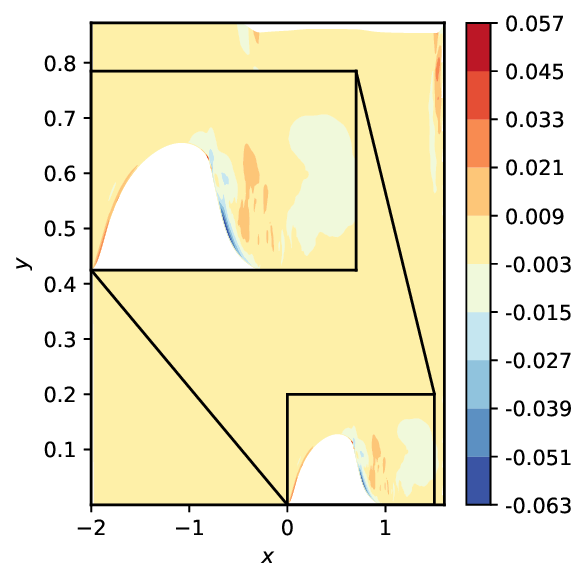}
        \caption{Error field for $V$ ($(V^{LES}(\bm{x})-V^{opt}(\bm{x}))/U_\infty$).}
        \label{fig:v_wh_impli}
    \end{subfigure}
    \caption{Data assimilation with $ \beta $ on Wall-Mounted Hump case handled with implicit layer: representation of the error fields for the velocities at the final iteration with respect to the LES solutions, normalized by the reference freestream velocity $U_\infty$. It is not possible to show the whole computational domain depicted in Figure~\ref{fig:geomWH} since the LES solution is available only in the central region of the computational mesh. The same scales as Figure~\ref{fig:vel_wh_sa} were kept to highlight the decrease in errors. We can see that the $\beta$ parameter is not sufficiently flexible to accurately reproduce the high-fidelity data in the recirculation region. \citet{franceschini2020mean} showed that this may be understood by the rigidity of the chosen control parameter $ \beta $. As the region of primary interest is the recirculation zone downstream of the hump, the figure shows the error profiles with a zoom on this area.}
    \label{fig:vel_wh_impli}
\end{figure}

We denote by $\beta(\bm{x})~=~\beta^{opt}(\bm{x})$ the value of the optimized spatial parameter after the final iteration. It is then possible to study the error fields for the $U$ and $V$ components of the velocity. Figure~\ref{fig:vel_wh_sa} and Figure~\ref{fig:vel_wh_impli} show the local errors on the discretized mesh for the baseline and the optimized models, respectively. It is important to note that not all of the domain is shown in the figures since the LES reference values were available only in the depicted region.

\begin{figure}
    \centering
    \begin{subfigure}[t]{\textwidth}
        \centering
        \includegraphics[width=\textwidth]{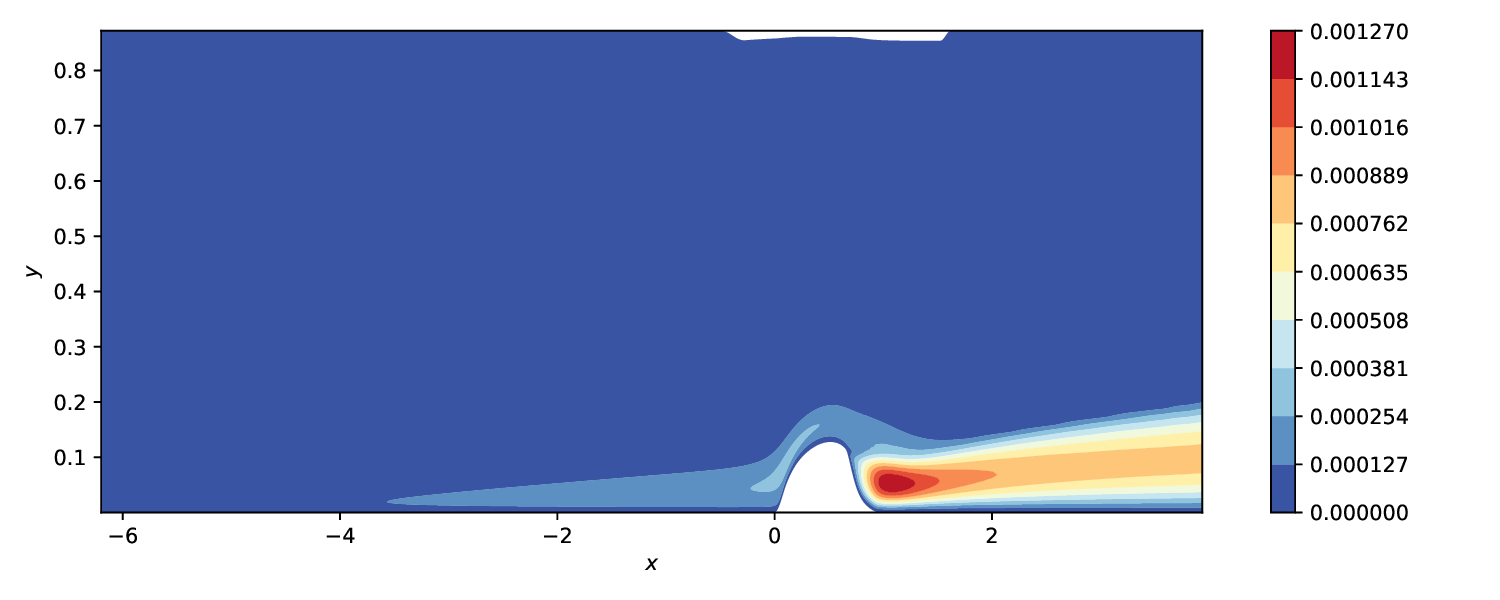}
        \caption{Eddy viscosity $\mu_t^{SA}(\bm{x})$ obtained with the baseline SA model.}
        \label{fig:mut_wh_SA}
    \end{subfigure}
    
    \begin{subfigure}[t]{\textwidth}
        \centering
        \includegraphics[width=\textwidth]{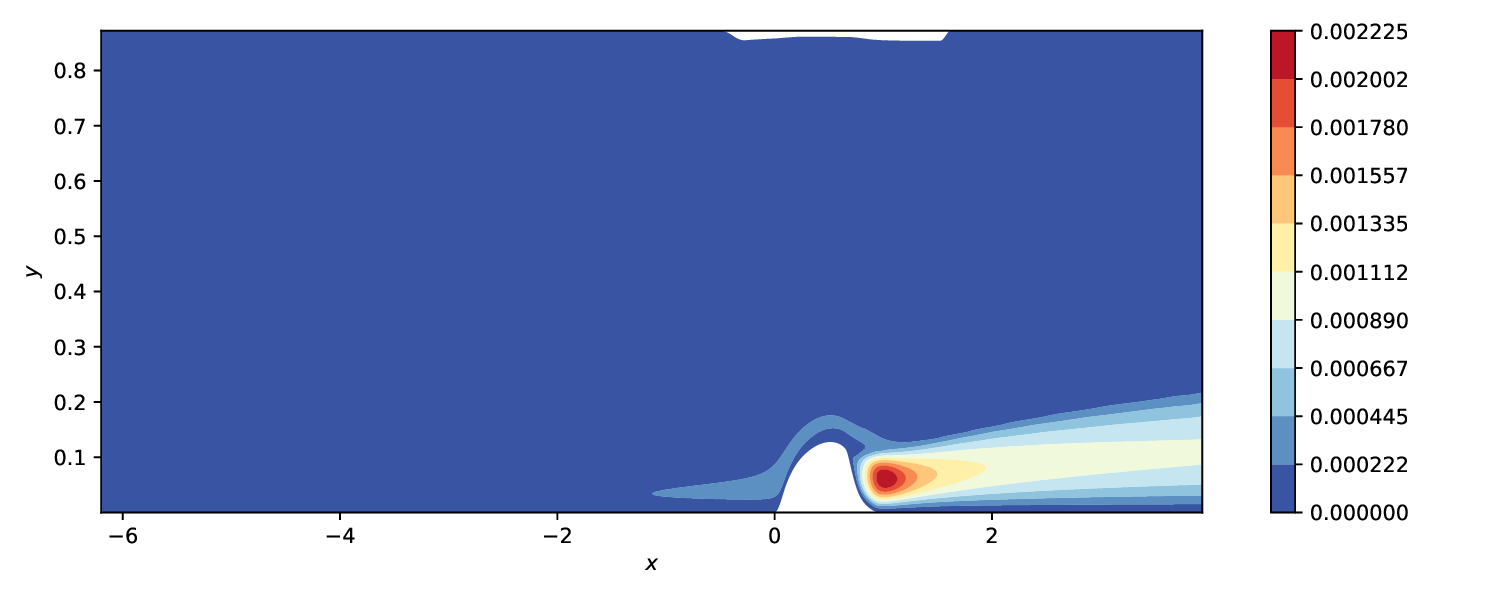}
        \caption{Corrected eddy viscosity $ \mu_t^{opt}(\bm{x}) $ obtained considering $(1+\beta^{opt}(\bm{x}))P(\bm{w})$ as production term.}
        \label{fig:mut_wh_opt_beta}
    \end{subfigure}
    
    \begin{subfigure}[t]{\textwidth}
        \centering
        \includegraphics[width=\textwidth]{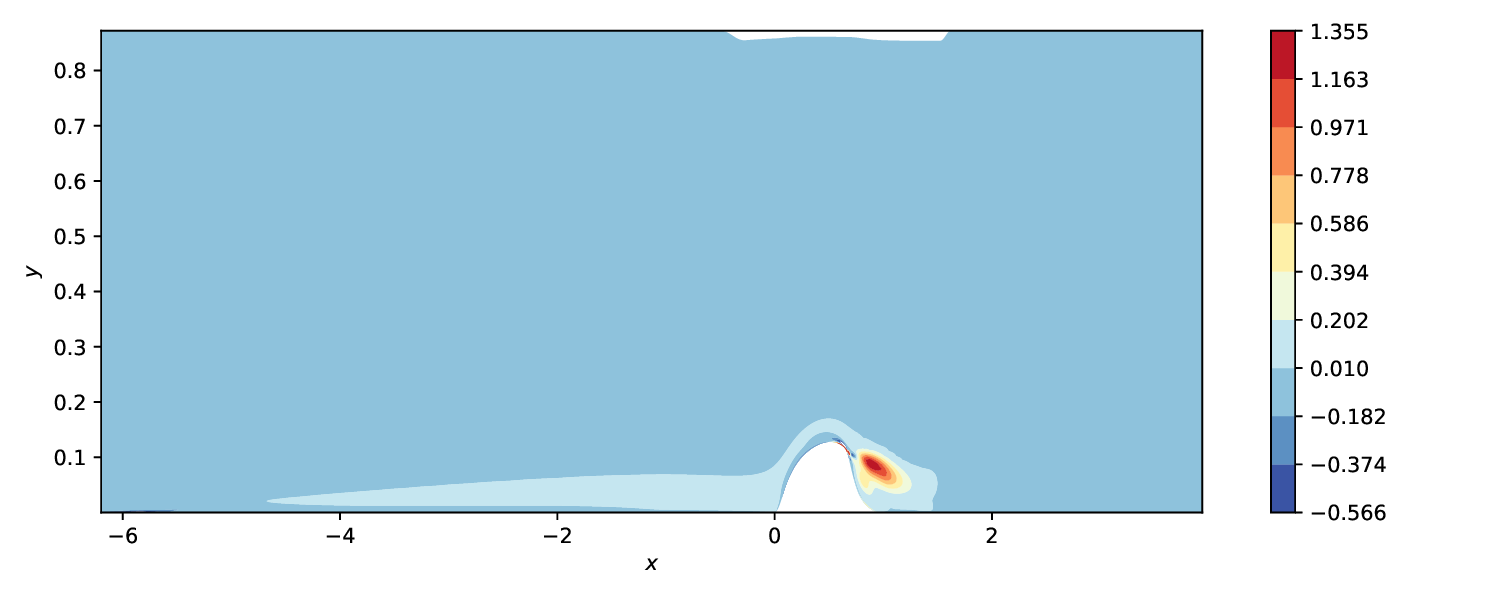}
        \caption{Optimal correction field $\beta^{opt}(\bm{x})$.}
        \label{fig:beta_opt_impli}
    \end{subfigure}
    \caption{Data assimilation with $ \beta $ on Wall-Mounted Hump case handled with implicit layer.}
    \label{fig:beta_wh_impli}
\end{figure}

Moreover, the optimization behavior can be assessed by analyzing the final distribution of $\beta^{opt}(\bm{x})$ and its impact on the eddy viscosity. In particular, we can compare the eddy viscosity obtained for the baseline model $\beta(\bm{x})~=~0$ with that resulting from $\beta^{opt}(\bm{x})$.
Figure~\ref{fig:beta_wh_impli} shows three quantities: the initial eddy viscosity computed with the SA turbulence model, the corresponding eddy viscosity $\mu_t$ obtained using $\beta^{opt}$, and the optimized correction field $\beta^{opt}(\bm{x})$. 

\begin{figure}
    \begin{subfigure}[t]{0.5\textwidth}
        \centering
        \includegraphics[width=\textwidth]{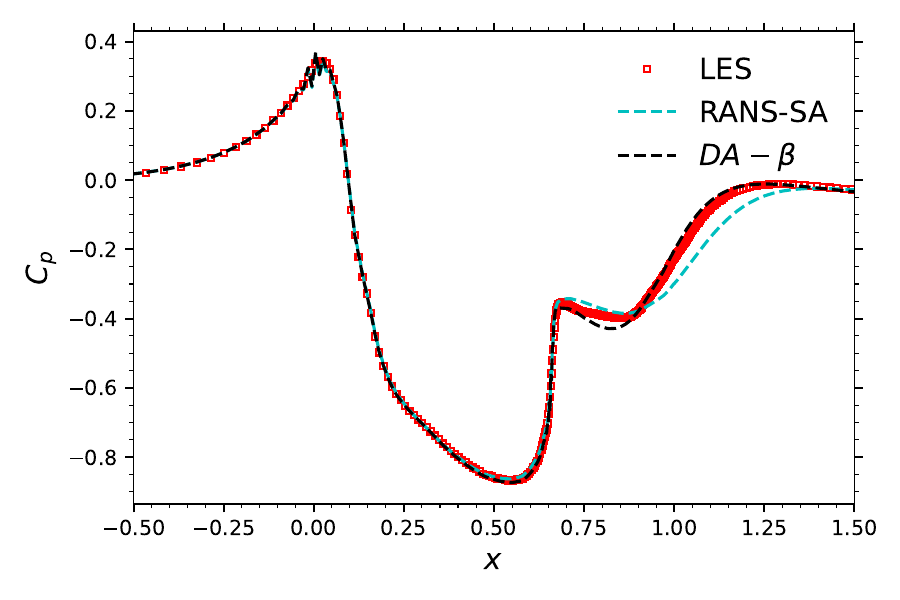}
        \caption{$C_p$.}
        \label{fig:Cp}
    \end{subfigure}
    \begin{subfigure}[t]{0.5\textwidth}
        \centering
        \includegraphics[width=\textwidth]{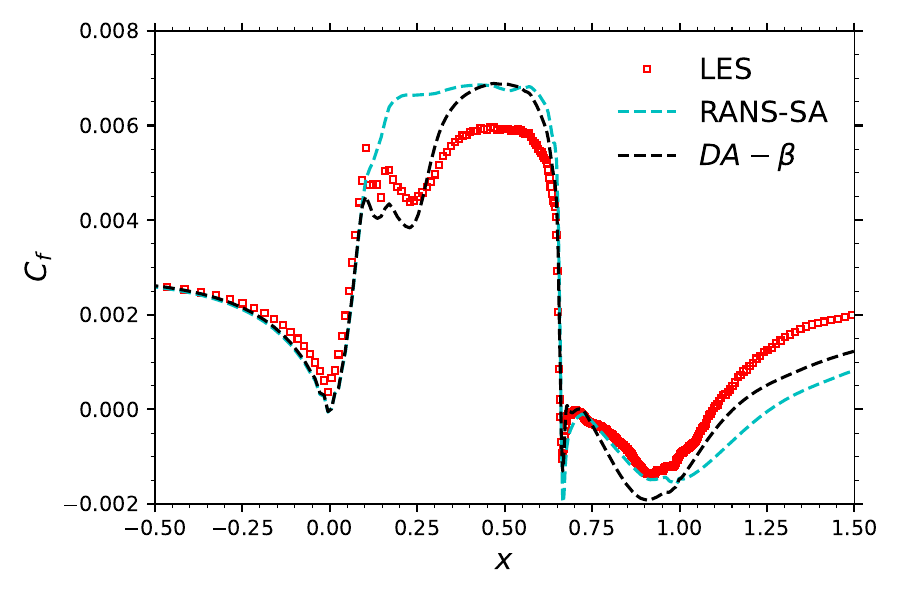}
        \caption{$C_f$.}
        \label{fig:Cf}
    \end{subfigure}
    \caption{Data assimilation with $ \beta $ on Wall-Mounted Hump case handled with implicit layer: comparison between wall-quantities computed from LES data \cite{doi:10.2514/1.J056397}, RANS using the baseline SA model and the corrected RANS model using the optimized $\beta$.}
    \label{fig:Coefficients}
\end{figure}

We can see that the turbulent viscosity computed with the optimized value of $\beta$ has been increased in the recirculation bubble. These results demonstrate the potential of the approach and help identify areas for further refinement and optimization.
One can also compare different models in terms of the $C_f$ skin-friction coefficient and the $C_p$ pressure coefficient. Figure~\ref{fig:Coefficients} shows that the obtained results are closer to the LES values than those of the baseline SA turbulence model, thereby confirming that the optimization has yielded more accurate predictions.

It is important to note that, for this study, at every iteration performed by the optimizer a FP problem must be solved using a Newton method, after which another linear system must be inverted to compute the adjoint solution, as detailed in Section~\ref{sec:FixedPoint}. Each iteration using the BROADCAST solver takes approximately one minute, depending on the hardware used, and benefits from multiprocessing to construct the sparse Jacobian matrices.

\subsubsection{Data assimilation with \texorpdfstring{$\beta$}{b}: full-state measurements and explicit layer}
\label{sec:ResultsWHexplicit}

In order to fully validate the developed framework, the explicit residual minimization strategy is applied to this same case.
As mentioned above, such a strategy requires full-state information $ \bm{w}_m $.
We still consider the corrected model presented in \eqref{eq:NS_SA} but we now apply the explicit optimization framework described in  \crefrange{opt:forward_explicit}{opt:backward_explicit}. The full-state  $\bm{w}_m $ is taken as the one obtained in the last section for the optimized value of $\beta$. The optimization algorithm is initialized with $\beta=0$ in the whole domain. Results are obtained using the Stochastic Gradient Descent (SGD) PyTorch optimizer \citep{ruder2016overview}, a tolerance threshold of $10^{-6}$ being fixed for the variation in the normalized loss function.
 
\begin{figure}
    \centering
    \includegraphics[width=0.5\textwidth]{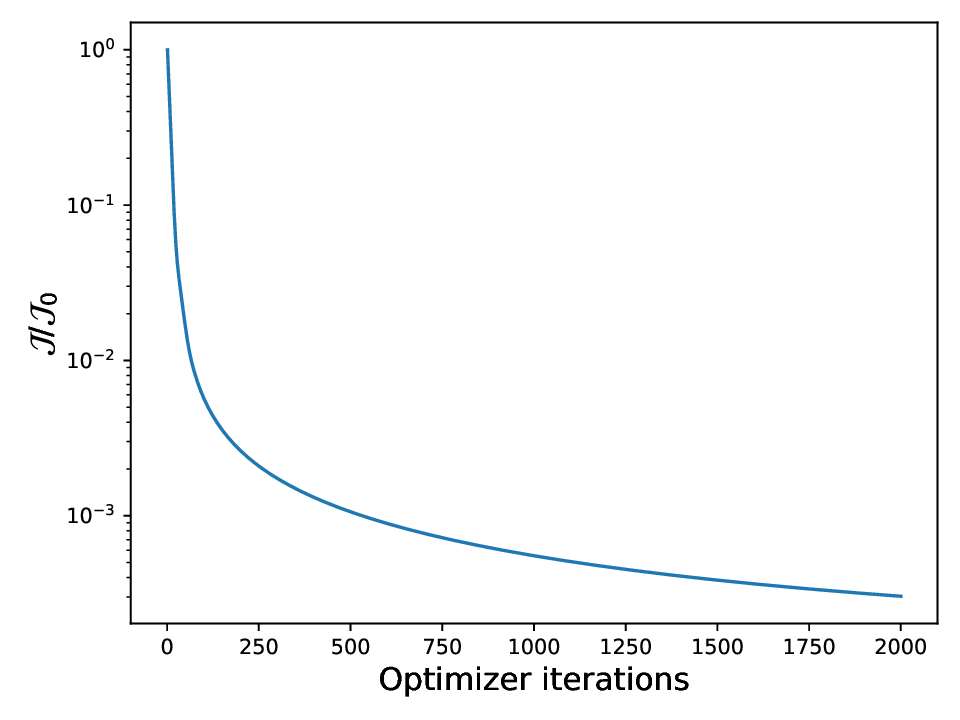}
    \caption{Data assimilation with $ \beta $ on Wall-Mounted Hump case handled with explicit layer: behavior of the normalized cost-function $ \mathcal{J}/\mathcal{J}_0$ consisting here in the residual of the corrected RANS system for 2000 optimizer iterations using the SGD PyTorch optimizer. In this case, the proposed optimization procedure consistently reduces this residual by several orders of magnitude, reflecting a drastically improved consistency between the data and the corrected model.}
    \label{fig:loss_beta_explicit}
\end{figure}

This strategy is considerably faster than the implicit one, since instead of performing several Newton iterations, only a single evaluation of a Fortran function is needed. Figure~\ref{fig:loss_beta_explicit} shows the behavior of the normalized objective function.

\begin{figure}
    \centering
    \includegraphics[width=\textwidth]{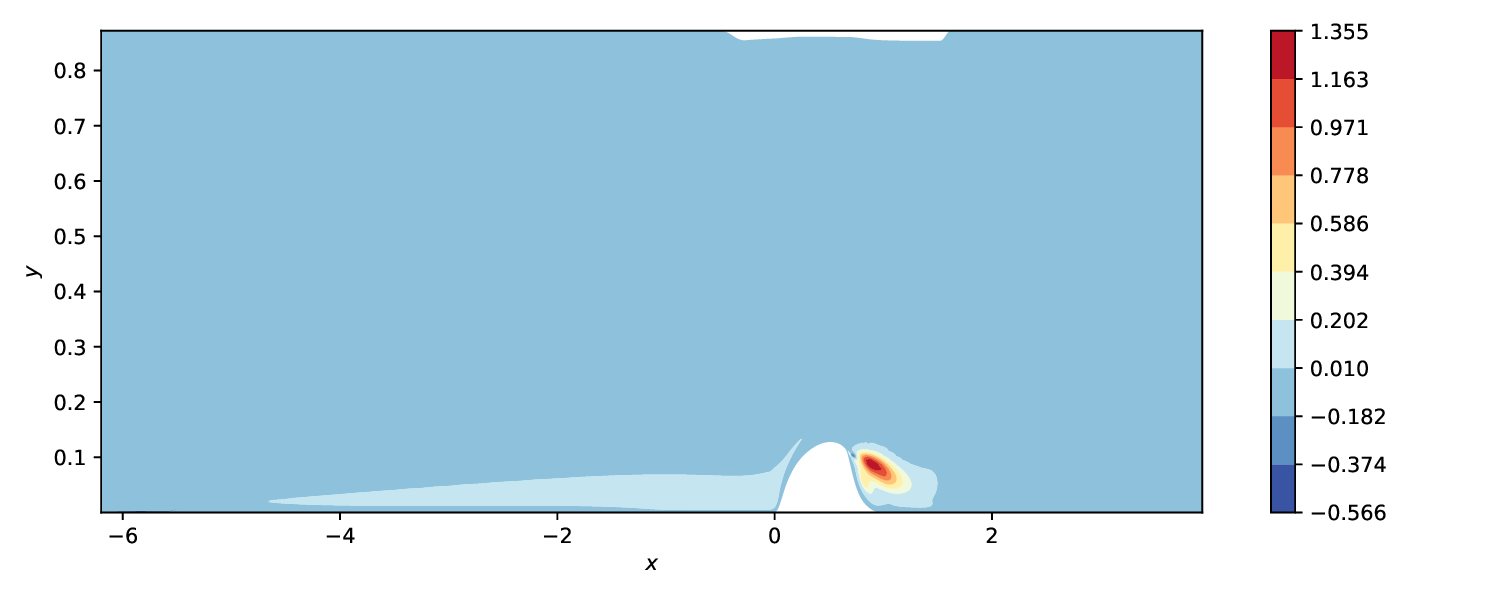}
    \caption{Data assimilation with $ \beta $ on Wall-Mounted Hump case handled with explicit layer: $\beta^{opt}$ obtained by minimizing the residual of RANS equations. It is possible to see that it is practically exact with respect to the results showed in Figure~\ref{fig:beta_wh_impli}.}
    \label{fig:beta_explicit}
\end{figure}

The final values of $\beta$ are shown in Figure~\ref{fig:beta_explicit}.

While the previous approach required solving many linear systems, this one only requires repeated calls to explicit functions of the CFD Fortran code compiled for Python. Therefore, the optimization using this strategy is much faster, performing thousands of iterations within minutes; however, it requires the full state field.

\subsection{Transonic flow through the VKI LS-59 turbine blade cascade: Validation of BROADCAST}
\label{sec:VKI}

We also consider a transonic flow passing between the blades of the VKI LS-59 turbine cascade \citep{Hercus2011}. The LS-59 geometry is a well-studied linear turbine blade cascade used both in experiments and numerical simulations to analyze aerodynamic losses, boundary-layer behavior, shock-boundary layer interactions, wake turbulence, and transition phenomena \citep{LS59exe, 10.1115/1.4067438, CINNELLA20161}. For this case, either a C-mesh or an H-mesh is commonly used in the literature. These mesh types allow for proper discretization of the boundary layer around the airfoil. 

The key aerodynamic phenomena to be addressed include:

\begin{itemize}
  \item Shock-wave formation on the suction side when the flow becomes locally supersonic, and the associated interaction with the boundary layer, possibly leading to separation or shock-boundary layer interaction.  
  \item Wake development downstream of the trailing edge, including turbulence intensity, wake spreading, and mixing losses. These wake effects influence the overall cascade loss and are sensitive to both Mach number and incoming turbulence.
  \item Influence of varying the outlet isentropic Mach number (subsonic vs transonic) on the flow features above (shock strength, separation, wake behavior) and on the losses.
\end{itemize}

\begin{figure}
    \centering
    \includegraphics[width=\textwidth]{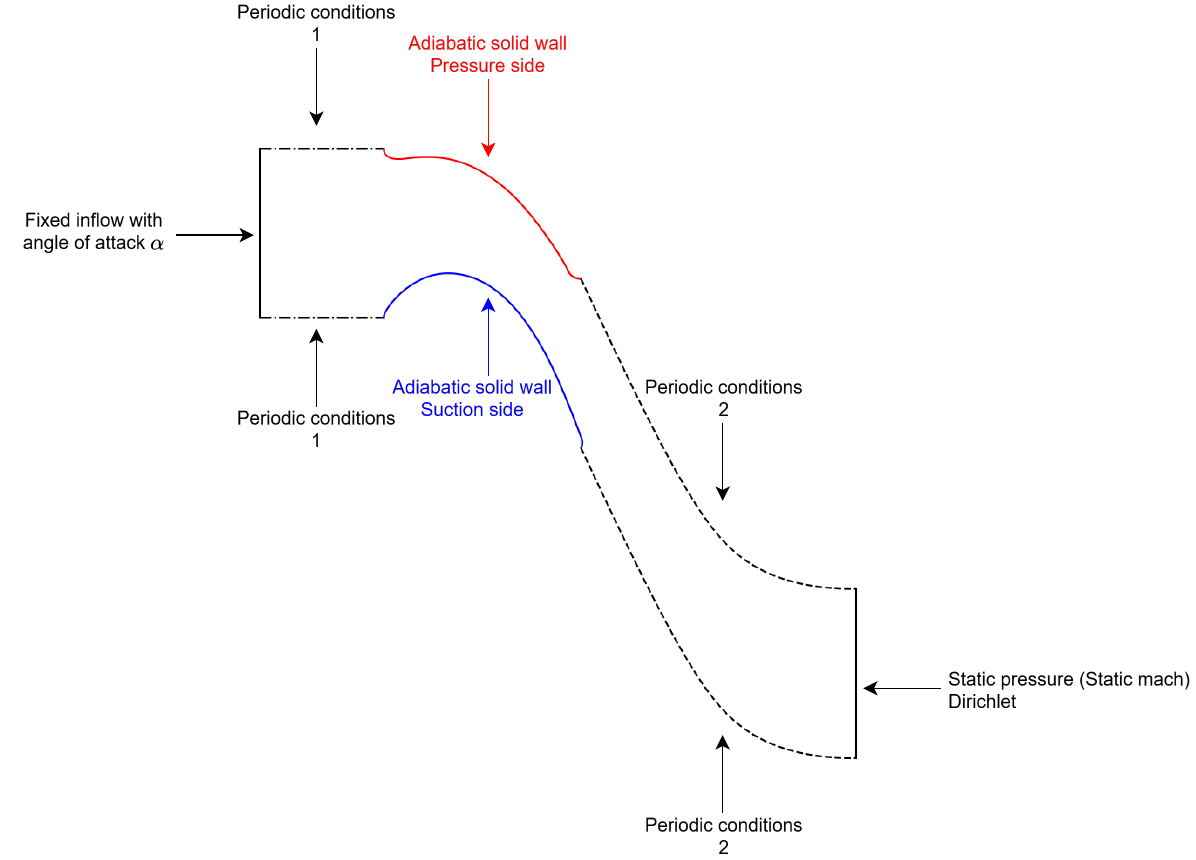}
    \caption{Representation of the outline of the VKI LS-59 structured mesh with the boundary conditions imposed in the CFD solver.}
    \label{fig:geomVKI}
\end{figure}

A mesh is generated to run a RANS simulation on this case. The BROADCAST solver requires structured meshes, as mentioned earlier; therefore, a single-block structured mesh of the inter-blade space is constructed, as shown in Figure~\ref{fig:geomVKI}. The boundary conditions are set as in the transonic case studied in \cite{LS59exe}, as depicted in Figure~\ref{fig:geomVKI}.

The following parameters were fixed: (i) mesh sizes and geometry: $n_i~=~300$, $n_j~=~120$ ($n~=~n_i n_j$), geometry from DOE, (ii) scheme: FE-MUSCL of order 7, (iii) turbulence model: SA, (iv) physical parameters: $Mach_{in}~=~0.275$, $Mach_{out}$ from DOE, $\alpha$ (angle of attack) from DOE, $Re_{in}~=~2.68~\times~10^5$, $Re_{out}~=~6.8~\times~10^5$, $T_{ext}~=~300~K$, $P_{ext}~=~1.01~\times~10^5~Pa$.

\begin{figure}
    \centering
    \includegraphics[width=0.5\textwidth]{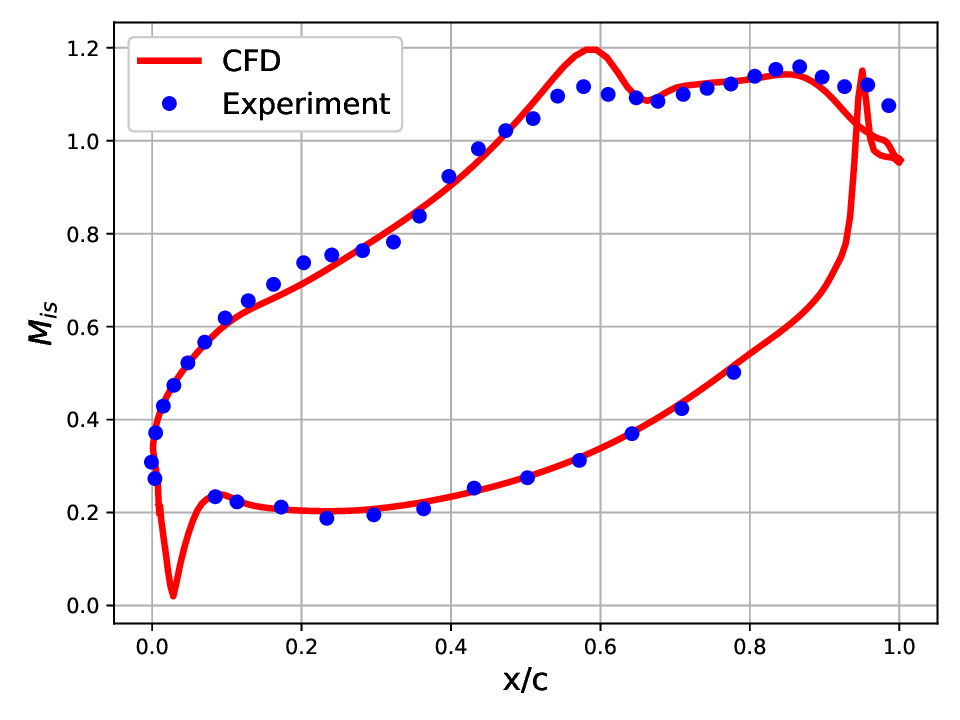}
    \caption{Solution of baseline SA model on VKI LS-59 case with Newton-based relaxation method: isentropic Mach Number at the wall computed with the BROADCAST solver with respect to experimental results extracted from \cite{LS59exe}.}
    \label{fig:VKIMachIS}
\end{figure}

By then solving the RANS equations using the relaxation method (see \S~\ref{sec:newton}), it is possible to compare the results with reference or experimental data. To this end, the isentropic Mach number is computed and compared with the values reported in \cite{LS59exe}. The results are shown in Figure~\ref{fig:VKIMachIS}.

\subsection{Transonic flow through the VKI LS-59 turbine blade cascade: Data assimilation with \texorpdfstring{$\mu_t$}{mt} with full-state measurements and explicit layer}
\label{sec:ResultsVKI}

We use the corrected model presented in \eqref{eq:NS} and the explicit optimization framework described in \crefrange{opt:forward_explicit}{opt:backward_explicit}. The considered correction parameter is the eddy viscosity:
    \begin{equation}
       \mu_{t,{\bm{\vartheta}}}(\bm{w}) ={\bm{\vartheta}} \, ,
    \end{equation}
    
As previously explained, if the full velocity field is available, it is in principle possible to use the residual minimization technique to predict turbulent parameters of the flow. In the present case, given the full state, can we correctly predict the eddy viscosity? To check this, a trainable parameter tensor of the size of the mesh is initialized with a constant value throughout the domain and provided as $\mu_t$ in the routine. The loss function is composed of two terms: first, the norm of the full residual (laminar baseline model plus turbulent forcing) with the state $ \bm{w}_m $ corresponding to the solution computed in the previous section (see Fig. \ref{fig:VKIMachIS}); second, a regularization term defined as $\gamma \| \mu_t \|^2_M$, where $\gamma > 0$, to minimize the values of $ \mu_t $ if not needed (the initial condition starts from a non-zero positive $\mu_t$ value and should tend to zero in the free-stream for instance).

\begin{figure}
    \begin{subfigure}[b]{\textwidth}
        \centering
        \includegraphics[width=0.5\textwidth]{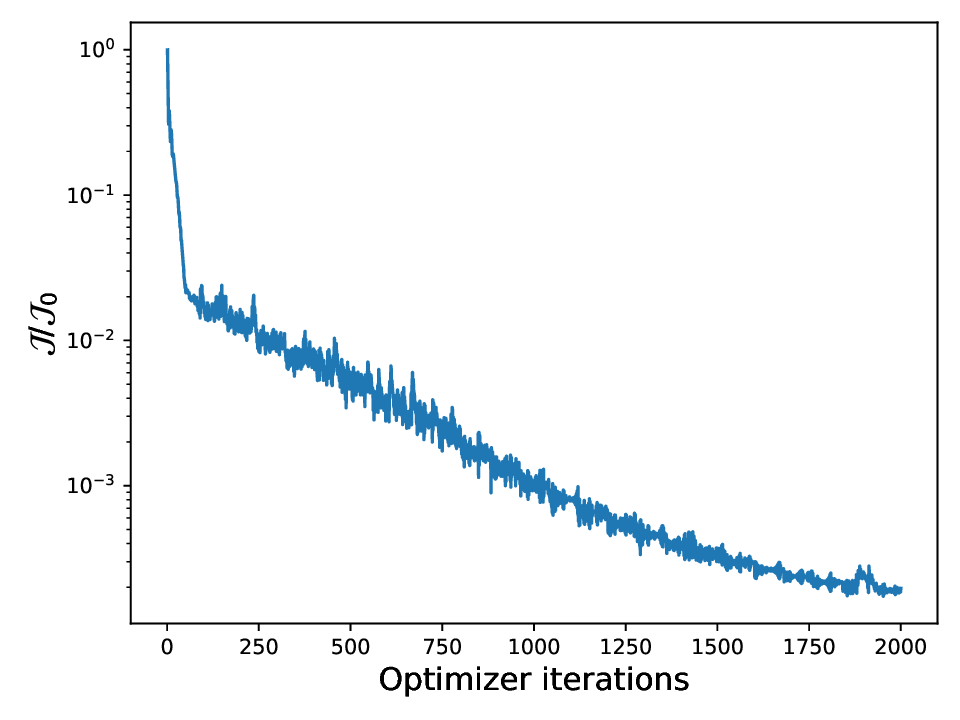}
        \caption{Behavior of the normalized cost-function $ \mathcal{J}/\mathcal{J}_0$ for 2000 optimizer iterations using the SGD PyTorch optimizer. Also in this industrial case, the optimization procedure consistently reduces the residual of the modified RANS equations by 3 orders of magnitude, reflecting an improved consistency of the corrected model.}
        \label{fig:loss_mut_implicit}
    \end{subfigure}
    \begin{subfigure}[t]{0.495\textwidth}
        \centering
        \includegraphics[height=7cm]{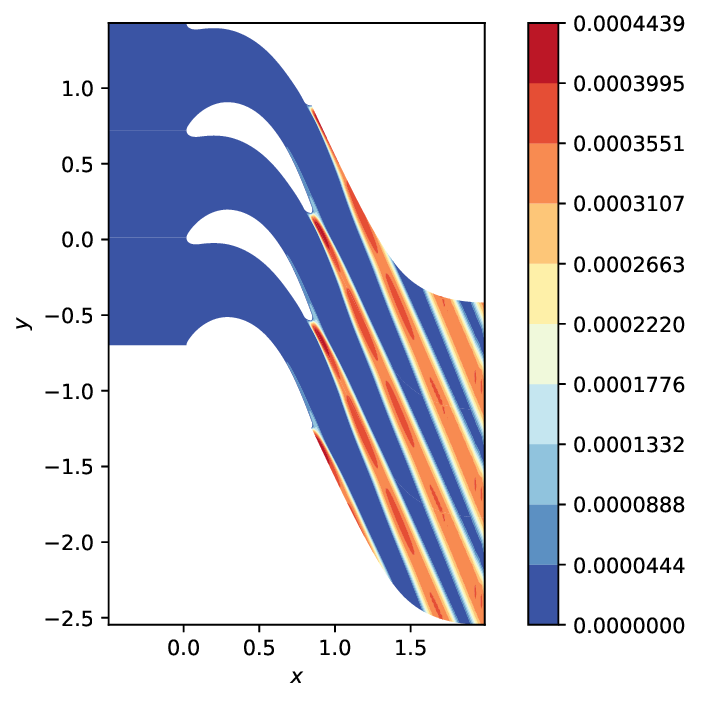}
        \caption{Predicted eddy viscosity after 2000 epochs.}
        \label{fig:mut_implicit}
    \end{subfigure}
    \begin{subfigure}[t]{0.495\textwidth}
        \centering
        \includegraphics[height=7cm]{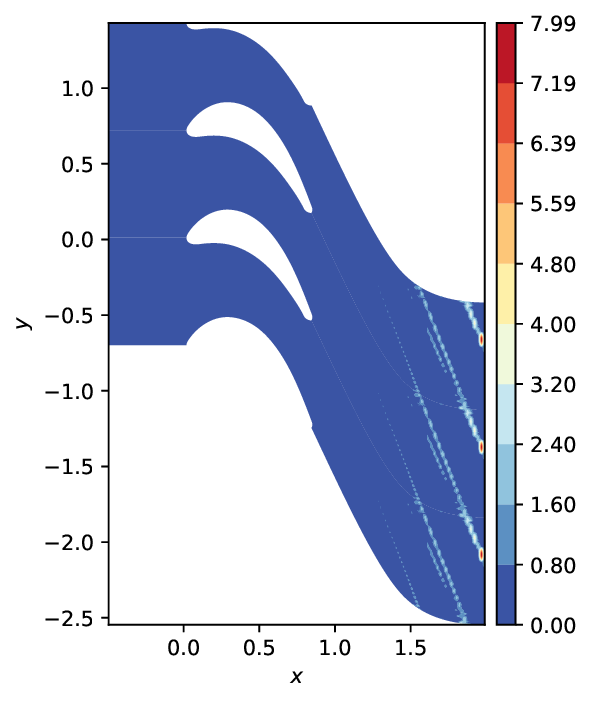}
        \caption{Pointwise absolute error \% with respect to the SA solution (normalized by the maximum value of the SA solution).}
        \label{fig:mut_SA}
    \end{subfigure}
    \caption{Data assimilation with $ \mu_t $ on VKI LS-59 case handled with explicit layer: results of the residual minimization strategy on the VKI LS-59 case with respect to SA results.}
    \label{fig:VKI_mut_implicit}
\end{figure}

By running the routine for 2000 optimizer iterations using SGD, both the behavior of the loss function and the final parameter field demonstrate the capabilities of this strategy.
Figure~\ref{fig:VKI_mut_implicit} shows the evolution of the normalized residual norm (fig. a), the final $\mu_t$ field (fig. b) and the pointwise absolute error of this solution w.r.t the to the SA one, normalized by the maximal value of the SA solution (fig. c).

We can see that, for this case, the optimization framework accurately recovers the eddy viscosity. Since the state variables used are the result of the SA model, the recovered $\mu_t$ exhibits the same features expected from that turbulence model. Indeed, the retrieved $\mu_t$ distribution reproduces the expected behavior, correctly capturing the decay within the viscous sublayers near the walls and the destruction in the outer regions of the boundary layers. In the free-stream, the eddy viscosity is advected downstream without destruction, a characteristic behavior of the SA model.
The largest relative errors ($>7\%$) are localized in the wake region downstream of the blade, where thin convective structures and strong eddy viscosity gradients increase the sensitivity of the prediction. Nevertheless, the overall distribution and magnitude of the eddy viscosity field remain well reproduced.

\subsection{Turbine flow modeling with a CNN for \texorpdfstring{$\mu_t(\bm{w})$}{mt(w)}}
\label{sec:Dataset}

As outlined previously, our objective is to augment RANS simulations with data-driven corrections. To this end, a representative dataset of simulations is required for training the proposed model, with the VKI LS-59 cascade selected as the baseline configuration. In this section, we first present the dataset (\S \ref{sec:DATAGen}), then use the explicit strategy to learn a predictive model for the eddy viscosity based on full-state fields $ \bm{w}_m $ from the dataset (\S \ref{sec:ResultsDATA}).

\subsubsection{Generation of the dataset}
\label{sec:DATAGen}

When using an ML model, special care must be taken, especially if spatial convolutions are required. To balance the requirements imposed by the solver, a single-block structured mesh discretizing the blade passage is considered. For all parameters, the topology and the number of cells in each direction are kept constant. The suction and pressure sides of the blade are split, positioning them in the upper and lower parts of the computational domain, respectively.
This solution addresses the challenges of using ML models but requires particular attention during mesh generation to: (i) correctly discretize the boundary layer, (ii) avoid overly stretched cells at the leading and trailing edges of the blade, and (iii) adapt the domain shape to better align the flow with the generalized computational coordinates.
To achieve this, a custom elliptic mesh generator has been implemented. Specifically, by varying the control points used to define the camber line of the blade and the angle of rotation, many different meshes can be generated with the same tool.

\begin{figure}
    \centering
    \includegraphics[width=\textwidth]{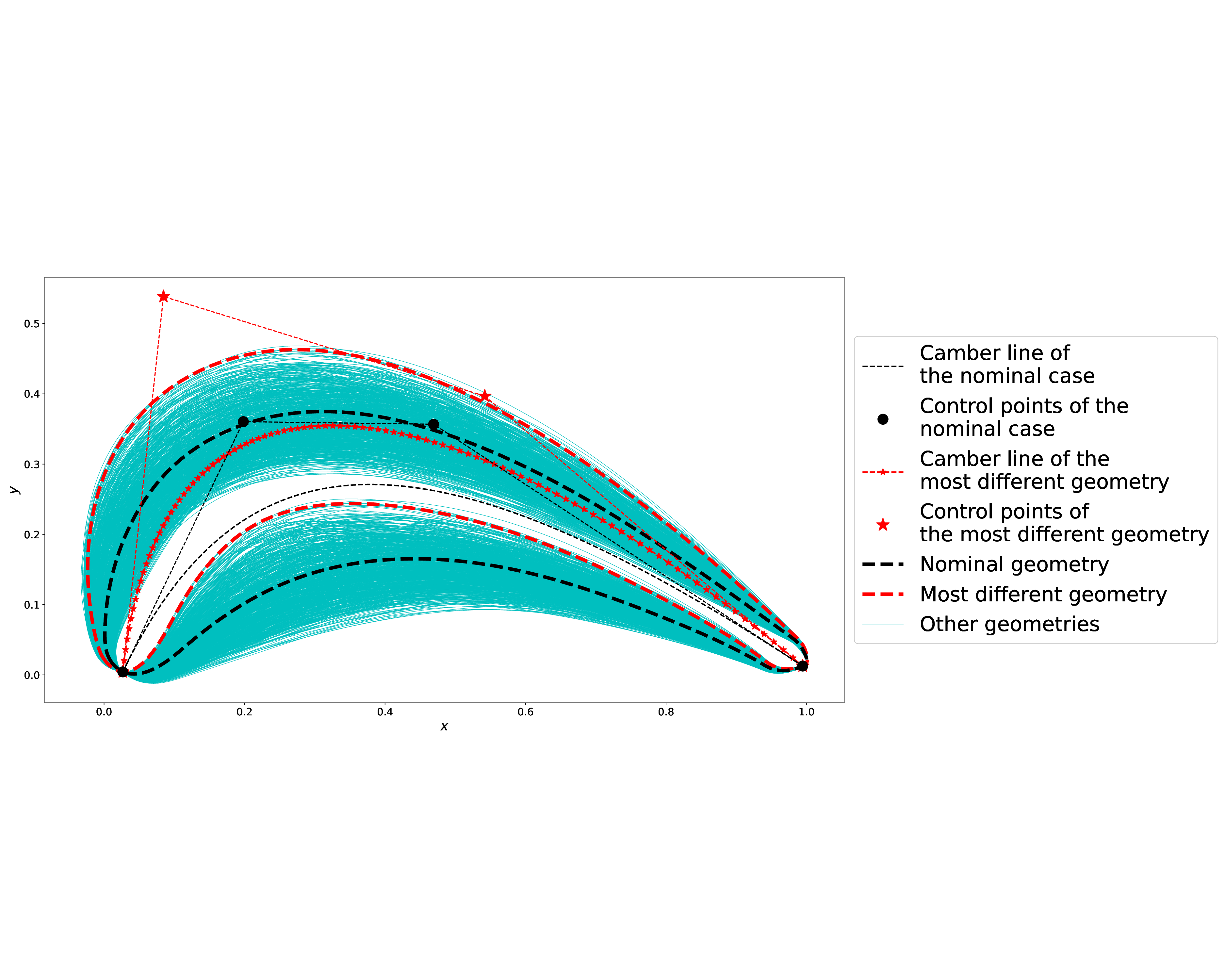}
    \caption{Superposition of all generated blade geometries with the original blade profile and control points (from \cite{Hercus2011}).}
    \label{fig:blades}
\end{figure}

To illustrate the effectiveness of the mesh generator, Figure~\ref{fig:blades} presents the collection of blade profiles generated, in comparison with the original blade geometry and its corresponding control points. These variations highlight the flexibility of the parametrization strategy in exploring a wide design space around the reference configuration. The original profile and control points, extracted from \cite{Hercus2011}, are shown for reference.

The resulting dataset, comprising all generated geometries and corresponding simulation data, has been made openly accessible to the research community and is available at \cite{bucci_2025_14840512}. This resource aims to foster reproducibility and further investigations in the field of turbo-machinery optimization and data-driven turbulence modeling. Moreover, this dataset, together with several others structured in the Plaid format, has been used to assess the performance of different surrogate models. A comprehensive analysis of these applications is provided in~\cite{casenave2025physicslearningaidatamodelplaid}.

To create the dataset, a Design of Experiments (DOE) was generated using the maximum projection algorithm \citep{MaxProj}; Table~\ref{tab:DOE} shows the intervals chosen for the parameters\footnote{Further details can be found at \cite{bucci_2025_14840512}.}, and the remaining parameters were fixed as detailed in Section~\ref{sec:VKI}.

\begin{table}
    \centering
    \caption{Parameters and ranges used to generate the DOE.}
    \label{tab:DOE}
    \small
    \begin{tabular}{lccc}
        \toprule
        Parameter & Minimal Value & Maximal Value \\
        \midrule
        Distance of the 2$^{\circ}$ point from the 1$^{\circ}$ control point & -0.15  & 0.15 \\
        Angle of attack & 10$^{\circ}$ & 50$^{\circ}$ \\
        $\Delta x$ of the 3$^{\circ}$ control point from the original geometry & -0.2 & 0.2 \\
        $\Delta y$ of the 3$^{\circ}$ control point from the original geometry & -0.2 & 0.2 \\
        Pitch (inter-blades distance) & 0.68 & 1.1 \\
        Mach number at outflow & 0.75 & 0.98 \\
        \bottomrule
    \end{tabular}
\end{table}

In Appendix~\ref{sec:complete} the dataset is analyzed further, showing its completeness and sensitivity with respect to the parameters used in the DOE; see Table~\ref{tab:DOE}.

\subsubsection{Closure with \texorpdfstring{$\mu_t$}{mt} using a CNN and explicit layer}
\label{sec:ResultsDATA}

We still use the corrected model presented in \eqref{eq:NS} and the explicit optimization framework described in  \crefrange{opt:forward_explicit}{opt:backward_explicit}.

Optimization with respect to an eddy viscosity model depending on the state $\bm{w}$ is now performed:
\begin{equation}
    \mu_{t,{\bm{\vartheta}}}(\bm{w}) =\mathcal{M}_{{\bm{\vartheta}}}(\bm{w})\, ,
\end{equation}
The model $\mathcal{M}_{{\bm{\vartheta}}}$ can be any kind of NN. For this study, a CNN is selected since we are dealing with structured meshes. CNNs exploit spatial correlations through convolutional filters, making them particularly effective at capturing local flow features such as gradients, shear layers, and boundary-layer structures. Moreover, compared to fully connected or recurrent architectures, CNNs require fewer parameters for high-dimensional inputs, improving both training efficiency and generalization.

The dataset generated from BROADCAST simulations in the last section has been split into training (80\%) and validation (20\%) subsets. This division ensures that the model can be optimized effectively while monitoring its generalization performance on unseen data.

Different optimization algorithms available in PyTorch were tested; the L-BFGS optimizer was ultimately selected to produce the reported results, using a 
mini-batch size of 100. During training, the cost functional $\mathcal{J}/\mathcal{J}_0$ decreases by approximately two orders of magnitude, reaching $\mathcal{J}_f/\mathcal{J}_0 = (4.75 \pm 1.39)\times 10^{-3}$ over 10 independent runs (mean $\pm$ standard deviation). The corresponding validation value is $\mathcal{J}_f^{\mathrm{val}}/\mathcal{J}_0 = (6.35 \pm 0.83)\times 10^{-3}$, indicating limited sensitivity
to the random initialization of the network weights and to the mini-batch training procedure. Although L-BFGS is classically formulated in a deterministic full-batch setting, mini-batch variants have also been investigated in the optimization literature~\citep{berahas2019robustmultibatchlbfgsmethod}. Although these final values are of the same order of magnitude as those obtained in Fig.~\ref{fig:loss_mut_implicit}, the latter corresponds to the direct optimization of $\mu_t$ for a single configuration, for which nearly perfect convergence could be achieved. In the present case, the optimization is carried out over a diverse set of configurations, so that the network parameters must provide a global compromise rather than an exact optimum for a single case.

\begin{figure}
    \begin{subfigure}[b]{0.5\textwidth}
        \centering
        \includegraphics[height=7cm]{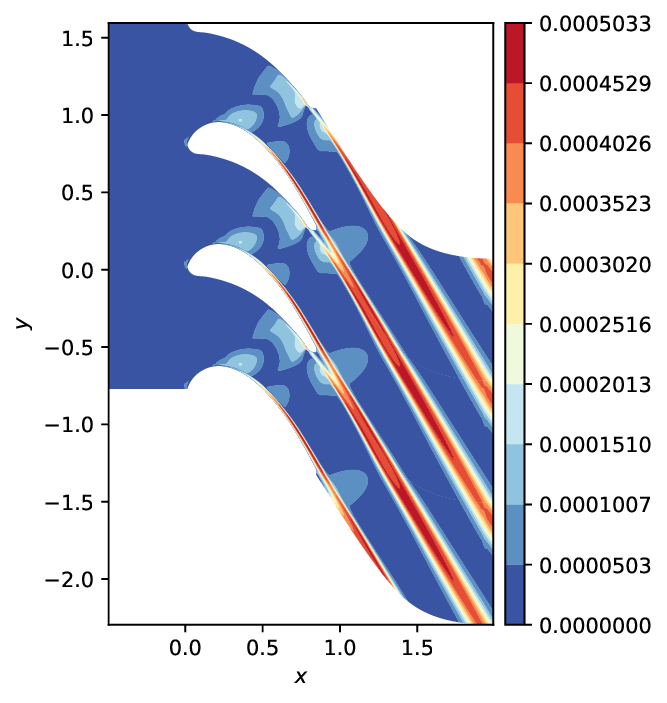}
        \caption{Predicted eddy viscosity $\mu_{t,{\bm{\vartheta}}}$ on case 676.}
        \label{fig:data_mut_explicit_676}
    \end{subfigure}
    \begin{subfigure}[b]{0.5\textwidth}
        \centering
        \includegraphics[height=7cm]{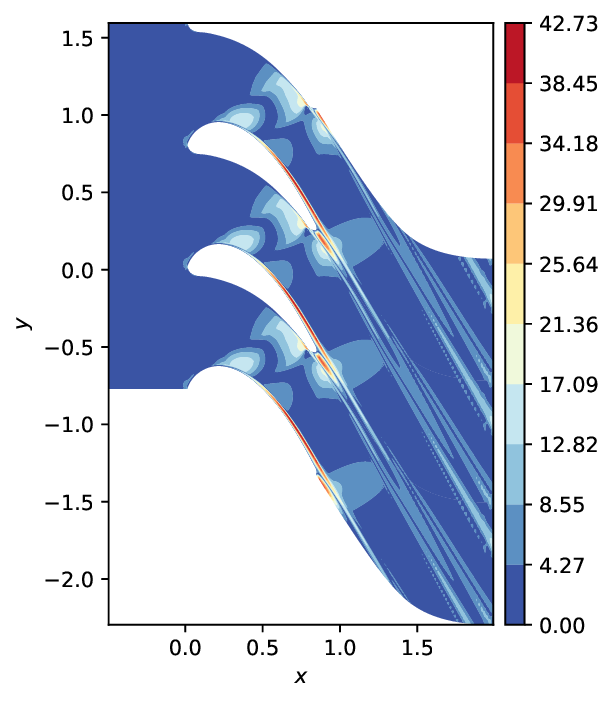}
        \caption{Error \% with respect to the SA turbulence model on case 676.}
        \label{fig:data_err_explicit_676}
    \end{subfigure}\\
    \begin{subfigure}[b]{0.5\textwidth}
        \centering
        \includegraphics[height=7cm]{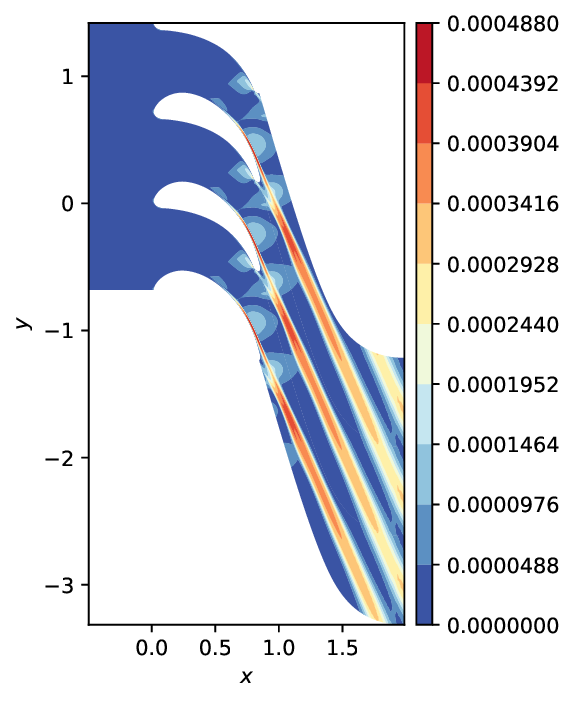}
        \caption{Predicted eddy viscosity $\mu_{t,{\bm{\vartheta}}}$ on case 679.}
        \label{fig:data_mut_explicit_679}
    \end{subfigure}
    \begin{subfigure}[b]{0.5\textwidth}
        \centering
        \includegraphics[height=7cm]{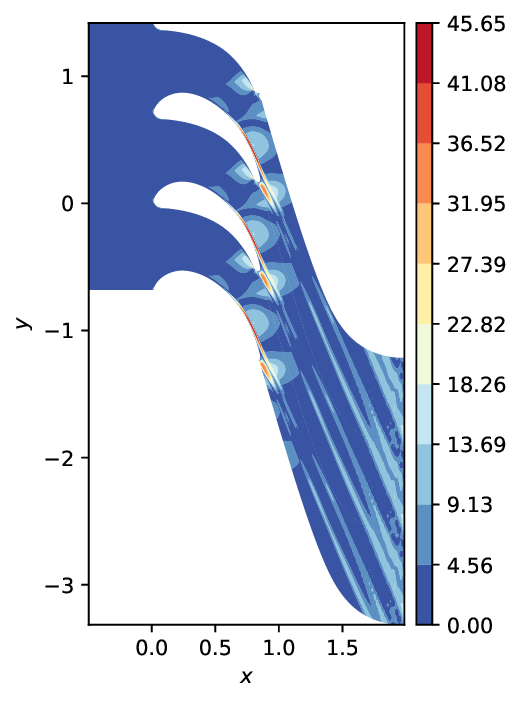}
        \caption{Error \% with respect to the SA turbulence model on case 679.}
        \label{fig:data_err_explicit_679}
    \end{subfigure}
    \caption{Closure with a CNN for $ \mu_t $ on custom dataset handled with explicit layer: eddy viscosity prediction on 2 cases extracted from the validation set.}
    \label{fig:data_mut_explicit}
\end{figure}

Figure~\ref{fig:data_mut_explicit} provides a visualization of the model predictions for two randomly selected configurations (cases 676 and 679, whose simulation parameters are presented in Table~\ref{tab:2cases}) drawn from the validation set. For each configuration, two fields are shown to offer a thorough assessment of the model's performance. The first field represents the predicted eddy viscosity distribution, denoted as $\mu_{t,{\bm{\vartheta}}}(\bm{w})$, across the entire domain. The second field illustrates the absolute pointwise percentage error in the eddy viscosity with respect to the SA eddy viscosity obtained with the BROADCAST solver, $ |\mu_{t,{\bm{\vartheta}}}(\bm{w})(\bm{x}) - \mu_{t}^{SA}(\bm{x})|/\max_{\bm{x}}{\mu_{t}^{SA}(\bm{x})}$.
\begin{table}
    \centering
    \caption{Simulation parameters for cases 676 and 679.}
    \label{tab:2cases}
    \small
    \begin{tabular}{lccc}
        \toprule
        Parameter & 676 & 679 \\
        \midrule
        $\Delta x$ of the 3$^{\circ}$ control point from the original geometry & -0.1468 & 0.1799 \\
        $\Delta y$ of the 3$^{\circ}$ control point from the original geometry & -0.1329 & 0.1289 \\
        Pitch (inter-blades distance) & 0.7893 & 0.7 \\
        Distance of the 2$^{\circ}$ point from the 1$^{\circ}$ control point & -0.1467  & -0.1491 \\
        Angle of attack & 31$^{\circ}$ & 35$^{\circ}$ \\
        Mach number at outflow & 0.8645 & 0.7743 \\
        \bottomrule
    \end{tabular}
\end{table}

The results show that the CNN-based algebraic turbulence model, despite its locality constraint, is capable of accurately reconstructing the eddy viscosity in the wake of profiles. This can be attributed to the fact that the CNN does not rely on explicit quantities such as the distance to the closest wall, in contrast to classical algebraic turbulence models. Furthermore, the errors are small in the attached regions of the boundary layers, while they remain non-negligible on the suction side of the profiles near the trailing edge, where flow separation occurs. This clearly indicates that a local algebraic turbulence model does not possess sufficient degrees of freedom to accommodate the full range of flow regimes, including attached boundary layers, separated boundary layers, and wake regions.

To further assess the robustness of the proposed framework, we consider an ensemble of models differing only by random initialization and stochastic optimization. The ensemble is constructed from the same 10 runs used in the training and validation loss analysis, allowing us to quantify prediction variability and estimate epistemic uncertainty.

\begin{figure}
    \begin{subfigure}[b]{0.5\textwidth}
        \centering
        \includegraphics[height=7cm]{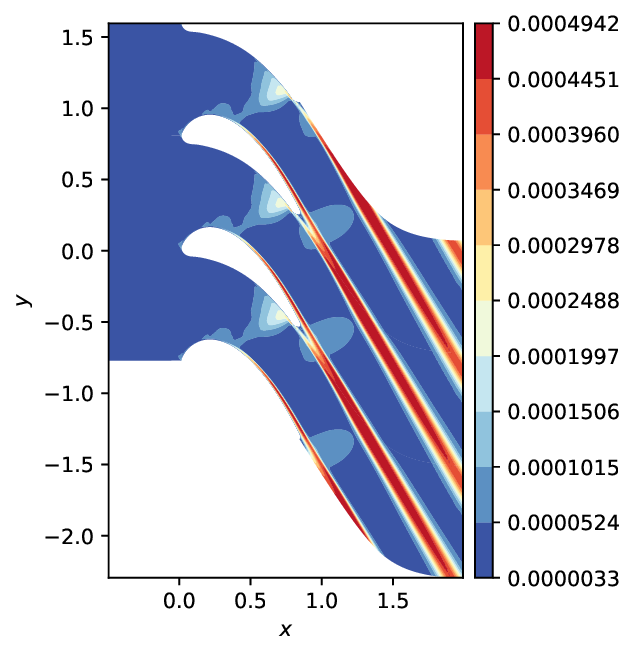}
        \caption{Ensemble mean.}
        \label{fig:mean_676}
    \end{subfigure}
    \begin{subfigure}[b]{0.5\textwidth}
        \centering
        \includegraphics[height=7cm]{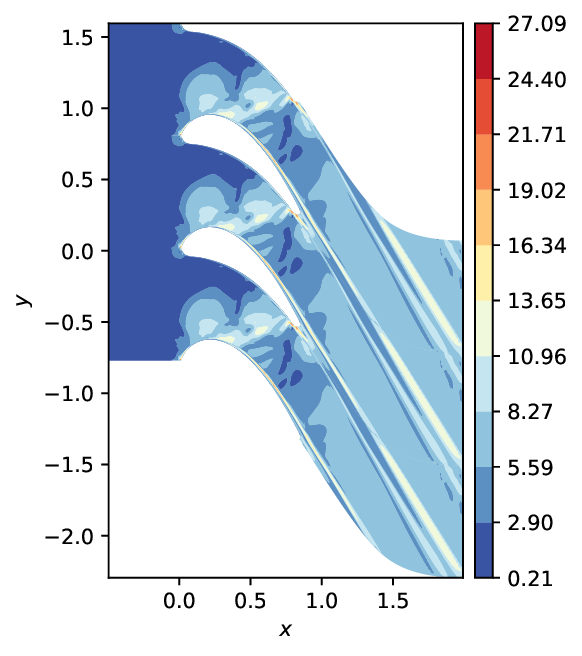}
        \caption{Ensemble standard deviation (\%).}
        \label{fig:std_676}
    \end{subfigure}
    \caption{Closure with a CNN for $ \mu_t $ on custom dataset handled with explicit layer: ensemble statistics of the predicted turbulent viscosity field for case 676. The mean field highlights the robustness of the prediction, while the standard deviation identifies regions of higher variability across models.}
    \label{fig:ensemble_676}
\end{figure}

For case 676, previously presented, we compute the ensemble mean and standard deviation of the predicted turbulent viscosity field. The ensemble mean, shown in Figure~\ref{fig:mean_676}, closely matches the previously reported prediction, demonstrating the consistency of the model across different training instances.

In contrast, the standard deviation, reported in Figure~\ref{fig:std_676}, reveals localized regions of increased variability. Notably, these regions correspond to areas where larger prediction errors were observed in the previous analysis. This correlation between ensemble spread and prediction error suggests that the standard deviation can serve as a meaningful indicator of model uncertainty and reliability.

Overall, this analysis highlights the benefit of ensemble-based approaches for identifying regions of reduced confidence in the predictions, providing additional insight that is not accessible from a single model realization.

Several directions could be explored to further improve the proposed framework. These include the use of richer architectures (e.g., increased channel capacity or alternative field embeddings such as implicit neural representations), the incorporation of tags in the input fields to explicitly encode the boundary conditions information and modifications of the loss function, for instance through importance sampling based on velocity gradients or by adding a regularization term directly on the turbulent viscosity. Finally, much care must be taken to ensure that the trained model behaves robustly when embedded within a Newton solver, in particular with respect to FP stability. These aspects will be addressed in future work.

\section{Discussions and conclusions}
\label{sec:Conclusion}

The main objective of this work was to demonstrate the capability of the proposed framework to perform end-to-end optimization within a fully differentiable and physically consistent pipeline.

After introducing the optimization problem, the reference configurations, and the training dataset, the first two test cases show that both optimization strategies are fully operational. In particular, the explicit residual minimization and the implicit Newton-like solver correctly propagate gradients through the entire computational chain in a truly end-to-end fashion. This validates the correctness of the proposed formulation and its seamless integration within a differentiable programming environment.

The results further confirm the effectiveness of the proposed approach in predicting the eddy viscosity directly from the conservative flow variables. These findings highlight the potential of the framework as a reliable tool for data-assisted turbulence modeling and provide a solid basis for future developments aimed at improving accuracy, robustness, and scalability.

While demonstrated here for RANS turbulence modeling, the proposed framework is general and extends naturally to PDE-constrained systems with embedded closure models. It thus establishes a consistent and scalable paradigm for integrating trainable components within physics-based solvers via differentiable formulations and adjoint-based optimization.

The most natural extension of this work is the application of the implicit (Deep Equilibrium) strategy to the developed dataset, still targeting the prediction of the eddy viscosity in the Spalart-Allmaras turbulence model. Further extensions may include the adoption of more complex neural architectures, the treatment of different closure problems, and the investigation of transfer capabilities across distinct flow regimes. These directions will contribute to strengthening the synergy between numerical modeling and data-driven methodologies in computational physics.

\begin{Backmatter}

\paragraph{Acknowledgments}
We would like to thank Pedro Stefanin Volpiani for all the helpful discussions and advice during the development of this work.

\paragraph{Funding Statement}
This work was partially supported by the French National Research Agency under project ANR-22-FAI2-0002-01.

\paragraph{Competing Interests}
The authors declare none.

\paragraph{Data Availability Statement}
The presented VKI dataset \citep{bucci_2025_14840512}, comprising all generated geometries and corresponding simulation data, has been made openly accessible to the research community and is available at \url{https://doi.org/10.5281/zenodo.14840512}. The open-source CFD solver BROADCAST \citep{Poulain2023} is available on GitHub at \url{https://github.com/onera/Broadcast}. The LES data of the 2D NASA Wall-Mounted Hump can be accessed at \url{https://turbmodels.larc.nasa.gov/Other_LES_Data/nasa_hump_uzun_2017.html}.

\paragraph{Ethical Standards}
The research meets all ethical guidelines, including adherence to the legal requirements of the study country.

\paragraph{Author Contributions}
Conceptualization: M.A.B., C.C., D.S.; Software: L.S., G.F., C.C; Validation: L.S., G.F.; Methodology: L.S.,M.A.B., C.C.; Data curation: L.S., M.A.B., G.F.; Data visualisation: L.S.; Writing original draft: L.S.; Supervision: M.A.B., C.C., D.S.; All authors approved the final submitted draft.

\printbibliography

@article{crivellini2011implicit,
  title={An implicit matrix-free discontinuous Galerkin solver for viscous and turbulent aerodynamic simulations},
  author={Andrea Crivellini and Francesco Bassi},
  journal={Computers \& fluids},
  volume={50},
  number={1},
  pages={81--93},
  year={2011},
  publisher={Elsevier}
}

@article{franceschini2020mean,
  title={Mean-flow data assimilation based on minimal correction of turbulence models: Application to turbulent high Reynolds number backward-facing step},
  author={Franceschini, Lucas and Sipp, Denis and Marquet, Olivier},
  journal={Physical Review Fluids},
  volume={5},
  number={9},
  pages={094603},
  year={2020},
  publisher={APS}
}

@article{talagrand1997assimilation,
  title={Assimilation of observations, an introduction},
  author={Talagrand, Olivier},
  journal={Journal of the Meteorological Society of Japan},
  volume={75},
  number={1B},
  pages={191--209},
  year={1997}
}

@book{chapman1990mathematical,
  title={The mathematical theory of non-uniform gases: an account of the kinetic theory of viscosity, thermal conduction and diffusion in gases},
  author={Chapman, Sydney and Cowling, Thomas George},
  year={1990},
  publisher={Cambridge university press}
}

@book{huang2008statistical,
  title={Statistical Mechanics, 2nd Ed},
  author={Huang, K.},
  isbn={9788126518494},
  url={https://books.google.fr/books?id=ZHl8HLk-K3AC},
  year={2008},
  publisher={Wiley India Pvt. Limited}
}

@book{torquato2002random,
  title={Random heterogeneous materials: microstructure and macroscopic properties},
  author={Torquato, Salvatore and others},
  volume={16},
  year={2002},
  publisher={Springer}
}

@article{End-to-end,
   author = {Carlos A. Michelén Ströfer and Heng Xiao},
   doi = {10.1016/j.taml.2021.100280},
   issn = {20950349},
   issue = {4},
   journal = {Theoretical and Applied Mechanics Letters},
   title = {{End-to-end differentiable learning of turbulence models from indirect observations}},
   volume = {11},
   year = {2021},
}

@article{Wu2018,
   author = {Jin Long Wu and Heng Xiao and Eric Paterson},
   doi = {10.1103/PhysRevFluids.3.074602},
   issn = {2469990X},
   issue = {3},
   journal = {Physical Review Fluids},
   title = {{Physics-informed machine learning approach for augmenting turbulence models: A comprehensive framework}},
   volume = {7},
   year = {2018},
}

@inproceedings{Hercus2011,
   author = {Samuel J. Hercus and Paola Cinnella},
   doi = {10.1115/AJK2011-05007},
   issue = {PARTS A, B, C, D},
   booktitle = {ASME-JSME-KSME 2011 Joint Fluids Engineering Conference, AJK 2011},
   title = {Robust shape optimization of uncertain dense gas flows through a plane turbine cascade},
   volume = {1},
   year = {2011},
}

@inproceedings{Tapenade,
   author = {Laurent Hascoët},
   booktitle = {ECCOMAS 2004 - European Congress on Computational Methods in Applied Sciences and Engineering},
   title = {Tapenade: A tool for automatic differentiation of program},
   year = {2004},
}

@article{AutomaticD,
   author = {Laurent Hascoët},
   journal = {INRIA Sophia-Antipolis, TROPICS team. Available on-line},
   title = {Automatic differentiation by program transformation},
   year = {2007},
}

@article{Poulain2023,
   author = {Arthur Poulain and Cédric Content and Denis Sipp and Georgios Rigas and Eric Garnier},
   doi = {https://doi.org/10.1016/j.cpc.2022.108557},
   issn = {0010-4655},
   journal = {Computer Physics Communications},
   pages = {108557},
   title = {{BROADCAST: A high-order compressible CFD toolbox for stability and sensitivity using Algorithmic Differentiation}},
   volume = {283},
   url = {https://www.sciencedirect.com/science/article/pii/S0010465522002764},
   year = {2023},
}

@article{Bai2019,
   author = {Shaojie Bai and J. Zico Kolter and Vladlen Koltun},
   journal = {Advances in Neural Information Processing Systems},
   title = {Deep equilibrium models},
   volume = {32},
   year = {2019},
}

@article{LS59exe,
title = {Experimental investigation of subsonic and transonic flows through a linear turbine cascade},
journal = {European Journal of Mechanics - B/Fluids},
volume = {103},
pages = {182-192},
year = {2024},
issn = {0997-7546},
doi = {https://doi.org/10.1016/j.euromechflu.2023.10.002},
url = {https://www.sciencedirect.com/science/article/pii/S0997754623001437},
author = {Idalia Jagodzińska and Bartosz Olszański and Konrad Gumowski and Sławomir Kubacki}
}

@misc{spalart,
author = {NASA - Christopher Rumsey},
year = {2020},
title = {{The Spalart-Allmaras Turbulence Model}},
note = {Accessed: 2023-07-03},
howpublished = {\url{https://turbmodels.larc.nasa.gov/spalart.html}}
}

@inproceedings{spalart1992one,
  title={A one-equation turbulence model for aerodynamic flows},
  author={Philippe Spalart and Steven Allmaras},
  booktitle={30th aerospace sciences meeting and exhibit},
  pages={439},
  year={1992}
}

@phdthesis{boussinesq1877,
  author       = {Joseph Boussinesq},
  title        = {{Essai sur la théorie des eaux courantes}},
  year         = {1877},
  school       = {Académie des Sciences},
  type         = {Mémoires présentés par divers savants},
  volume       = {23},
  number       = {1},
  pages        = {1--680},
  address      = {Paris, France}
}

@inproceedings{ADPyTorch,
  title={{Automatic differentiation in PyTorch}},
  author={Adam Paszke and Sam Gross and Soumith Chintala and Gregory Chanan and Edward Yang and Zachary DeVito and Zeming Lin and Alban Desmaison and Luca Antiga and Adam Lerer},
  booktitle={NIPS-W},
  year={2017}
}

@misc{bucci_2025_14840512,
  author       = {Michele Alessandro Bucci and Luca Saverio and Fabien Casenave},
  title        = {{VKI-LS59: a 2D internal aero CFD RANS dataset,
                   under geometrical variations
                  }},
  month        = feb,
  year         = 2025,
  publisher    = {Zenodo},
  doi          = {10.5281/zenodo.14840512},
  url          = {https://doi.org/10.5281/zenodo.14840512},
}

@article{MaxProj,
 ISSN = {00063444},
 URL = {http://www.jstor.org/stable/43908541},
 author = {V. Roshan Joseph and Evren Gul and Shan Ba},
 journal = {Biometrika},
 number = {2},
 pages = {371--380},
 publisher = {[Oxford University Press, Biometrika Trust]},
 title = {Maximum projection designs for computer experiments},
 urldate = {2025-02-24},
 volume = {102},
 year = {2015}
}

@article{CINNELLA20161,
title = {High-order implicit residual smoothing time scheme for direct and large eddy simulations of compressible flows},
journal = {Journal of Computational Physics},
volume = {326},
pages = {1-29},
year = {2016},
issn = {0021-9991},
doi = {https://doi.org/10.1016/j.jcp.2016.08.023},
url = {https://www.sciencedirect.com/science/article/pii/S0021999116303734},
author = {P. Cinnella and C. Content}
}

@misc{lbfgs,
      title={{A Regularized Limited Memory BFGS method for Large-Scale Unconstrained Optimization and its Efficient Implementations}}, 
      author={Hardik Tankaria and Shinji Sugimoto and Nobuo Yamashita},
      year={2021},
      eprint={2101.04413},
      archivePrefix={arXiv},
      primaryClass={math.OC},
      url={https://arxiv.org/abs/2101.04413}, 
}

@misc{casenave2023mmgpmeshmorphinggaussian,
      title={{MMGP: a Mesh Morphing Gaussian Process-based machine learning method for regression of physical problems under non-parameterized geometrical variability}}, 
      author={Fabien Casenave and Brian Staber and Xavier Roynard},
      year={2023},
      eprint={2305.12871},
      archivePrefix={arXiv},
      primaryClass={cs.LG},
      url={https://arxiv.org/abs/2305.12871}, 
}

@misc{casenave2025physicslearningaidatamodelplaid,
      title={{Physics-Learning AI Datamodel (PLAID) datasets: a collection of physics simulations for machine learning}}, 
      author={Fabien Casenave and Xavier Roynard and Brian Staber and William Piat and Michele Alessandro Bucci and Nissrine Akkari and Abbas Kabalan and Xuan Minh Vuong Nguyen and Luca Saverio and Raphaël Carpintero Perez and Anthony Kalaydjian and Samy Fouché and Thierry Gonon and Ghassan Najjar and Emmanuel Menier and Matthieu Nastorg and Giovanni Catalani and Christian Rey},
      year={2025},
      eprint={2505.02974},
      archivePrefix={arXiv},
      primaryClass={cs.LG},
      url={https://arxiv.org/abs/2505.02974}, 
}

@article{RAISSI2019686,
title = {{Physics-informed neural networks: A deep learning framework for solving forward and inverse problems involving nonlinear partial differential equations}},
journal = {Journal of Computational Physics},
volume = {378},
pages = {686-707},
year = {2019},
issn = {0021-9991},
doi = {https://doi.org/10.1016/j.jcp.2018.10.045},
url = {https://www.sciencedirect.com/science/article/pii/S0021999118307125},
author = {M. Raissi and P. Perdikaris and G.E. Karniadakis}
}

@misc{oshea2015introductionconvolutionalneuralnetworks,
      title={{An Introduction to Convolutional Neural Networks}}, 
      author={Keiron O'Shea and Ryan Nash},
      year={2015},
      eprint={1511.08458},
      archivePrefix={arXiv},
      primaryClass={cs.NE},
      url={https://arxiv.org/abs/1511.08458}, 
}

@article{10.1115/1.4067438,
    author = {Xavier Gloerfelt and Paola Cinnella},
    title = {{High-Fidelity Investigation of Vortex Shedding From a Highly Loaded Turbine Blade}},
    journal = {Journal of Turbomachinery},
    volume = {147},
    number = {9},
    pages = {091009},
    year = {2025},
    month = {02},
    issn = {0889-504X},
    doi = {10.1115/1.4067438},
    url = {https://doi.org/10.1115/1.4067438},
    eprint = {https://asmedigitalcollection.asme.org/turbomachinery/article-pdf/147/9/091009/7415679/turbo-24-1257.pdf},
}

@TechReport{jespersen2016,
    author      = {Dennis C. Jespersen and Thomas H. Pulliam and Marissa L. Childs},
    title       = {{OVERFLOW Turbulence Modeling Resource Validation Results}},
    institution = {NASA Ames Research Center},
    number      = {NAS-2016-01},
    year        = {2016}
}

@misc{nasaNASAWallMounted,
	author = {NASA - Clark Pederson},
	title = {2{D} {N}{A}{S}{A} {W}all-{M}ounted {H}ump {S}eparated {F}low {V}alidation {C}ase --- turbmodels.larc.nasa.gov},
	howpublished = {\url{https://turbmodels.larc.nasa.gov/nasahump_val.html}},
	year = {2021},
	note = {[Accessed 12-11-2025]},
}

@article{FANIZZA2025115984,
title = {{Adjoint-based optimization for non-linear inverse problems with high-order discretization of the compressible RANS equations}},
journal = {Applied Mathematical Modelling},
volume = {142},
pages = {115984},
year = {2025},
issn = {0307-904X},
doi = {https://doi.org/10.1016/j.apm.2025.115984},
url = {https://www.sciencedirect.com/science/article/pii/S0307904X25000599},
author = {Bartolomeo Fanizza and Pedro Stefanin Volpiani and Florent Renac and Emeric Martin and Denis Sipp}
}

@article{CATO2023106054,
title = {{Comparison of different data-assimilation approaches to augment RANS turbulence models}},
journal = {Computers \& Fluids},
volume = {266},
pages = {106054},
year = {2023},
issn = {0045-7930},
doi = {https://doi.org/10.1016/j.compfluid.2023.106054},
url = {https://www.sciencedirect.com/science/article/pii/S0045793023002797},
author = {Arthur Shiniti Cato and Pedro Stefanin Volpiani and Vincent Mons and Olivier Marquet and Denis Sipp}
}

@article{doi:10.2514/1.J056397,
author = {Ali Uzun and Mujeeb R. Malik},
title = {{Large-Eddy Simulation of Flow over a Wall-Mounted Hump with Separation and Reattachment}},
journal = {AIAA Journal},
volume = {56},
number = {2},
pages = {715-730},
year = {2018},
doi = {10.2514/1.J056397},
URL = {https://doi.org/10.2514/1.J056397},
eprint = {https://doi.org/10.2514/1.J056397}
}

@inproceedings{baldwin1978thin,
  title={Thin-layer approximation and algebraic model for separated turbulentflows},
  author={Baldwin, Barrett and Lomax, Harvard},
  booktitle={16th aerospace sciences meeting},
  pages={257},
  year={1978}
}

@article{ruder2016overview,
  title={{An overview of gradient descent optimization algorithms}},
  author={Sebastian Ruder},
  journal={arXiv preprint arXiv:1609.04747},
  year={2016}
}

@article{duraisamy2019turbulence,
  title={Turbulence modeling in the age of data},
  author={Duraisamy, Karthik and Iaccarino, Gianluca and Xiao, Heng},
  journal={Annual review of fluid mechanics},
  volume={51},
  number={1},
  pages={357--377},
  year={2019},
  publisher={Annual Reviews}
}

@misc{Girimaji2023,
      title={{Turbulence closure modeling with machine learning approaches: A perspective}}, 
      author={Sharath S. Girimaji},
      year={2023},
      eprint={2312.14902},
      archivePrefix={arXiv},
      primaryClass={physics.flu-dyn},
      url={https://arxiv.org/abs/2312.14902}, 
}

@misc{Sanhueza2022,
      title={{Machine Learning for RANS Turbulence Modelling of Variable Property Flows}}, 
      author={Rafael Diez Sanhueza and Stephan Smit and Jurriaan Peeters and Rene Pecnik},
      year={2022},
      eprint={2210.15384},
      archivePrefix={arXiv},
      primaryClass={physics.flu-dyn},
      url={https://arxiv.org/abs/2210.15384}, 
}

@article{Chen2024,
author = {Zhuo Chen and Jian Deng},
year = {2024},
month = {03},
pages = {},
title = {{Data-driven RANS closures for improving mean field calculation of separated flows}},
volume = {12},
journal = {Frontiers in Physics},
doi = {10.3389/fphy.2024.1347657}
}

@article {Aly2024,
author = {A. M. Aly},
title = {{Deep Learning-Based Eddy Viscosity Modeling for Improved RANS Simulations of Wind Pressures on Bluff Bodies}},
journal = {Journal of Applied Fluid Mechanics},
volume = {17},
number = {12},
pages = {2514-2532},
year  = {2024},
publisher = {},
issn = {1735-3572}, 
eissn = {1735-3645}, 
doi = {10.47176/jafm.17.12.2770},	
url = {https://www.jafmonline.net/article_2512.html},
eprint = {https://www.jafmonline.net/article_2512_f958e5979d404a47b904e6f736c36eed.pdf}
}

@article{Schlichter2024,
title = {{Surrogate model benchmark for $k\omega$-SST RANS turbulence closure coefficients}},
journal = {Journal of Wind Engineering and Industrial Aerodynamics},
volume = {246},
pages = {105678},
year = {2024},
issn = {0167-6105},
doi = {https://doi.org/10.1016/j.jweia.2024.105678},
url = {https://www.sciencedirect.com/science/article/pii/S0167610524000412},
author = {Philipp Schlichter and Michaela Reck and Jutta Pieringer and Thomas Indinger}
}

@misc{Catalani2025,
      title={{Towards scalable surrogate models based on Neural Fields for large scale aerodynamic simulations}}, 
      author={Giovanni Catalani and Jean Fesquet and Xavier Bertrand and Frédéric Tost and Michael Bauerheim and Joseph Morlier},
      year={2025},
      eprint={2505.14704},
      archivePrefix={arXiv},
      primaryClass={physics.flu-dyn},
      url={https://arxiv.org/abs/2505.14704}, 
}

@article{ZHAO2020109413,
title = {{RANS turbulence model development using CFD-driven machine learning}},
journal = {Journal of Computational Physics},
volume = {411},
pages = {109413},
year = {2020},
issn = {0021-9991},
doi = {https://doi.org/10.1016/j.jcp.2020.109413},
url = {https://www.sciencedirect.com/science/article/pii/S002199912030187X},
author = {Yaomin Zhao and Harshal D. Akolekar and Jack Weatheritt and Vittorio Michelassi and Richard D. Sandberg}
}

@misc{Zhu2021,
author = {Linyang Zhu and Weiwei Zhang and Guohua Tu},
year = {2021},
month = {06},
journal = {Advances in Aerodynamics},
title = {{Generalization Enhancement of Artificial Neural Network For Turbulence Closure By Feature Selection}},
doi = {10.21203/rs.3.rs-594074/v2}
}

@article{ZHANG2023108632,
title = {Physical interpretation of neural network-based nonlinear eddy viscosity models},
journal = {Aerospace Science and Technology},
volume = {142},
pages = {108632},
year = {2023},
issn = {1270-9638},
doi = {https://doi.org/10.1016/j.ast.2023.108632},
url = {https://www.sciencedirect.com/science/article/pii/S1270963823005291},
author = {Xin-Lei Zhang and Heng Xiao and Solkeun Jee and Guowei He}
}

@misc{nasaTurbulenceModeling,
	author = {NASA - Clark Pederson},
	title = {{T}urbulence {M}odeling {R}esource --- turbmodels.larc.nasa.gov},
	howpublished = {\url{https://turbmodels.larc.nasa.gov/}},
	year = {2025},
	note = {[Accessed 18-11-2025]},
}

@misc{quattromini2025activelearningdataassimilationclosures,
      title={{Active learning of data-assimilation closures using Graph Neural Networks}}, 
      author={Michele Quattromini and Michele Alessandro Bucci and Stefania Cherubini and Onofrio Semeraro},
      year={2025},
      eprint={2303.03806},
      archivePrefix={arXiv},
      primaryClass={physics.flu-dyn},
      url={https://arxiv.org/abs/2303.03806}, 
}

@misc{quattromini2025meanflowdataassimilation,
      title={{Mean flow data assimilation using physics-constrained Graph Neural Networks}}, 
      author={Michele Quattromini and Michele Alessandro Bucci and Stefania Cherubini and Onofrio Semeraro},
      year={2025},
      eprint={2411.09476},
      archivePrefix={arXiv},
      primaryClass={physics.flu-dyn},
      url={https://arxiv.org/abs/2411.09476}, 
}

@inproceedings{maugars:hal-03759125,
  TITLE = {{Algorithmic Differentiation for an efficient CFD solver}},
  AUTHOR = {Bruno Maugars and S{\'e}bastien Bourasseau and C{\'e}dric Content and Bertrand Michel and B{\'e}renger Berthoul and Jorge Nunez Ramirez and Pascal Raud and Laurent Hasco{\"e}t},
  URL = {https://hal.science/hal-03759125},
  BOOKTITLE = {{ECCOMAS 2022 - 8th European Congress on Computational Methods in Applied Sciences and Engineering}},
  ADDRESS = {Oslo, Norway},
  YEAR = {2022},
  MONTH = Jun,
  HAL_ID = {hal-03759125},
  HAL_VERSION = {v1},
}

@article{Kutz2017,
author = {J. Kutz},
year = {2017},
month = {03},
pages = {1-4},
title = {{Deep learning in fluid dynamics}},
volume = {814},
journal = {Journal of Fluid Mechanics},
doi = {10.1017/jfm.2016.803}
}

@misc{chu2024physicsconstraineddeeplearning,
      title={{Physics Constrained Deep Learning For Turbulence Model Uncertainty Quantification}}, 
      author={Minghan Chu and Weicheng Qian},
      year={2024},
      eprint={2405.16554},
      archivePrefix={arXiv},
      primaryClass={physics.flu-dyn},
      url={https://arxiv.org/abs/2405.16554}, 
}

@misc{nishijima2021universalapproximationtheoremneural,
      title={{Universal Approximation Theorem for Neural Networks}}, 
      author={Takato Nishijima},
      year={2021},
      eprint={2102.10993},
      archivePrefix={arXiv},
      primaryClass={cs.LG},
      url={https://arxiv.org/abs/2102.10993}, 
}

@article{Newton,
author = {Boris Polyak},
year = {2007},
month = {09},
pages = {1086-1096},
title = {Newton’s method and its use in optimization},
volume = {181},
journal = {European Journal of Operational Research},
doi = {10.1016/j.ejor.2005.06.076}
}

@article{Pope_1975,
title={A more general effective-viscosity hypothesis},
volume={72},
DOI={10.1017/S0022112075003382},
number={2},
journal={Journal of Fluid Mechanics},
author={S. B. Pope},
year={1975},
pages={331–340}
}

@misc{deromémont2024datadrivenlearneddiscretizationapproach,
      title={A data-driven learned discretization approach in finite volume schemes for hyperbolic conservation laws and varying boundary conditions}, 
      author={Guillaume de Romémont and Florent Renac and Jorge Nunez and Francisco Chinesta},
      year={2024},
      eprint={2412.07541},
      archivePrefix={arXiv},
      primaryClass={math.NA},
      url={https://arxiv.org/abs/2412.07541}, 
}

@misc{brantner2024volumepreservingtransformerslearningtime,
      title={Volume-Preserving Transformers for Learning Time Series Data with Structure}, 
      author={Benedikt Brantner and Guillaume de Romemont and Michael Kraus and Zeyuan Li},
      year={2024},
      eprint={2312.11166},
      archivePrefix={arXiv},
      primaryClass={math.NA},
      url={https://arxiv.org/abs/2312.11166}, 
}

@misc{deromémont2025datadrivenadaptivegradientrecovery,
      title={Data-Driven Adaptive Gradient Recovery for Unstructured Finite Volume Computations}, 
      author={G. de Romémont and F. Renac and F. Chinesta and J. Nunez and D. Gueyffier},
      year={2025},
      eprint={2507.16571},
      archivePrefix={arXiv},
      primaryClass={math.NA},
      url={https://arxiv.org/abs/2507.16571}, 
}

@article{MUMPS:1,
   title   = {A Fully Asynchronous Multifrontal Solver Using Distributed Dynamic Scheduling},
   author  = {P.R. Amestoy and I. S. Duff and J. Koster and J.-Y. L'Excellent},
   journal = {SIAM Journal on Matrix Analysis and Applications},
   volume  = {23},
   number  = {1},
   year    = {2001},
   pages   = {15-41}
 }

@article{MUMPS:2,
  title = {{Performance and Scalability of the Block Low-Rank Multifrontal
  Factorization on Multicore Architectures}},
  author = {P.R. Amestoy and A. Buttari and J.-Y. L'Excellent and T. Mary},
  journal = {ACM Transactions on Mathematical Software},
  volume = 45,
  issue = 1,
  pages = {2:1--2:26},
  year={2019},
}

@article{GMRes,
author = {Youcef Saad and Martin H. Schultz},
title = {{GMRES: A Generalized Minimal Residual Algorithm for Solving Nonsymmetric Linear Systems}},
journal = {SIAM Journal on Scientific and Statistical Computing},
volume = {7},
number = {3},
pages = {856-869},
year = {1986},
doi = {10.1137/0907058},
URL = {https://doi.org/10.1137/0907058},
eprint = {https://doi.org/10.1137/0907058}
}

@article{PCA,
author = {Karl Pearson},
title = {{LIII. On lines and planes of closest fit to systems of points in space}},
journal = {The London, Edinburgh, and Dublin Philosophical Magazine and Journal of Science},
volume = {2},
number = {11},
pages = {559-572},
year  = {1901},
publisher = {Taylor & Francis},
doi = {10.1080/14786440109462720},
}

@article{Atkinson2025,
doi = {10.21105/joss.07602},
url = {https://doi.org/10.21105/joss.07602},
year = {2025},
publisher = {The Open Journal},
volume = {10},
number = {107},
pages = {7602},
author = {Jack Atkinson and Athena Elafrou and Elliott Kasoar and Joseph G. Wallwork and Thomas Meltzer and Simon Clifford and Dominic Orchard and Chris Edsall},
title = {{FTorch: a library for coupling PyTorch models to Fortran}},
journal = {Journal of Open Source Software}
}

@incollection{lecun:hal-05083427,
  TITLE = {{Convolutional networks for images, speech, and time series}},
  AUTHOR = {Yann LeCun and Yoshua Bengio},
  URL = {https://hal.science/hal-05083427},
  BOOKTITLE = {{The handbook of brain theory and neural networks}},
  PUBLISHER = {MIT Press},
  YEAR = {1998},
  MONTH = Oct,
  DOI = {10.5555/303568.303704},
  HAL_ID = {hal-05083427},
  HAL_VERSION = {v1},
}

@book{bookWilcox,
author = {David Wilcox},
year = {2006},
month = {01},
edition = {3},
title = {{Turbulence Modeling for CFD}},
publisher = {DCW Industries}
}

@misc{berahas2019robustmultibatchlbfgsmethod,
      title={A Robust Multi-Batch L-BFGS Method for Machine Learning}, 
      author={Albert S. Berahas and Martin Takáč},
      year={2019},
      eprint={1707.08552},
      archivePrefix={arXiv},
      primaryClass={math.OC},
      url={https://arxiv.org/abs/1707.08552}, 
}

\begin{appendix}

\section{PyTorch routines}
\label{sec:code}

All routines described in this work are implemented in Python, while the underlying CFD solver, including its numerical schemes, is implemented in Fortran and compiled (e.g., using Intel or \texttt{gfortran} compilers) as Python-callable libraries. These are interfaced with PyTorch via \texttt{ctypes} and wrapped using \texttt{torch.autograd.Function}. This hybrid approach combines the robustness of established numerical solvers with the flexibility and differentiability of modern ML frameworks.

This modular structure allows not only precise control over each component of the solver, but also facilitates generalization to other PDE systems. This strategy, used for the boundary condition layer (\textbf{BC}) in Listing~\ref{lst:Autograd}, can be replicated for other components of a numerical scheme, making the framework extensible and adaptable to a wide range of physics-informed learning tasks.

\begin{lstlisting}[language=Python, basicstyle=\ttfamily\scriptsize, caption=BC Autograd Function., label=lst:Autograd]
class BC(torch.autograd.Function):
    @staticmethod
    def forward(u):
        return bc.f90(u)
    @staticmethod
    def backward(grad):
        return bc_adj.f90(grad)
    @staticmethod
    def jvp(u_lin):
        return bc_lin.f90(u_lin)
\end{lstlisting}

Once the individual computational blocks have been defined and wrapped as differentiable layers, they can be composed into a single \texttt{torch.nn.Module} that encapsulates the entire forward computation. When the backward pass is triggered—either during training or optimization—PyTorch automatically applies the chain rule across the composed operations. This ensures that the gradients of the loss function with respect to the trainable parameters are correctly propagated through each step of the numerical solver.

In the specific case of the RANS equations, the complete differentiable module is constructed by sequentially stacking the custom layers for boundary condition imposition (\textbf{BC}), flux computation (\textbf{NS}), and turbulence forcing (\textbf{FORCE}). The resulting \texttt{torch.nn.Module}, shown in Listing~\ref{lst:Module}, acts as a differentiable surrogate for the physical solver and can be directly integrated into the learning loop.

\begin{lstlisting}[language=Python, basicstyle=\ttfamily\scriptsize, caption=RANS Module., label=lst:Module]
class RANS(nn.Module):
    def __init__(self, params):
        super().__init__()
        self.BC_layer = BC()
        self.NS_layer = NS()
        self.F_layer = F()
        self.parameters = params
    def forward(self, u):
        u_bc = self.BC_layer(u)
        return self.NS_layer(u_bc) + self.F_layer(u_bc, self.parameters)
\end{lstlisting}

Finally, once an iterative solver is selected and the action of the Jacobian of the system can be evaluated (either explicitly or in a matrix-free manner), it becomes possible to define a custom \texttt{torch.autograd.Function} for the entire solution process. In this formulation, the system of governing equations is solved implicitly, and the \texttt{backward} method is overridden to compute gradients via the adjoint system, rather than backpropagating through each individual nonlinear iteration and its associated linear solves. This significantly reduces memory consumption and improves computational efficiency, as it avoids the need to store or differentiate through the full unrolled iteration path.

This concept is closely related to the philosophy behind Deep Equilibrium Models (DEQs)~\citep{Bai2019}, where FP equations are treated as implicit layers and differentiated through using the implicit function theorem. In the present work, this strategy is adopted to compute gradients consistently with the physics-based residual, while maintaining full compatibility with the PyTorch optimization ecosystem. The implemented FP module is shown (in pseudo-code) in Listing~\ref{lst:FPModule}.

\begin{lstlisting}[language=Python, basicstyle=\ttfamily\scriptsize, caption=Newton Solver Module., label=lst:FPModule]
class FIXED_POINT(torch.autograd.Function):
    @staticmethod
    def forward(ctx, x, parameters, config_rans, config_FP):
        with torch.no_grad():
            x_sol, J = NEWTON_solver(x, parameters, config_rans, config_FP)
        ctx.save_for_backward(x_sol, parameters)
        ctx.JT = torch.transpose(J, 0, 1)
        return x_sol
    @staticmethod
    def backward(ctx, grad_output):
        saved = ctx.saved_tensors
        x_sol, parameters = saved
        JT = ctx.JT
        grad_x = ADJOINT_solver(JT, grad_output)
        f = F(x_sol, parameters)
        grads_params = torch.autograd.grad(outputs=f, inputs=parameters, grad_outputs=grad_x)
        return grad_x, grads_params, None, None

\end{lstlisting}

\section{Completeness and sensitivity of the dataset}
\label{sec:complete}

The DOE procedure described in Section~\ref{sec:DATAGen} generated a total of 1500 distinct parameter sets. Each of these sets was used to run a numerical simulation, resulting in 840 successfully converged cases, having defined a tolerance for the value of the RANS residual. To assess the representativeness and coverage of the sampled parameter space, a univariate statistical analysis was performed. Specifically, a histogram was constructed for each parameter, illustrating the distribution of the converged cases within the original design space, as shown in Figure~\ref{fig:converged} for every parameter.

\begin{figure}[ht]
    \begin{subfigure}[b]{0.325\textwidth}
        \centering
        \includegraphics[width=\textwidth]{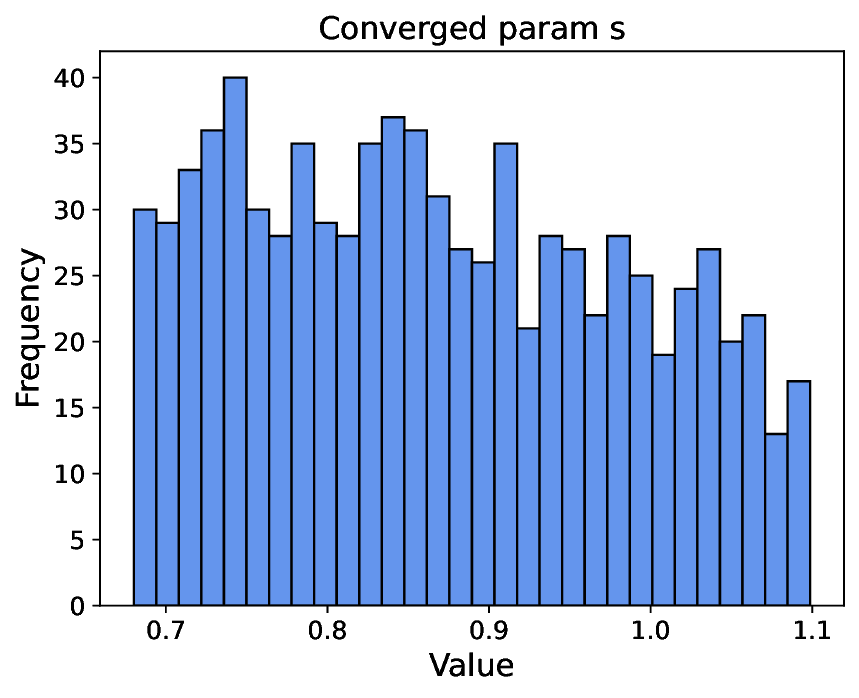}
        \caption{Converged values for the pitch (inter-blade distance).}
        \label{fig:convergeds}
    \end{subfigure}
    \begin{subfigure}[b]{0.325\textwidth}
        \centering
        \includegraphics[width=\textwidth]{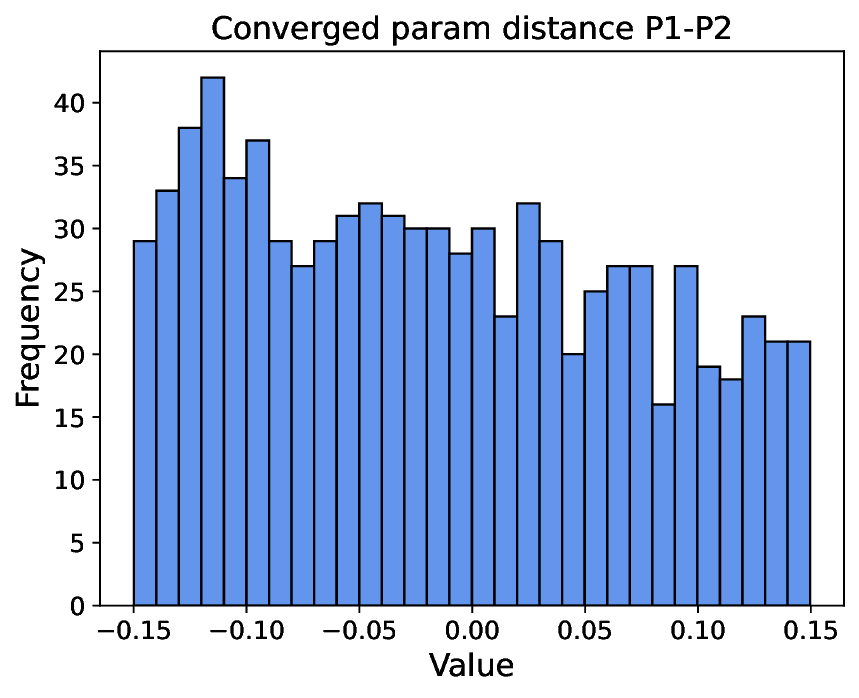}
        \caption{Converged values for the distance between the first 2 control points.}
        \label{fig:convergeddist1-2}
    \end{subfigure}
    \begin{subfigure}[b]{0.325\textwidth}
        \centering
        \includegraphics[width=\textwidth]{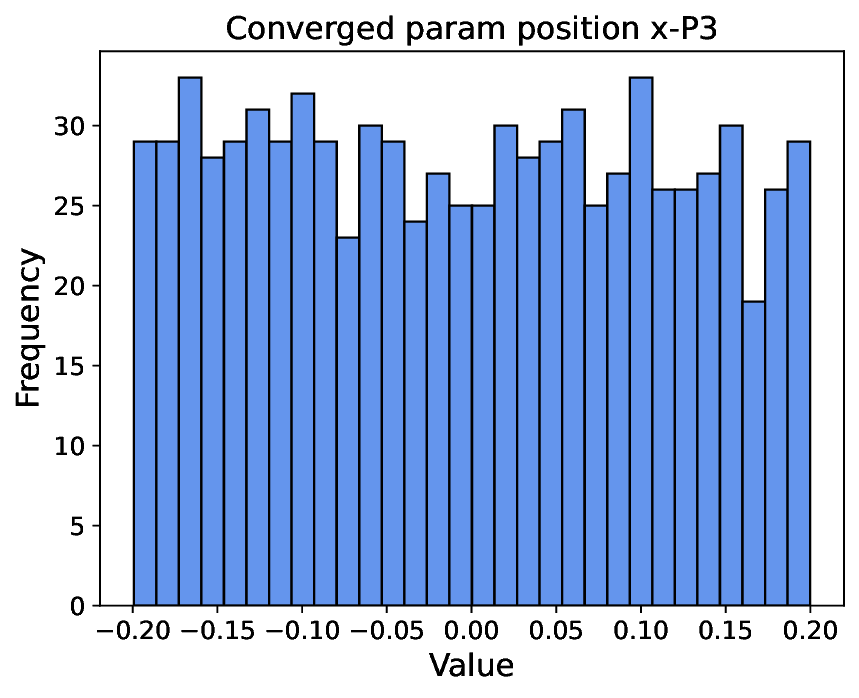}
        \caption{Converged values for the $\Delta x$ of the 3$^{\circ}$ control point.}
        \label{fig:convergedx3}
    \end{subfigure}
    \begin{subfigure}[b]{0.325\textwidth}
        \centering
        \includegraphics[width=\textwidth]{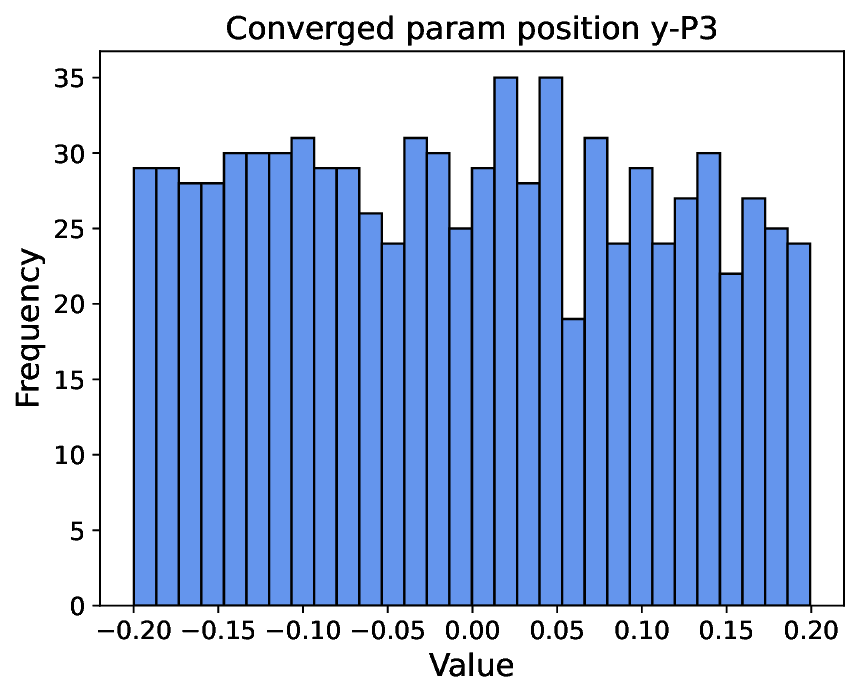}
        \caption{Converged values for the $\Delta y$ of the 3$^{\circ}$ control point.}
        \label{fig:convergedy3}
    \end{subfigure}
    \begin{subfigure}[b]{0.325\textwidth}
        \centering
        \includegraphics[width=\textwidth]{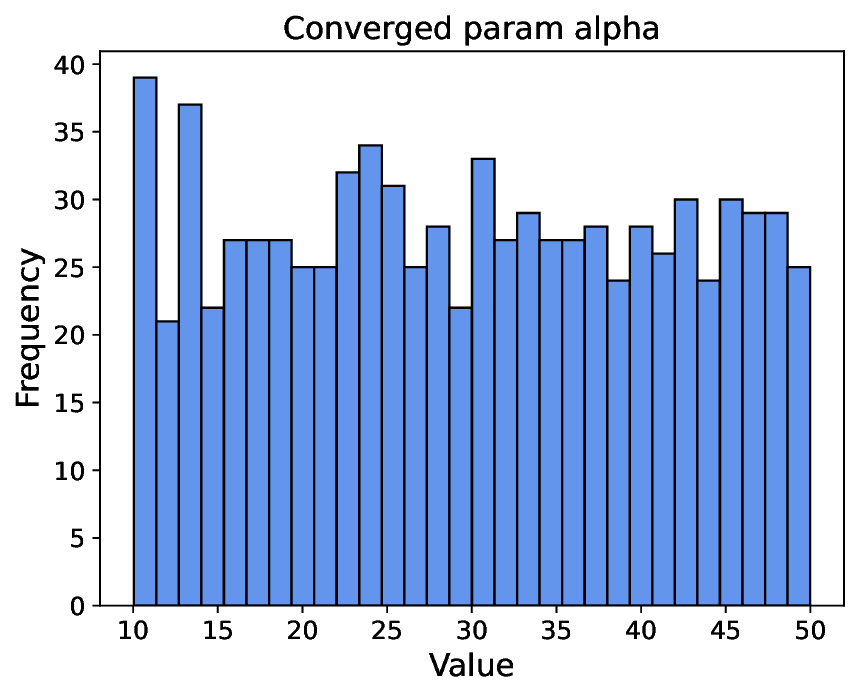}
        \caption{Converged values for the angle of attack.}
        \label{fig:convergedalpha}
    \end{subfigure}
    \begin{subfigure}[b]{0.325\textwidth}
        \centering
        \includegraphics[width=\textwidth]{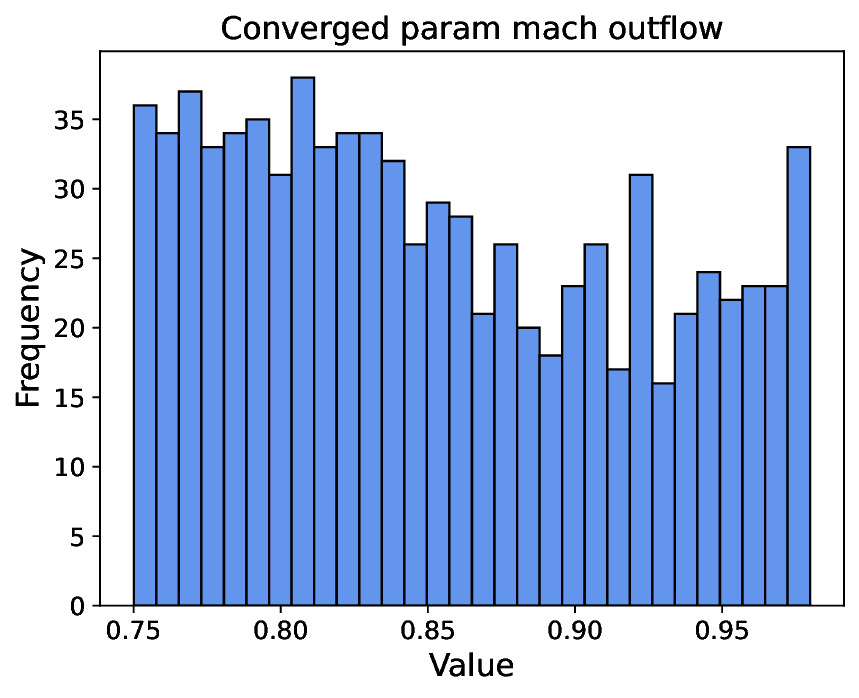}
        \caption{Converged values for the Mach number at outflow.}
        \label{fig:convergedmach}
    \end{subfigure}
    \caption{Histograms of the input parameters for the converged simulations. Despite the removal of some cases due to lack of convergence, the retained simulations remain well distributed across the prescribed parameter ranges, ensuring a representative coverage of the design space.}
    \label{fig:converged}
\end{figure}

To further investigate the sensitivity of the output variables with respect to the input parameters, a variance-based global sensitivity analysis was conducted using Sobol indices. To this end, a dimensionality reduction was applied to both the geometrical data and the physical simulation outputs using Principal Component Analysis (PCA)~\citep{PCA}.

For the geometrical data, consisting of the $x$ and $y$ coordinates of the computational mesh nodes, a PCA was performed to identify the dominant deformation modes. By retaining 99.9\% of the total variance, only 4 principal components were required. This result confirms that the selected geometrical design parameters are sufficient to span the space of observed shape variations.

Similarly, a PCA was applied to the set of conservative state variables (e.g., $\rho$, $\rho u$, $\rho v$, $\rho E$, $\mu_t$) at convergence. In this case, 50 principal components were necessary to capture the same amount of variance, reflecting the higher complexity of the solution fields.

Next, a Gaussian Process (GP) regression model \citep{casenave2023mmgpmeshmorphinggaussian} was trained to approximate the mapping from the reduced input space (comprising the 4 geometrical modes and 2 physical parameters) to the reduced output space defined by the 50 conservative modes. This non-intrusive surrogate model enables efficient evaluation of the model response across the input domain.

Finally, zero-order and first-order Sobol indices were computed using the trained GP model, allowing for a quantitative assessment of the influence of each input parameter on the variance of the output. These indices provide insight into both the individual effects and the interactions among parameters. Figure~\ref{fig:Sobol} shows the Sobol indices for each parameter for the first 4 and last 4 modes.

\begin{figure}[ht]
    \begin{subfigure}[b]{0.49\textwidth}
        \centering
        \includegraphics[width=\textwidth]{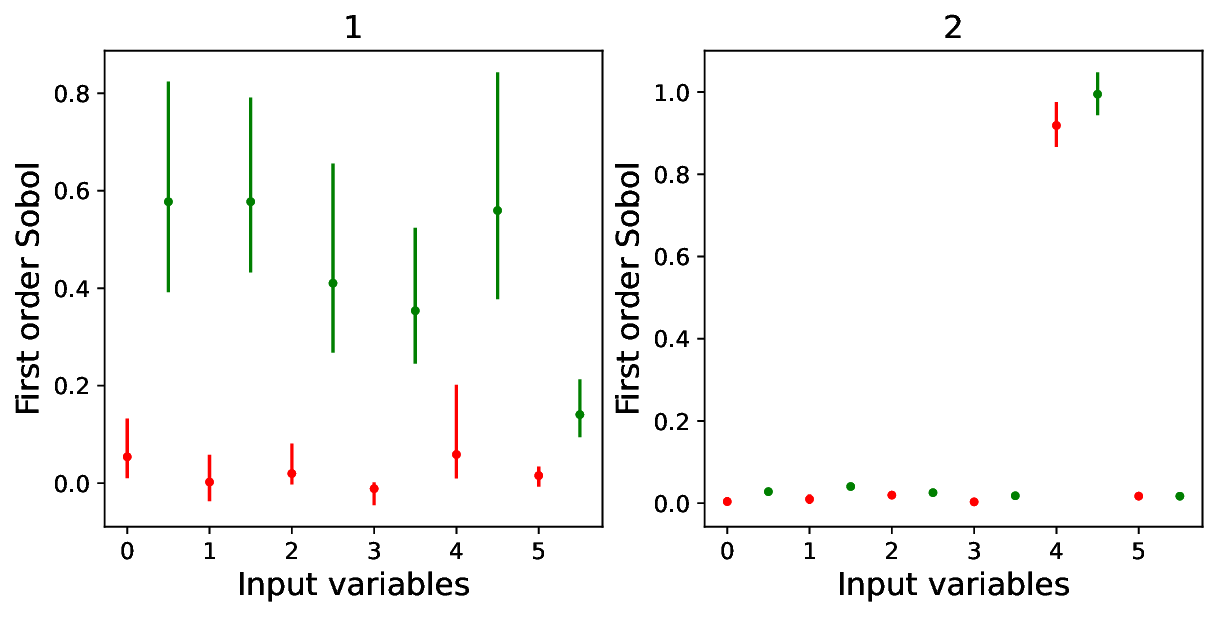}
        \caption{First-order and total Sobol indices for the $1^{st}$ and $2^{nd}$ principal modes.}
        \label{fig:Sobol1}
    \end{subfigure}
    \begin{subfigure}[b]{0.49\textwidth}
        \centering
        \includegraphics[width=\textwidth]{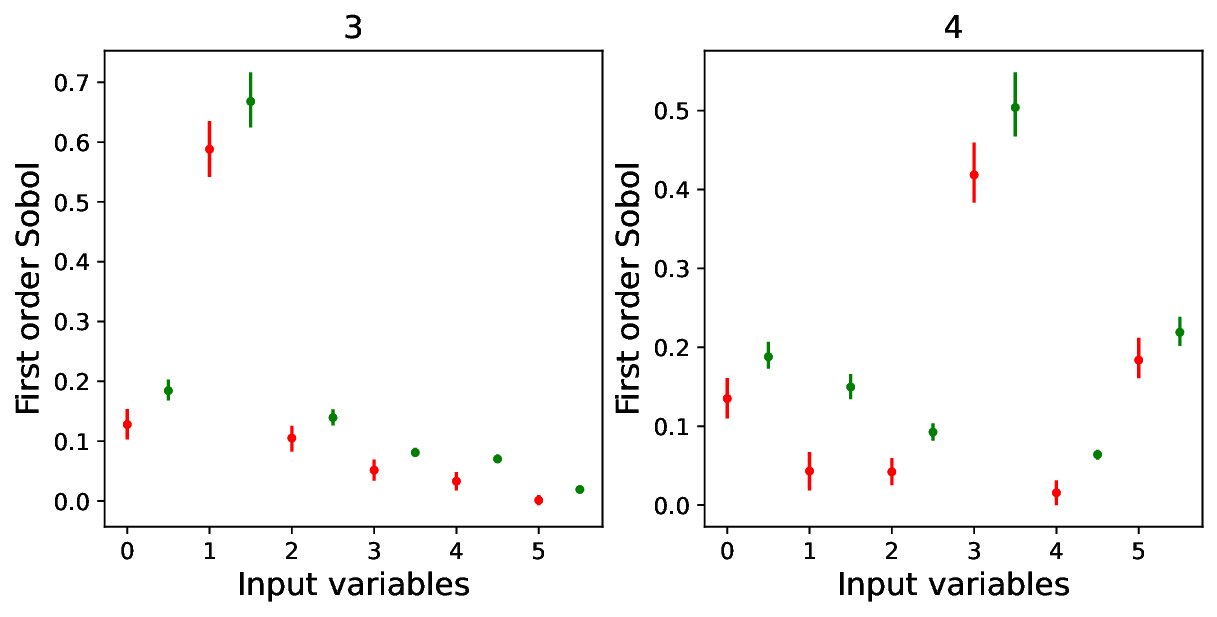}
        \caption{First-order and total Sobol indices for the $3^{rd}$ and $4^{th}$ principal modes.}
        \label{fig:Sobol2}
    \end{subfigure}
    \begin{subfigure}[b]{0.49\textwidth}
        \centering
        \includegraphics[width=\textwidth]{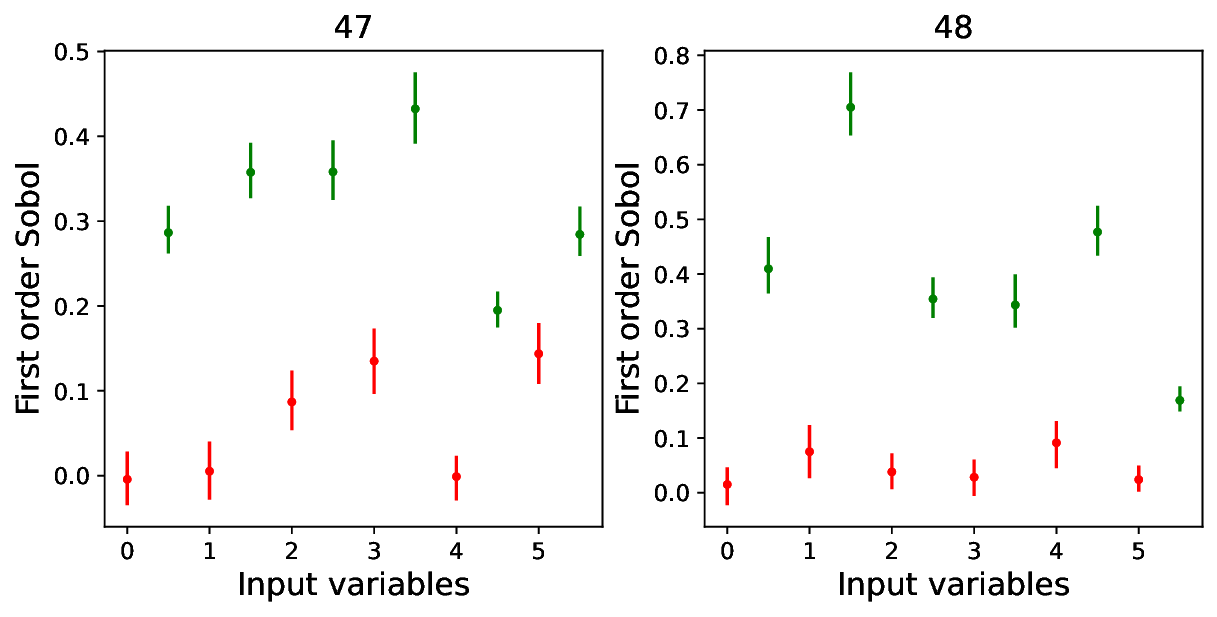}
        \caption{First-order and total Sobol indices for the $47^{th}$ and $48^{th}$ principal modes.}
        \label{fig:Sobol3}
    \end{subfigure}
    \begin{subfigure}[b]{0.49\textwidth}
        \centering
        \includegraphics[width=\textwidth]{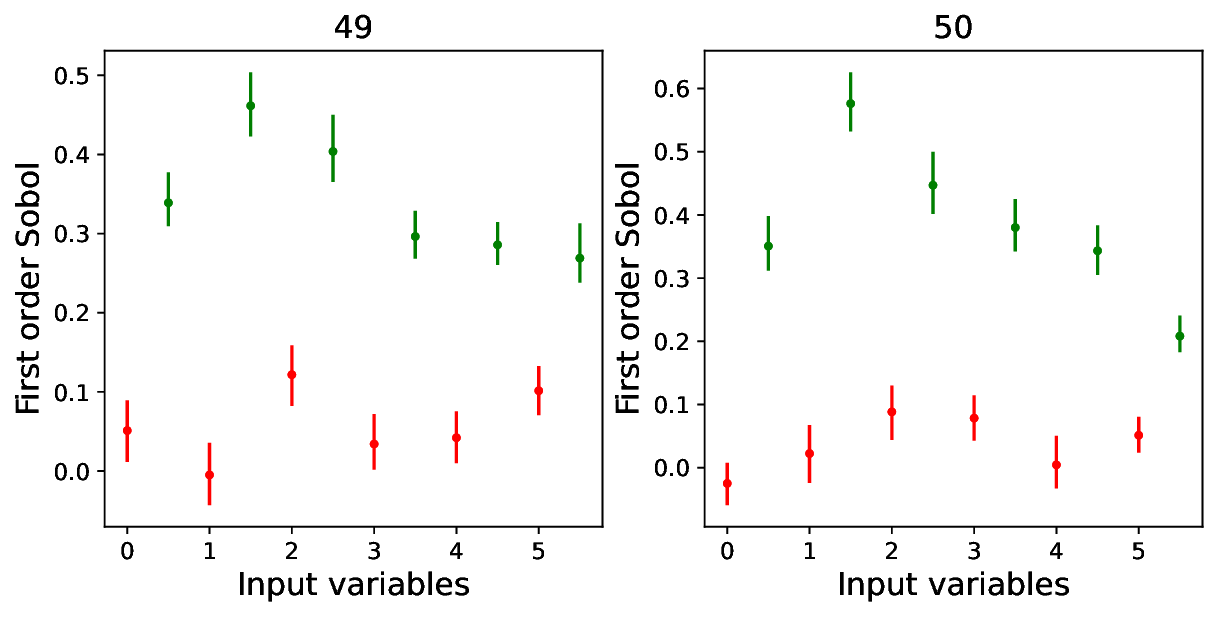}
        \caption{First-order and total Sobol indices for the $49^{th}$ and $50^{th}$ principal modes.}
        \label{fig:Sobol4}
    \end{subfigure}
    \caption{Sobol sensitivity indices for selected principal modes of the conservative variables. In red the $0^{th}$ order indices and in green the $1^{st}$ order. The first four modes capture dominant flow features, while  subsequent modes highlight the contribution of input parameters to higher-order variations.}
    \label{fig:Sobol}
\end{figure}

\section{Computing the gradients for the backward operation in PyTorch}
\label{sec:grads}

Section~\ref{sec:FixedPoint} details how the gradients must be computed to correctly implement the backward operation. In practice, what is done in the code is a change of variable $\tilde{{\bm{\vartheta}}} = \bm{N} {\bm{\vartheta}}$ at the definition of the problem:

\begin{equation}
\begin{cases}
    \min_{\tilde{{\bm{\vartheta}}}} \mathcal{J}(\bm{w}(\tilde{{\bm{\vartheta}}})) \\
    \text{s.t. } \Rd(\bm{w}) + \bm{f}_{\tilde{{\bm{\vartheta}}}}(\bm{w}) = 0
\end{cases}\, \text{, since } {\bm{\vartheta}} = \bm{N}^{-1}\tilde{{\bm{\vartheta}}}.
\end{equation}

We can then directly compute the partial derivative of the Lagrangian with respect to $\tilde{{\bm{\vartheta}}}$ as:

\begin{equation}
    \partial_{\tilde{{\bm{\vartheta}}}} \mathcal{L} \delta \tilde{{\bm{\vartheta}}} = \left(\partial_{\bm{\vartheta}} \mathcal{J} + \bm{\lambda}\T \partial_{\bm{\vartheta}} \bm{f}_{{\bm{\vartheta}}}\right) \bm{N}^{-1} \delta \tilde{{\bm{\vartheta}}} = \delta \mathcal{J}\, ,
\end{equation}
then, since $\bm{N}$ is a symmetric matrix\footnote{Since the mass matrix $\bm{M}$ is diagonal, the associated Cholesky factor $\bm{N}=\sqrt{\bm{M}}$ is diagonal and thus symmetric.}, the differential of the loss function is computed as:

\begin{equation}
    \delta \mathcal{J} = \left< \bm{N}^{-1} \left( \left( \partial_{\bm{\vartheta}} \mathcal{J} \right)\T + \left( \partial_{\bm{\vartheta}} \bm{f}_{{\bm{\vartheta}}} \right)\T \bm{\lambda} \right), \delta \tilde{{\bm{\vartheta}}} \right>_2 \,,
\end{equation}
which is precisely as computed in \eqref{eq:dL}, but with respect to the new variable $\tilde{{\bm{\vartheta}}}$ (see also \eqref{eq:dJN}).

Therefore, if done correctly, PyTorch can compute the gradients automatically. This is achieved by first applying the change of variable in a \texttt{torch.nn.Module}, and then applying the \texttt{torch.autograd.Function}; in this case, the gradients of these operations are concatenated correctly by the compiler, without any further correction by the user. This operation can be observed in Listing~\ref{lst:Changeofvariable}. It is nonetheless essential to define the loss function correctly, as shown in Listing~\ref{lst:Loss}:

\begin{lstlisting}[language=Python, basicstyle=\ttfamily\scriptsize, caption=Loss definition in Python and call to the backward mode., label=lst:Loss]
vol = torch.from_numpy(self.config_deq["vol"])
loss = 0.5 * torch.sum(vol.unsqueeze(-1)*(out-reference)**2)

# vol is a tensor of shape (im, jm),
# while out and reference have shape (im, jm, em)

loss.backward()

\end{lstlisting}

\begin{lstlisting}[language=Python, basicstyle=\ttfamily\scriptsize, caption=Change of variable in the forward pass of the outer \texttt{torch.nn.Module}., label=lst:Changeofvariable]
class RANSLayer(nn.Module):
    def __init__(self, theta_tilde_init, config_rans, config_FP):
        super().__init__()
        self.theta_tilde = theta_tilde_init # parameters
        self.config_rans =  config_rans
        self.config_FP =  config_FP

    def forward(self, x):
        Nm1 = torch.from_numpy(self.config_FP["vol"])**(-0.5)
        theta = Nm1 * self.theta_tilde
        return FIXED_POINT.apply(x, theta, self.config_rans, self.config_FP)

\end{lstlisting}

\section{Architecture of the developed Convolutional Neural Network}
\label{sec:CNN}

The NN employed in this work (see \S~\ref{sec:ResultsDATA}) is designed to approximate the relationship between the mean-flow conservative variables and the turbulent eddy viscosity field, $\mu_t(\bm{w})$. The input is given by the volumes of the discretized mesh and the discrete flow field $\bm{w}$, including ghost cells (see \S~\ref{sec:PyTorchB}), while the output is defined only on the physical cells, consistently with the definition of the closure term.

The architecture, as shown in Table~\ref{tab:cnn_arch}, combines one-dimensional convolutional layers with fully connected (linear) layers and nonlinear activation functions. The convolutional part is used to extract local flow features, while the fully connected layers map these features to the final correction. The choice of one-dimensional convolutions is guided by the structured nature of the mesh and ensures invariance with respect to rotations aligned with the grid directions, as the same convolutional filter is applied along each coordinate direction (as shown in Figure~\ref{fig:CNN}). Furthermore, it avoids the need to define ghost-cell values at corner locations, where boundary conditions from different directions may overlap, thus preventing potential inconsistencies in the treatment of boundary regions.

A key aspect of the design concerns the spatial support of the convolutional kernels. In particular, the kernel size is kept smaller than the stencil width of the underlying numerical scheme. This ensures that the dependence of $\mu_t(\bm{w})$ on the state remains local and does not introduce additional couplings beyond those already present in the discretization. As a result, the sparsity pattern of the Jacobian $\partial_{\bm{w}}\!\left(\bm{R} + \bm{f}_{{\bm{\vartheta}}}\right)$ is preserved, which is crucial for both the consistency and the efficiency of the solver, especially when adjoint-based gradient computations are involved. This aspect becomes even more relevant when sparse linear solvers are used (see \S~\ref{sec:PyTorchB}).

From an implementation standpoint, care is taken to ensure consistency between input and output domains. Ghost cells are included in the input to provide boundary information to the convolutions, but are excluded from the output, so that the predicted eddy viscosity is defined only on physically meaningful cells. This is also done to avoid introducing artificial boundary effects and to maintain compatibility with the numerical solver.

\begin{table}
\centering
\begin{tabular}{lll}
\hline
\textbf{Layer} & \textbf{Type} & \textbf{Specification} \\
\hline
1 & Conv1D & kernel size = 5 \\
2 & Activation & SiLU \\
3 & Conv1D & kernel size = 5 \\
4 & MLP & linear layers with SiLU activation \\
\hline
\end{tabular}
\caption{Architecture of the CNN used in this work.}
\label{tab:cnn_arch}
\end{table}

\begin{figure}
    \centering
    \includegraphics[width=\textwidth]{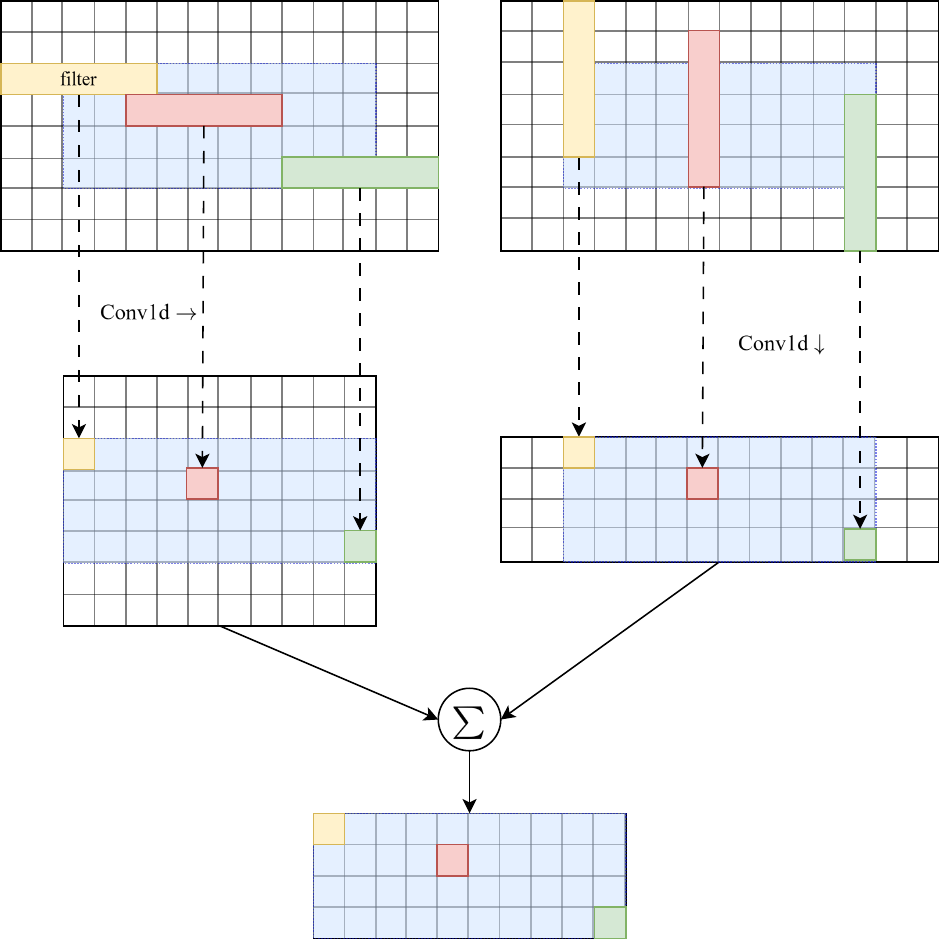}
    \caption{Schematic illustration of the convolutional kernel. The light blue cells denote the physical domain and the white cells the ghost cells. The convolutional kernels (yellow, red, and green) are chosen smaller than the stencil of the underlying numerical scheme, ensuring locality and preventing the introduction of additional couplings. This preserves the sparsity pattern of the Jacobian matrix. Moreover, the kernels applied along the different directions share the same weights (indicated by identical colors), enforcing rotational invariance with respect to the grid. The contributions from the different convolutional directions are then combined in the physical domain.}
    \label{fig:CNN}
\end{figure}

\end{appendix}

\end{Backmatter}

\end{document}